\documentclass[fleqn,usenatbib,twocolumn]{mnras}

\usepackage[T1]{fontenc}
\usepackage{ae,aecompl}

\usepackage{etoolbox}
\makeatletter
\patchcmd\@combinedblfloats{\box\@outputbox}{\unvbox\@outputbox}{}{%
}%
\makeatother

\usepackage{graphicx}	
\usepackage{amsmath}	
\usepackage{amssymb}	
\usepackage{float}
\usepackage{verbatim}
\usepackage{xcolor}
\usepackage{multicol}
\usepackage{epigraph}
\usepackage{tikz}
\usepackage{orcidlink}
\usepackage{newtxtext,newtxmath}
\usepackage{fontawesome5}
\DeclareRobustCommand{\VAN}[3]{#2}
\let\VANthebibliography\thebibliography
\def\thebibliography{\DeclareRobustCommand{\VAN}[3]{##3}\VANthebibliography}

\newcommand{\civ}{\ion{C}{iv}}
\newcommand{\civline}{\ion{C}{iv}~$\lambda$1550}

\newcommand{\nv}{\ion{N}{V}}
\newcommand{\nvline}{\ion{N}{V}~$\lambda$1240}

\newcommand{\thecode}{\textsc{python}}
\newcommand{\cloudy}{\textsc{cloudy}}
\newcommand{\cmfgen}{\textsc{cmfgen}}
\newcommand{\tardis}{\textsc{tardis}}
\newcommand{\xcode}{\textsc{Sirocco}}
\newcommand{\chianti}{\textsc{chianti}}
\defcitealias{shlosman93}{SV93}
\defcitealias{matthews15}{M15}
\defcitealias{knigge95}{KWD95}
\defcitealias{long02}{LK02}

\title
[{\em SIROCCO:} A MCRT Code for Outflows]
{{\em SIROCCO:} A Publicly Available Monte Carlo Ionization 
and Radiative Transfer Code for Astrophysical Outflows
}

\author[J. H. Matthews et al.]
{James~H.~Matthews$^{\orcidlink{0000-0002-3493-7737}}$,$^{1}$\thanks{james.matthews@physics.ox.ac.uk}
Knox S. Long$^{\orcidlink{0000-0002-4134-864X}}$,$^{2,3}$\thanks{long@stsci.edu\newline 
\faGithub \href{https://github.com/sirocco-rt/sirocco}{~Github}\newline 
\faBook \href{https://sirocco-rt.readthedocs.io}{~ReadTheDocs}\newline
\faChartLine \href{https://github.com/sirocco-rt/release-models}{~Plots}
}
Christian Knigge$^{\orcidlink{0000-0002-1116-2553}}$,$^{4}$
Stuart A. Sim$^{\orcidlink{0000-0002-9774-1192}}$,$^{5}$
Edward J. Parkinson$^{\orcidlink{0000-0003-3902-052X}}$,$^{4}$
\newauthor
Nick Higginbottom$^{\orcidlink{0000-0001-7560-4747}}$,$^{4}$
Samuel W. Mangham$^{\orcidlink{0000-0001-7511-5652}}$,$^{4}$
Nicolas Scepi,$^{\orcidlink{0000-0003-3909-2486}}$,$^{6}$
Austen Wallis$^{\orcidlink{0000-0003-0770-9015}}$,$^{4}$
Henrietta A. Hewitt $^{\orcidlink{}}$$^{5}$
\newauthor
Amin Mosallanezhad$^{\orcidlink{0000-0002-4601-7073}}$,$^{4}$
\\$^1$Department of Physics, Astrophysics, University of Oxford, Denys Wilkinson Building, Keble Road, Oxford, OX1 3RH, UK\\
$^{2}$Space Telescope Science Institute, 3700 San Martin Drive, Baltimore, MD, 21218, USA\\
$^{3}$Eureka Scientific Inc., 2542 Delmar Avenue, Suite 100, Oakland, CA, 94602-3017, USA\\
$^{4}$School of Physics \& Astronomy, University of Southampton, Southampton SO17 1BJ, UK\\
$^{5}$School of Mathematics and Physics, Queen's University Belfast, University Road, Belfast, BT7 1NN, UK\\
$^{6}$Univ. Grenoble Alpes, CNRS, IPAG, 38000 Grenoble, France
}

\date{\today}

\pubyear{2024}

\begin{document}
\label{firstpage}
\pagerange{\pageref{firstpage}--\pageref{lastpage}}
\maketitle

\begin{abstract}
Outflows are critical components of many astrophysical systems, including accreting compact binaries and active galactic nuclei (AGN). These outflows can significantly affect a system's evolution and alter its observational appearance by reprocessing the radiation produced by the central engine.  \xcode\  (Simulating Ionization and Radiation in Outflows Created by Compact Objects -- or ``the code formerly known as \emph{Python}'') is a Sobolev-based Monte Carlo ionization and radiative transfer code. It is designed to simulate the spectra produced by any system with an azimuthally-symmetric outflow, from spherical stellar winds to rotating, biconical accretion disc winds. Wind models can either be parametrized or imported, e.g. from hydrodynamical simulations. The radiation sources include an optically thick accretion disc and various central sources with flexible spectra and geometries. The code tracks the ``photon packets'' produced by the sources in any given simulation as they traverse and interact with the wind. The code assumes radiative near-equilibrium, so the thermal and ionization state can be determined iteratively from these interactions. Once the physical properties in the wind have converged, \xcode\ can be used to generate synthetic spectra at a series of observer sightlines. Here, we describe the physical assumptions, operation, performance and limitations of the code. We validate it against \tardis, \cmfgen\ and \cloudy, finding good agreement, and present illustrative synthetic spectra from disc winds in cataclysmic variables, tidal disruption events, AGN and X-ray binaries. \xcode\ is publicly available on {\em GitHub}, alongside its associated data, documentation and sample input files covering a wide range of astrophysical applications.
\end{abstract}



\section{Introduction} \label{sec:intro}

Outflows are ubiquitous among astrophysical systems on all scales. They can be (quasi-)spherical, with or without significant rotation, such as the solar-type winds driven from low-mass stars, the line-driven winds from high-mass stars or the outflows resulting from supernova (SN) explosions. They can also be collimated and/or bipolar, such as the accretion disc winds driven from young stellar objects (YSOs), compact binary systems and active galactic nuclei (AGN).

All of these outflows modify the emergent spectra, sometimes dramatically so. Perhaps the best-known wind signatures are the classic blue-shifted absorption or full-blown P~Cygni profiles seen in the ultraviolet (UV) resonance lines of systems including hot stars \citep[e.g.][]{puls2008, sander2024}, cataclysmic variables \citep[CVs;][]{cordova82,greenstein_rw_1982} and quasars \citep{weymann_comparisons_1991,hewett_frequency_2003}. Similar signatures are sometimes observed in other bands as well, from X-ray (e.g. in X-ray binaries [\citealt{lee2002,miller2004,miller2006a,miller2006b,ponti_ubiquitous_2012}] and AGN [\citealt{pounds_high-velocity_2003,gofford_suzaku_2013}]) to optical/infrared wavelengths (e.g. in YSOs and Wolf-Rayet stars, as well as in some CVs and quasars). However, outflows can also completely change the overall appearance of a system, including its broad-band continuum and emission line spectrum. If the wind absorbs and reprocesses a sufficient amount of the radiation produced by the central engine, it will generate its own continuum and/or line emission, typically mainly via collisional and/or recombination processes. This wind emission can dominate the observed spectrum over significant parts of the observable wavelength range. If the wind is highly optically thick -- as can be the case for outflows from tidal disruption events (TDEs) and SNe, for example -- essentially all of the observable radiation is reprocessed.

In addition to their importance for the interpretation of observations, outflows act as a sink of 
mass, energy and (angular/linear) momentum for the underlying system, and as a source of these quantities for its environment. Both of these functions are astrophysically important. For example, in young stellar objects \citep{konigl2000,zhang2017} and low-luminosity CVs \citep{scepi2019}, accretion may only proceed at all because magnetic disc winds carry away the excess angular momentum of the material injected into the disc. In X-ray binaries \citep[e.g.][]{ponti_ubiquitous_2012,higginbottom19}, TDEs \citep[e.g.][]{Strubbe2009,wu2018,fu2023} and ultraluminous X-ray sources \citep{fabrika2015,middleton2014,middleton2022}, a significant fraction -- or even the majority -- of this material may actually be ejected, so that the accretion rate onto the central object can be much less than that supplied to the outer disc. The wider impact of outflows is perhaps best illustrated by the discs winds and jets driven from AGN and quasars. These provide a mechanism by which SMBHs can interact with their environment on galactic and cluster scales. This ``feedback'' is a critical ingredient in our understanding of galaxy evolution \citep{silk_quasars_1998,king_black_2003,fabian_observational_2012,morganti2017,harrison2018}.

\begin{figure*}
    \centering
	\includegraphics[width=1.0\linewidth]{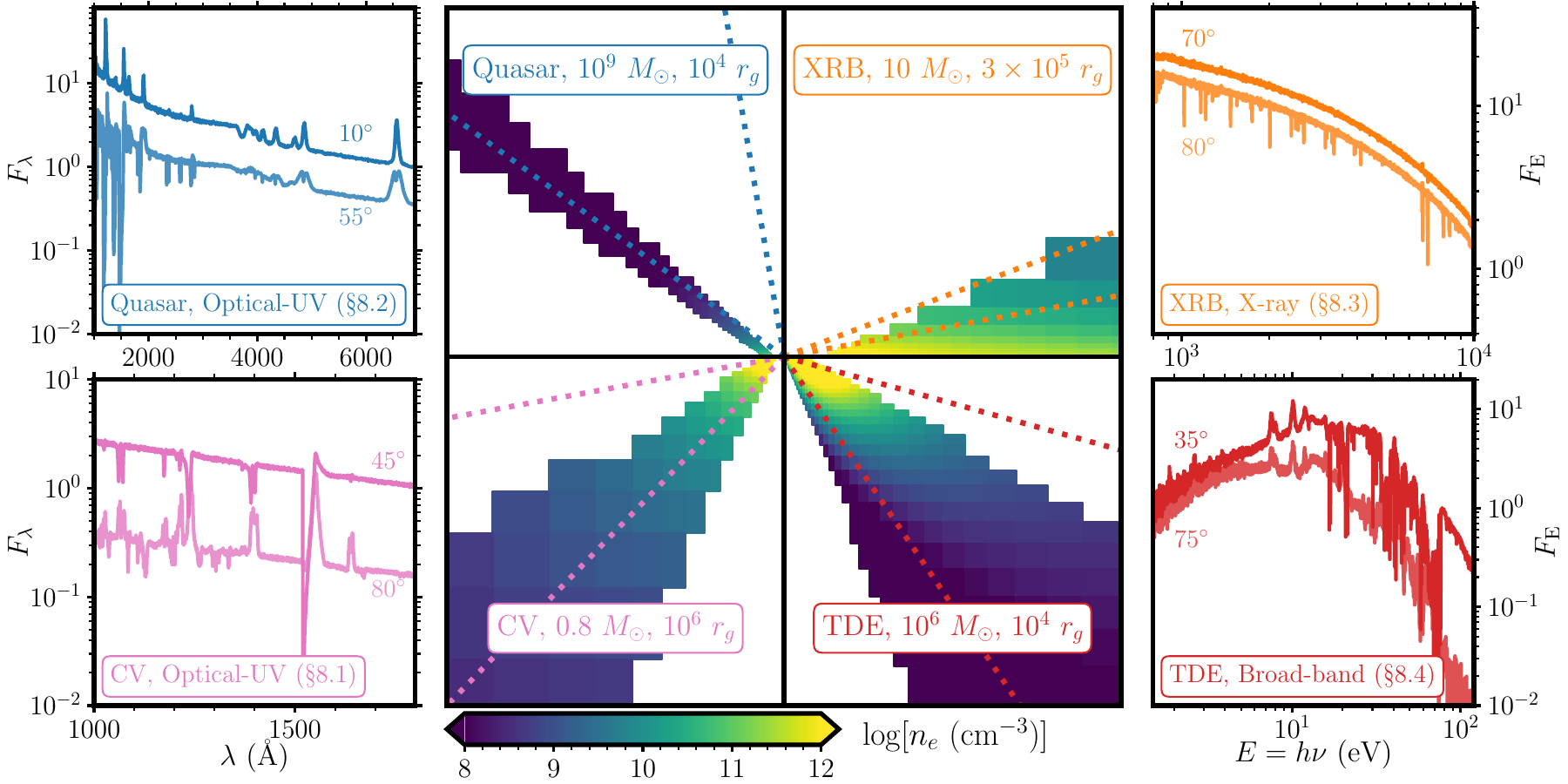}
    \caption[Graphical summary]{
    A graphical summary of the multi-wavelength, multi-scale modelling capabilities of \xcode. We present parametrized models for a quasar, CV, XRB and TDE; the inner four panels show a slice of the wind density structure in linear space, in the $r_{\rm cyl},z$ plane, with two sightlines marked as dotted lines. The white space around the wind is treated as vacuum. In each case, the mass of the compact object, and the size of the plotted domain in $r_g = G M / c^2$, is given in the label. The spectra computed with \xcode\ at these viewing angles from the same wind model are shown in the outer panels, colour-coded accordingly; the model with higher continuum level corresponds to the lower inclination sightline. Further details on each of these illustrative models can be found in section~\ref{sec: demo_models}, with wind parameters given in Table~\ref{tab:demo_model_parameters}.    
    }
    \label{fig:full_demo}
\end{figure*}

\defcitealias{long02}{LK02}
In all of the above astrophysical settings and contexts, the ability to model the effect of the outflow on the emergent spectrum is crucial. More specifically, it is required to reliably infer the physical parameters of both the outflow and the underling central engine. \xcode\ ({\sl Simulating Ionization and Radiation in Outflows Created by Compact Objects}) is a time-independent Monte Carlo ionization and radiative transfer code designed to provide this ability. It was originally developed by \citet[][hereafter \citetalias{long02}]{long02} to model the UV spectra of CVs.
Since then, numerous improvements and extensions have allowed \xcode\  to be used in a wide variety of settings, including  CVs \citep{noebauer10,matthews15}, YSOs \citep{sim05,milliner19}, XRBs \citep{higginbottom17}, TDEs \citep{parkinson2020, parkinson2021} and AGN/quasars \citep{higginbottom13,higginbottom14,matthews16,matthews20,matthews23}. A demonstration of some of these capabilities is shown in Fig.~\ref{fig:full_demo}. The inclusion of light travel time information allows \xcode\ to predict the reverberation signatures of disc winds, such as the velocity-delay maps used to study the structure of the broad-line region (BLR) in AGN \citep{mangham17,mangham19}. The software tools required to create velocity-delay maps for disc wind models are packaged with the standard release of the code. Most recently, \xcode\  has been used in radiation-hydrodynamics simulations that employ {\sc pluto} \citep{mignone_pluto:_2007} as the hydrodynamics module and \xcode\  as the radiation module \citep{higginbottom18,higginbottom19,higginbottom20,higginbottom24}. 
Monte Carlo radiative transfer codes, and spectral synthesis and photoionization codes more generally, are frequently employed in astrophysical modelling. Photoionization codes such as \cloudy\ \citep{ferland_2013_2013,cloudy} and \textsc{xstar} \citep{kallman2001}, are widely used to calculate, usually in 1D, the physical conditions of, and emitted spectrum from, a plasma illuminated by a radiation source. There are also a whole host of Monte Carlo radiative transfer (MCRT) codes with varying limitations, applications and assumptions. For example, in the supernova and transients community MCRT has been used extensively to model SN spectra; for example, by \citep{mazzali93}, and with the codes \textsc{sedona} \citep{kasen2006}, \tardis\ \citep{kerzendorf_spectral_2014}, \textsc{artis} \citep{kromer09,shingles2020}) and \textsc{jekyll} \citep{ergon_monte-carlo_2018}. These methods can even be used to constrain the presence of heavy elements in kilonovae \citep{tanaka2013,gillanders2022}. Other notable MCRT codes, typically used to model outflows, nebulae or \ion{H}{ii} regions composed of gas (and often dust) include \textsc{Torus} \citep{harries2019}, \textsc{Mocassin} \citep{ercolano_mocassin_2003,ercolano_dusty_2005}, \textsc{Skirt}  \citep{camps_skirt_2015} and \textsc{Hyperion} \citep{robitaille_hyperion_2011}. MCRT techniques have also been applied in much stronger gravity regimes, for example in \textsc{GR-Monty} \citep{dolence_grmonty_2009}. \xcode\ is also not the only program used for multidimensional MCRT modelling of accretion disc winds; there are a number of other codes that have been applied to AGN \citep{sim_modelling_2005,sim_multidimensional_2008,sim_multidimensional_2010,hagino_origin_2015}, CVs \citep{kusterer2014}  and T Tauri stars \citep{kurosawa2011}. Overall, MCRT is a powerful technique, and there is now a relatively mature -- albeit, inevitably, somewhat heterogeneous -- suite of codes which can be used to synthesize spectra from astrophysical systems \citep[see review by][]{noebauer19}.

At the time Long and Knigge first named their radiative transfer code \thecode, we were not aware of the programming language of the same name, and we were certainly not aware of how dominant it would become. To reduce the confusion associated with the name, we are taking the opportunity of this updated description to rename the code \xcode.

The purpose of the present paper is to provide an current description of \xcode\ (the code formerly known as \thecode) and release it as an open-source tool available to the astrophysical community. Section~\ref{sec:overview} provides a high-level overview of the code. In Section~\ref{sec:rt}, we describe the radiative transfer methods including a discussion of how our assumptions of radiative, thermal and statistical equilibria are actually enforced. We describe the radiation sources and generation of MC representations of radiation in section~\ref{sec:radiation_sources}. The detailed code operation, including both the iterative calculation of a converged plasma state and the creation of the detailed spectrum is described in section~\ref{sec:operation}.
Section~\ref{sec:wind} presents the main parametrized kinematic wind models built into the code and provides some information on how one can use imported models from, e.g. hydrodynamic simulations. 
In section~\ref{sec:tests}, we validate \xcode\ against {\sc cmfgen} \citep{hillier98} and \tardis\ \citep{kerzendorf_spectral_2014}, for a hot stellar wind and supernova model, respectively, as well as presenting a comparison to ionization calculations carried out with \cloudy\ \cite{ferland_2013_2013}. In section~\ref{sec: demo_models}, we illustrate the capabilities of the code for a wide range of astrophysical systems (including CVs, XRBs, TDEs and AGN/quasars) and applications (including the generation of synthetic X-ray, UV and optical spectra). In section~\ref{sec:limitations}, we describe various limitations and caveats, before concluding by summarizing \xcode's capabilities in section~\ref{sec:conclusions}.
\begin{figure*}
    \centering
	\includegraphics[width=0.9\linewidth]{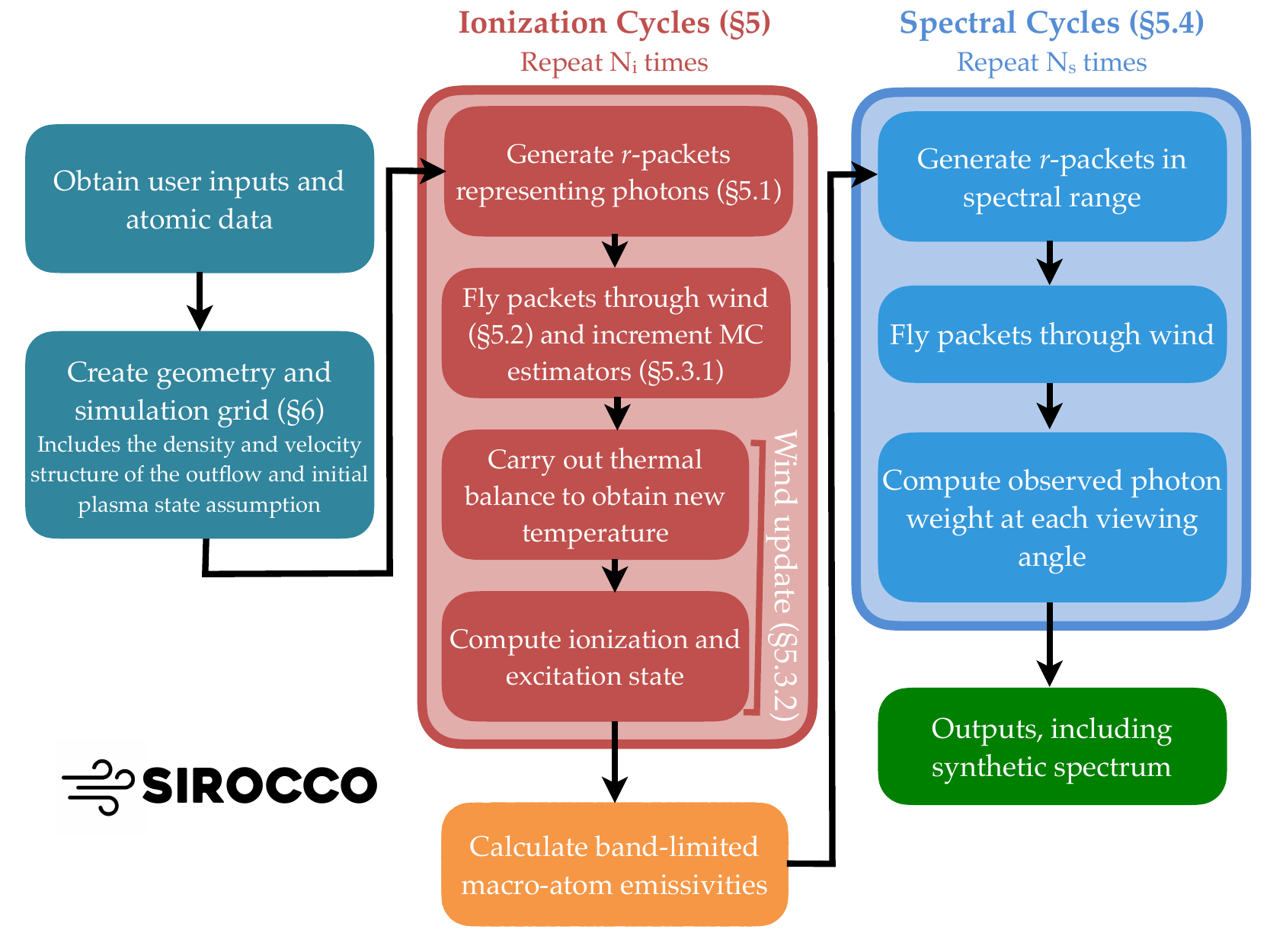}
    \caption[Flowchart]{
    A flowchart showing the basic operation of \xcode.  A high-level code overview is provided in section~\ref{sec:overview}, with a more detailed description of code operation given in section~\ref{sec:operation}. Where applicable, the relevant section number for each step of the flowchart is given in parentheses.}
    \label{fig:flowchart}
\end{figure*}

\section{Code Overview}
\label{sec:overview}
\xcode\ is a Monte Carlo code designed to calculate the ionization structure and emergent spectrum of a wind illuminated by radiation sources, typically a central object and an accretion disc (and, depending on how the radiation transfer is being calculated, the wind itself). The density and velocity structure of the wind, the nature of the emission from the radiation sources and an initial guess at the temperature of the wind are inputs to any calculation.  Once setup is complete, the calculation proceeds in two stages, as shown in Fig.\  \ref{fig:flowchart}: one to calculate the ionization, excitation and temperature structure of the wind ({\em Ionization cycles}; section~\ref{sec:ionization_cycles}), and one to calculate detailed spectra over the wavelength range of interest and for a specific set of viewing angles ({\em Spectral cycles}; section~\ref{sec:spectrum}).  

The ionization and thermal structure of the wind is calculated iteratively during the ionization cycles. During each cycle, a large number of photon bundles (typically $\gtrsim 10^6$) are generated which are allowed to traverse and interact with the wind through various processes, such as electron scattering, line scattering and photo-absorption.  The frequency distribution during the ionization cycles is broad, typically covering $10^{14}-10^{20}~{\rm Hz}$ (approximately $0.41$\,eV to $410$\,keV), reflecting the spectral range where the bulk of the radiated energy emerges in our chosen applications. During the interactions, various MC estimators are incremented, which are used as estimates of all relevant rates that depend on the mean intensity (e.g. the rates of photoionization, radiative heating, radiative excitation). After the photons have traversed the wind, the density of all of the ions in the calculation, their level populations, and the electron temperature of the wind are all updated to reflect the radiation field in each region of the wind. At that point, a new set of photons is generated, and the process repeats for a user-specified number of cycles, when, hopefully, the system has converged to an approximately constant plasma state in which temperatures and ion abundances are not changing significantly from cycle to cycle (see sections~\ref{sec:converge} and \ref{sec:converge_cv}).

Once the ionization calculation is complete, the state of the wind plasma is then fixed, and additional sets of photon bundles are generated and allowed to traverse the wind to compute spectra from different viewing angles. Usually, one is concerned with the spectrum within a narrow range, so here the wavelength range of the photons generated is restricted, although any range can be adopted. During this portion of the calculation, whenever a photon packet is generated or interacts inside the outflow, an appropriately reweighted version of the packet is extracted for each user-specified observer direction; this process is described in section~\ref{sec:spectrum}. Although we carry out the generation of the detailed spectrum in cycles, this is a computational convenience; each cycle contributes to the same spectrum, but the noise decreases as the spectra from each cycle build up over time.

\subsection{Key assumptions and concepts}
As noted above, \xcode\ iterates towards a converged plasma state in which the temperature and ionization state in each cell are no longer changing. This state is the solution for which spectra are then generated. The key assumption underlying this approach is that the entire flow is, {\em locally}, in both thermal and statistical equilibrium: the rates at which energy flows into and out of each cell should match, as should the transition rates into and out of any given atomic/ionic level. Moreover, {\em globally}, \xcode\ assumes that the external radiation sources that illuminate the flow are the {\em net} sources of energy. That is, the code assumes that the flow is in {\em radiative equilibrium}\footnote{Strictly speaking, this is not completely true. \xcode\ can account for adiabatic heating and/or cooling, i.e. the $PdV$ work associated with the divergence of the velocity field (see section~\ref{sec:sinks}). In principle, other non-radiative heating and/or cooling terms -- such as shock heating or magnetic dissipation -- could also be included via sub-grid recipes.}. \xcode's two main modes of operation, described in the next section, differ essentially in {\em how} radiative equilibrium is enforced.

\subsection{Notation and terminology}
Throughout, we follow the terminology of \cite{lucy_monte_2002} in using the phrase $r$-packets to refer to quanta of radiative energy and $k$-packets for quanta of kinetic (thermal or internal) energy, although we sometimes use the term photon more loosely or generically. The energies carried by these packets are often also referred to as weights, with the symbol $\epsilon$. \xcode\ is a grid code, meaning that the density and velocity fields are discretized into a series of volume elements which we refer to as {\em cells} throughout (see section~\ref{sec:wind} for more details). Each cell is defined by its location, and stores its own set of plasma conditions (temperatures, ion abundances, etc.) and Monte Carlo estimators recording during radiation transport. We use subscripted ${\cal N}$ to refer to numbers of quantities within the code (such as cycles, or $r$-packets), $N_i$ for ion densities, $n_e$ for electron density and $n_i$ for level populations. We use ${\cal H}$ and ${\cal C}$ to denote heating and cooling rates and $C$ for collisional rate coefficients. We use centimetre-gram-second units within both the code and this paper. 

\section{Radiative Transfer and Physical Processes}
\label{sec:rt}

\begin{figure}
    \centering
	\includegraphics[width=\linewidth]{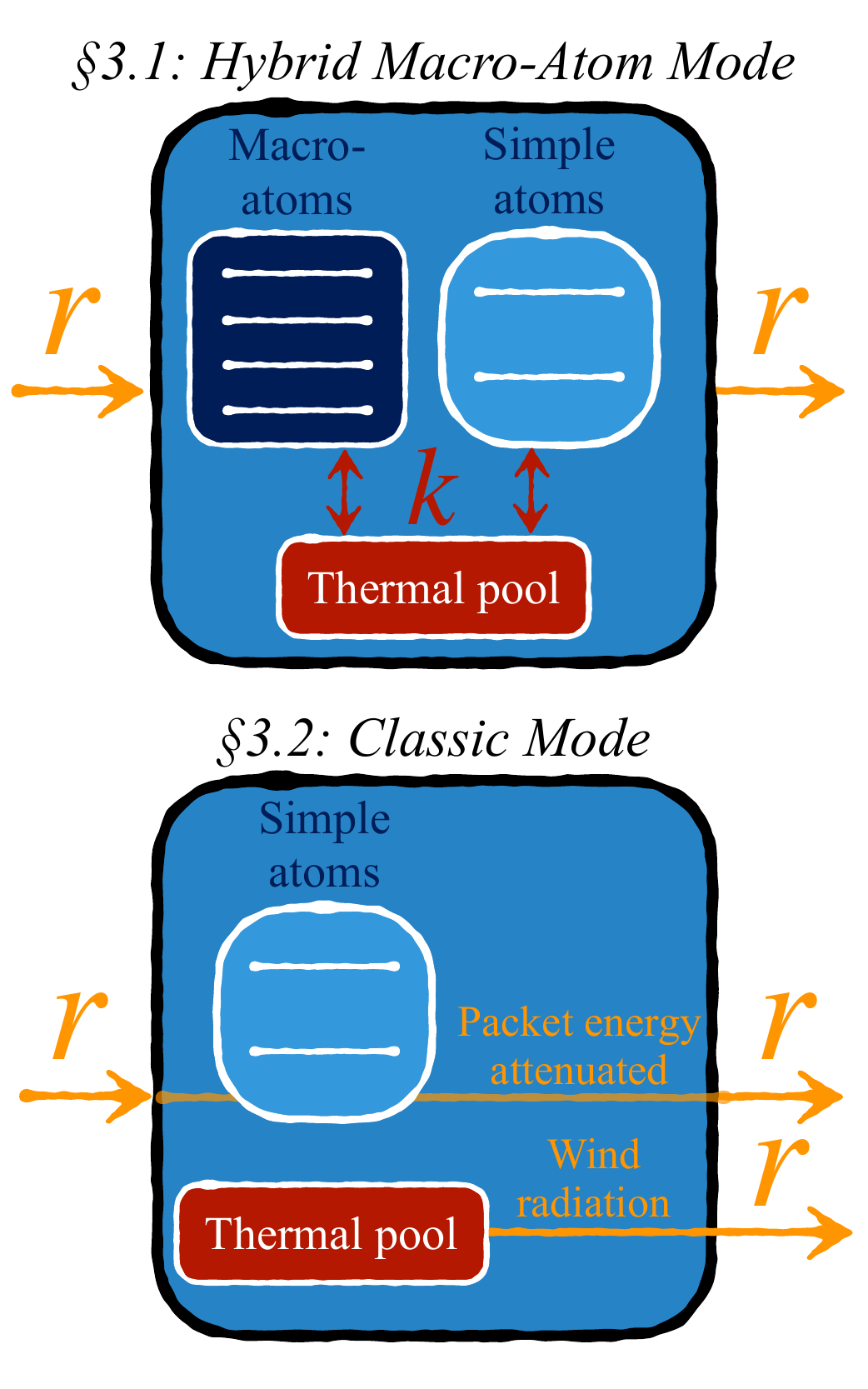}
    \caption[Radiative equilibrium schematic]{
    A schematic illustrating the two principal modes of operation of \xcode, showing how the two modes differ in the enforcement of radiative equilibrium and energy conservation.
    }
    \label{fig:rad_eq}
\end{figure}

\xcode\ has two modes of operation when it comes to radiative transfer: the first is the {\em hybrid macro-atom} mode, which is usually the more physically accurate; the second is the {\em classic} mode, which has been extensively used by our collaboration in the past and is still the more appropriate choice for certain applications. We will use this language throughout in describing the two modes of operation. Although both of these modes are designed with the same fundamental physical principles in mind, they differ firstly in how they enforce energy conservation and radiative equilibrium, and secondly in how they handle line transfer and the model atom. Below, we describe these two approaches in more detail and provide a schematic diagram in Fig.~\ref{fig:rad_eq}; we refer the reader to \cite{noebauer19} for a more complete review of different MCRT procedures. 

\subsection{Hybrid Macro-atom mode}
As recently reviewed by \cite{noebauer19}, most early MC radiative transfer codes followed photon packets through a structure and summed the photons that escaped  to obtain a simulated spectrum. As these packets proceeded through the wind, the energy or weight represented by the packet was reduced due to the effects of absorption. 

However, as was realized by \cite{abbott85} and \cite{lucy_computing_1999}, treating the quanta that are transported in a MC radiative transfer code as indivisible packets of co-moving frame energy -- rather than allowing the energy represented to decrease -- has significant advantages; not least of these advantages is that, throughout the calculation, it rigorously imposes (and therefore ensures) co-moving frame energy conservation at the point of interaction. Thus whenever a quantum of radiant energy (an $r$-packet) deposits energy into the wind during an interaction, an immediate (probabilistic) decision is made that determines how this energy reemerges. 

As an example, consider an $r$-packet undergoing a bound-free interaction. In a real photoionization event, some photon energy goes into liberating the electron from its atomic orbital (with threshold energy $h \nu_0$). The remainder, $h (\nu-\nu_0)$, goes into the kinetic energy of the electron, which ultimately shares this energy with the ``thermal pool'' of the plasma. The liberation of the electron represents an energy flow into the energy reservoir associated with the internal excitation/ionization energy of the species. 
In a steady state, this must be matched by an equal radiative energy flow out of the same excitation/ionization energy pool (e.g. via bound-bound or bound-free emission). 
Similarly, the kinetic energy of the liberated electron represents an energy flow into the thermal pool, and this must be matched by an equal radiative energy flow out of the thermal pool (via any allowed thermal emission process). 

In indivisible packet approaches, these energy flows are represented statistically. Thus each time there is a photoionization interaction, a choice is made to either (i) create a new $k$-packet (which represents kinetic or thermal energy) or (ii) interact with the relevant ion. These choices have relative probabilities $\nu_0/\nu$ and $(\nu-\nu_0)/\nu$, respectively. Eventually, a new $r$-packet is generated with the same co-moving frame energy, but not necessarily the same frequency, as before the interaction.

The vehicle for handling the interaction between packets of radiant energy and the atoms/ions in the gas is a construct called a {\em macro-atom} \citep{lucy_monte_2002,lucy_monte_2003}. Its purpose is to ensure that the energy flows through the ``excitation pool'' of the plasma are handled correctly. Within this framework, each atom is represented by a series of bound levels for all of the ions of a particular element. The levels are connected to one another by a series of transition probabilities, both radiative and collisional.  When a macro-atom is excited, the probability that it transitions to another state depends on the atomic properties of all connected states, on the current occupation numbers for the various levels of the element and on other local parameters of the plasma, such as the electron density and temperature. The levels within a macro-atom and the transition probabilities between them bear close analogy to real atomic energy levels and radiative/collisional transition rates. However, in the macro-atom formalism, the sampling of the transition probabilities is designed to model the energy flows through the macroscopic system. As shown by \cite{lucy_monte_2002}, this approach asymptotically reproduces the correct emissivity\footnote{The phrase ``asymptotically reproduces'' undersells the advantages of indivisible packet and macro-atom schemes. \cite{lucy_monte_2002} showed that the macro-atom scheme is quite insensitive to errors in level populations and can get remarkably close to the correct emissivities even before the calculation has converged (see his section 6). This property -- which ultimately comes from the enforcement of energy conservation and statistical equilibrium at the interaction point -- helps with the convergence rate, but also means a fairly reliable result can often be obtained even if the calculation is not fully self-consistent.} of the plasma and also allows one to obtain an accurate frequency distribution of the radiated energy, including all the important bound-free and bound-bound processes. 

The initial implementation of the macro-atom within \xcode\ was carried out by \cite{sim_modelling_2005}, with the hybrid macro-atom scheme originally described by \cite[][hereafter \citetalias{matthews15}]{matthews15}. The ``hybrid'' aspect of this scheme comes from the fact that \xcode\ can operate with some ions being treated as multi-level macro-atoms, and some within a simplified two-level atom framework; we use the phrase {\sl simple atoms}  to refer to the latter. Simple atoms do have associated levels, lines and photoionization cross-sections, but line transfer is carried out with a  two-level approximation, and recombination is modelled less accurately. In our modelling work to date, we have typically treated H and He as full macro-atoms and the metals as simple atoms. The latter choice is partly for convenience. In the past, there have been significant performance hits from using macro-atoms, particularly in dense regions, although this is less of a problem now due to the implementation of efficient Markov chain techniques. Nevertheless, developing full macro-atom data sets takes time and requires high-quality atomic data. A further limitation is that including macro-atoms can be rather memory intensive, because many Monte Carlo estimators must be stored for each grid cell. 

\subsection{``Classic'' mode}

In the original \cite{long02} code, we did not make use of indivisible packets, $k$-packets or macro-atoms. Instead, we used a scheme in which photon packets could lose energy. In this scheme, all atoms were treated as simple atoms, so that bound-bound interactions were only treated via a two-level atom and escape probability formalism. We have continued to use this mode of operation -- which we refer to as {\em classic mode} -- regularly over the years. In this approach, electron scattering and bound-bound interactions are treated as scattering processes, but bound-free and free-free interactions are treated as true absorption processes. As a result, as photons traverse the wind, some of their energy is absorbed and their weights are reduced. (For sufficiently high energy photons, electron scattering can also cause energy transfer from an $r$-packet to the plasma via the Compton effect.) For the calculation to be self-consistent, the  energy absorbed by the wind needs to be reradiated. Thus, in classic mode, the wind itself becomes a radiation source -- albeit, in a converged model, not a {\em net} one -- along with the external radiation sources, such as the star and the disc. 

The hybrid macro-atom mode is, in principle, more physically accurate than the classic mode. In particular, it is required to ensure adequate energy conservation in highly optically thick environments and/or whenever accurate recombination line fluxes are needed (e.g. for the Balmer series of H). The classic approach is typically sufficient when one is interested mainly in permitted lines linked to the ground-state, such as \ion{C}{iv}~1550~\AA. Even for elements that are {\em not} modelled as full macro-atoms, the ionization state is still calculated accurately (see section~\ref{sec:cloudy}). Differences between the two methods will be highlighted in the sections that follow.

\subsection{Radiative processes and interactions}
\xcode\ includes a number of different radiative processes, each of which constitutes a source of opacity and emissivity within the wind plasma. We do not provide an exhaustive list of equations here \cite[instead see e.g.][]{matthews_phd_2016}, but do provide a brief overview of the implementation of each of these processes within each mode of operation. Further discussion of the radiative transfer, rate estimators and ionization/thermal effects associated with these processes can be found in sections~\ref{sec:photon_transport}, \ref{sec:estimators} and \ref{sec:ionization}, respectively. 

\subsubsection{Bound-bound interactions (line transfer)}

Line transfer is calculated assuming the Sobolev approximation \citep{sobolev_diffusion_1957,sobolev_moving_1960,rybicki83}. No allowance is made for natural or thermal line broadening; this simplifies the identification of the Sobolev surfaces.  Collisional processes are included, using where possible tabulations of collisional cross sections from the \chianti\ database \citep{dere97,dere23}. If detailed collisional data is absent, the \cite{vanregemorter62} approximation is used. Photons can be reemitted isotropically, or more usually, preferentially in the direction of lowest optical depth in the line. The latter is achieved by Monte Carlo sampling of the direction-dependent local Sobolev optical depth. Optical depths of resonance lines can be large, and hence we use an escape probability formalism \cite[see, \citetalias{long02} and][]{rybicki83} when calculating relative radiative and collisional deexcitation rates. For example, the radiative deexcitation rate from level $u\to l$ is given by $R_{ul} = \beta_{ul} A_{ul} n_u$, with $\beta_{ul}$ the usual (angle-averaged) Sobolev escape probability for that transition, and $A_{ul}$ the corresponding Einstein coefficient. Similarly, the effective bound-bound radiative absorption rate from $l\to u$ is given by $R_{ul} = B_{lu} n_l J_{\rm est}$. Here $J_{\rm est}$ is the MC estimator for the mean intensity given by equation 4 of \cite{matthews15}, which includes the influence of direction-dependence in the Sobolev optical depth.

Bound-bound interactions of $r$-packets with macro-atom lines result in the activation of a macro-atom state, denoted $r \to {\cal A}^*$, with the level of activation corresponding to the upper level of the line transition. The Monte Carlo transition probabilities are then sampled -- either by a Monte Carlo `jumping' procedure \citep{lucy_monte_2002} or using a faster matrix scheme \citep{ergon_monte-carlo_2018} -- to determine the deactivation process. This is not strictly confined to bound-bound interactions, as the continuum state can interact with the bound levels within the macro-atom. If deactivation occurs via a bound-bound transition, a choice is made whether to create an $r$- or $k$-packet based on the relative collisional ($C_{ul}$) and radiative ($R_{ul}$) deexcitation rates. 

For simple atoms, line transfer is treated in the two-level approximation, allowing for the possibility of collisional deexcitation. In the classic mode of operation, the weight (i.e. energy) carried by an $r$-packets is typically reduced during each interactions. The remainder of the weight goes into the thermal pool, i.e. it heats the plasma. In hybrid macro-atom mode, $r$-packets are indivisible, i.e. their weight remains unchanged before and after each interactions. Instead, a probabilistic choice is made to create either a $k$-packet or an $r$-packet, based on the collisional and radiative rates.  Upper level populations are calculated using an on-the-spot approximation (as discussed in detail by \citetalias{long02}). 
For simple atoms -- i.e. non macro-atoms -- most of our included line transitions are from the ground state, but we do model some transitions from excited configurations. In this case, the lower level population is also calculated using an on-the-spot approximation. 

\subsubsection{Photoionization and recombination}

As an $r$-packet propagates through the plasma, \xcode\ keeps track of the cumulative optical depth associated with all of the photoionization opacities the packet encounters. 
In hybrid macro-atom mode, each photoionization event is immediately followed by the activation of a macro-atom or the generation of a $k$-packet. The probability of these outcomes are calculated from the corresponding energy flow rates. In this mode, the (co-moving frame) energy of the packet is strictly conserved, but there is no requirement that the $r$-packet that is eventually created has to be associated with the recombination cascade produced by the original photoionization event.\footnote{Such a requirement is imposed in the ``branching'' approach of \citealt{lucy_improved_1999}, for example.} However, in its sampling of the macroscopic energy flow rates through different atomic processes, the macro-atom approach does capture (asymptotically, but accurately) the recombination emission that can occur in reality from a recombining electron. This process is particularly important for H and He, both in terms of continuum reprocessing and observed line emission. We therefore usually model these elements with the "full" (multi-level) macro-atom formalism, while other elements are treated within the two-level atom approximation.

 In classic mode, photoionization is treated as a continuous absorption process, which reduces the weight of an $r$-packet as it passes through the plasma. In this mode, the radiative recombination cooling rate is used to generate photons in the wind (although at present dielectronic recombination cooling does not produce photons). In both modes, the energy absorbed via photoionizations 
 is logged as a heating process for thermal balance calculations. Finally, photoionizations and recombinations are also included in solving for the ionization state (for both simple atoms and macro-atoms) and the level populations (for macro-atoms). 

\subsubsection{Free-free (Bremsstrahlung)}
Free-free absorption and emission are included as both radiative and thermal (heating/cooling) processes. We use frequency-averaged Gaunt factors for free-free emission as tabulated by \cite{sutherland98}. In hybrid macro-atom mode, a free-free interaction creates a $k$-packet via an $r \to k$ transition (which is processed exactly as any other $k$-packet). Free-free cooling is also included as a $k \to r$ process, with the relative probability of this process chosen in proportion to the local free-free cooling rate. Conversely, in classic mode, free-free absorption is treated as a continuous absorption process, reducing $r$-packet weights. The free-free emissivity is then used to determine the number of $r$-packets created within the wind via free-free emission at the beginning of an ionization or spectral cycle. 

\subsubsection{Electron/Compton scattering}
As an $r$-packet moves through the wind, the Compton opacity it encounters can lead to both a scattering interaction and the transfer of energy to the plasma. 
The probability that an $r$-packet undergoes a Compton scattering interaction is calculated from 
the total Klein-Nishina cross section. When a Compton scattering event happens, the new $r$-packet direction is randomly selected based on the angular distribution of the differential Klein-Nishina cross section. The frequency of the scattered $r$-packet is reduced by an appropriate factor, and the weight of the $r$-packet is also reduced to maintain the correct photon number (in both hybrid macro-atom and classic modes; in the former case this represents one exception to the general rule of not changing the co-moving frame packet energy). This process produces a Compton-hump in appropriate conditions.

Compton heating is treated via an effective heating opacity that electrons present to the $r$-packet, which is proportional to 
the ``energy exchange'' cross section, as defined in the \cloudy\ documentation (see Hazy 3, eq. 6.6). 
This modification to the Thompson cross section takes account of the fact that the energy transferred from an $r$-packet to the electron pool depends on the energy of the photons. This energy is added to the heating rate in the cell, and in the simple-atom case the weight of the $r$-packet is reduced by the same amount.

Compton cooling (or inverse Compton scattering) is also included in the thermal balance, with the cooling rate given by an integral over the mean intensity and 
the inverse Compton cross-section $\beta_{\rm Comp} \sigma_T$. The latter takes account of the fact that the mean energy lost from the electrons to the photons depends on the photon frequency. 
The analytic expression we use for $\beta_{\rm Comp}(\nu)$ once again comes from the \cloudy\ documentation (see Hazy 3; eq 6.7). 
No $r$-packets are generated directly to match this cooling rate; instead, the energy loss is taken into account by reweighting during Compton scattering events.  

\section{Radiation Sources and Sinks}
\label{sec:radiation_sources}
\subsection{Radiation sources}
    
 \xcode\  is intended primarily to model accretion disc systems comprised of a central object, a star or BH,  and an optically thick disc (and in some cases a secondary star\footnote{The secondary star does not radiate, but is used to suggest an outer radius for the disc. It also blocks radiation in the calculation of the detailed spectrum as a function of orbital phase.}). The central object must have its mass and radius specified, and stars must also have a defined effective temperature.  The central object can radiate as a blackbody, as a bremsstrahlung source, or as a power law source (specified by a 2-10 keV luminosity and a spectral index). Spectra interpolated from grids of models, e.g Kurucz models, for stars and discs are also available. $r$-packets are generated from the entire central object surface with the angle of emission at any point following the Eddington approximation. For CV systems, emission from a boundary layer may be specified in terms of a luminosity and temperature, without the enforcement of the Stefan-Boltzmann law. The boundary layer is also assumed to radiate from the entire surface of the star. For BHs, a lamp-post model is implemented so that radiation can originating from a specified height above and below the origin. Finally, there is the flexibility to include any user-defined SED emanating from the central object, with a few options available for the angular distribution. 
    
The disc can be flat or vertically extended with a height $z_{\rm disc} \propto r^{\gamma_d}$). We normally adopt the standard temperature distribution for a steady state accretion disc \citep{shakura_black_1973} that transfers material at a fixed rate $\dot{m}_{\rm disc}$ onto a central object of a given mass $M$ and radius $R$. Arbitrary temperature distributions -- for example, from calculations including the effect of mass-loss or irradiation -- can also be imported from an input file. The disc can radiate as a collection of blackbody spectra or as collection of stellar atmospheres (or any set of model spectra specified by a temperature and a gravity). As in the case of radiation from a star, the Eddington approximation is used to determine the angular distribution of the emission from the surface of the disc. There is also an option to include a colour correction of the form given by \cite{done_intrinsic_2012}. 

\subsection{The wind as a radiation source}
\label{sec:wind_radiation}
In hybrid macro-atom mode, the wind is not source of radiant energy.  As $r$-packets pass through the wind,  they activate macro-atoms. In the process, an $r$-packet which originated from, for example, a thermal accretion disc continuum, can be converted to a recombination photon. This is achieved by sampling the macro-atom transition probabilities formulated according to statistical equilibrium. 

In classic mode, however, radiative equilibrium and energy conservation within the wind is not enforced at the point of interaction. Instead, as $r$-packets pass through the wind, they lose energy to all processes apart from resonant and electron scattering.  Photoionization and free-free emission are treated as pure absorption processes. The effects of collisional deexcitation from line interactions result in additional energy loss. To account for this deposited energy, the wind is a source of radiation in classic mode; $r$-packets are generated in each cell of the wind, depending upon the plasma state at the beginning of a cycle. The electron temperature is established as described in section~\ref{sec:ionization}. 

\subsection{Surfaces and boundary conditions}
\label{sec:surfaces}
Packets are tracked until they are destroyed, or until they reach the outer domain edge, at which point they are assumed to propagate to infinity along their current trajectory. The outer boundary can thus be thought of as a free-streaming boundary for $r$-packets, but there are also internal boundaries within the simulation: typically, a central compact object at the origin and an accretion disc at the midplane. How a packet is processed when it encounters one of these surfaces is determined by the user. There are three possibilities. First, the photon can be ``absorbed'', in which case the $r$-packet is simply lost and has no further effect on the rest of the simulation; this represents a net sink of radiation. Second, the photon can be used to heat the surface.  In this case the temperature of the surface will be increased in the next ionization cycle by the amount required to ensure that all of the energy absorbed in the previous cycle is reemitted. Third, the photon can be ``scattered'', i.e.  the photon is reflected from the surface with angular distribution given by the Eddington approximation.  This is most physically appropriate when surface interactions are dominated by electron scattering.

\subsection{Other sinks of radiation}
\label{sec:sinks}
At present, there are relatively few sinks of radiation in \xcode. As mentioned above, a surface, such as the disc or compact object, can be a sink of energy depending on the choice of surface reflection treatment. In addition, we choose to destroy energy packets if they scatter excessively (typically more than 2000 times) or if their weight drops below a critical threshold. However, perhaps the most common sink for radiative energy is adiabatic cooling of the gas, an energy loss that occurs when streamlines in a flow diverge ($\nabla \cdot \boldsymbol{v}>0$). This effect amounts to a loss of energy from the thermal pool. The rate at which energy is gained or lost is given by $P (\nabla \cdot \boldsymbol{v})$, where $P$ is the gas pressure and $\boldsymbol{v}$ is the local velocity. To account for adiabatic cooling, we destroy a certain proportion of packets when they become $k$-packets after an interaction, with a probability set by the adiabatic cooling rate divided by the total cooling rate (see equations~1 and 2 of \citealt{matthews16}). We do not currently generate additional $k$-packets due to adiabatic heating in a flow with negative velocity divergence, but this effect could in principle be included along with any non-radiative process. Accounting for adiabatic heating and cooling is slightly simpler in classic mode, because we generate wind photons at the beginning of each cycle, with the number of photons per cell being proportional to the emissivity.  In this case, the energy loss or gain is accounted for implicitly. This is because the total cooling is iteratively adjusted to match the total heating in a cell (see section~\ref{sec:ionization}), but without energy packets being explicitly generated or destroyed to represent adiabatic heating or cooling.

\section{Code Operation}
\label{sec:operation}
We describe the operation of \xcode\ in an approximate order of execution, following the flowchart shown in Fig~\ref{fig:flowchart} but starting from just after the setup phase. The creation of wind models and coordinate grids is described in section~\ref{sec:wind}.  

\subsection{Photon ({\it r}-packet) generation}
\label{sec:photon_generation}
We generate $r$-packets at the beginning of each cycle, and they are not destroyed until they leave the system or encounter one of the radiation sinks (section~\ref{sec:sinks}). The basic principle behind photon generation in \xcode\ and most time-independent MCRT codes is that $r$-packets represent quantised radiative energy with individual packet energies (or weights) given by 
\begin{equation}
\epsilon = \frac{\Delta E}{{\cal N}_\gamma} = \frac{\dot{E} \Delta t}{{\cal N}_\gamma} \, ,
\end{equation}
where ${\cal N}_\gamma$ is the number of $r$-packets generated, and $\Delta t$ does not represent any kind of time {\em dependence}, but is a somewhat arbitrary, pre-defined time interval. We set $\Delta t$ to one second such that $\dot{E}$ can be associated with the luminosity of the system. In simulations that include multiple radiation sources, indexed with $j$, the weight of each $r$-packet is similarly given by 
\begin{equation}
\epsilon = \Delta t \frac{\sum_j^\text{sources}  ~ \int_{\nu_{\text{min}}}^{\nu_{\text{max}}} L_{\nu, j} d\nu}{{\cal N}_\gamma} \, .
\end{equation}
In this case $\epsilon$ represents a single photon generation band from $\nu_{\text{min}}$ to $\nu_{\text{max}}$. The total luminosity of the photons generated corresponds to the luminosity of the external sources in the case of macro-atoms, and in addition to the luminosity of the wind in classic mode.  Note that in the latter case, we do not mean the luminosity of the photons that escape to infinity, but rather the emissivity of the wind integrated over the volume of the wind.  When wind photons are generated, the number of wind photons generated per cell is directly proportional to the volume times the emissivity of the cell.  

During spectral cycles, we do indeed use a single band to generate $r$-packets, typically a fairly narrow one corresponding to the wavelength range of interest. However, there is no necessity for each $r$-packet to have exactly the same weight, and indeed this is often sub-optimal in ionization cycles. Rather, the user can specify a variety of ways to apportion the number of photon bundles in various frequency ranges to approximate the spectrum of the radiation field within the grid, with their weights adjusted accordingly; this ``stratified sampling'' process works by defining a minimum fraction of photons $f_i$ in a given frequency band $i$ spanning frequencies $\nu_i,\nu_{i+1}$, such that the weight of $r$-packets generated in this band is given by 
\begin{equation}
\epsilon_{i}  = \Delta t \frac{\sum_j^\text{sources} ~ \int_{\nu_{i}}^{\nu_{i + 1}} L_{\nu, j} ~ d\nu}{f_{i} {\cal N}_\gamma} \, .
\end{equation}
If for any reason the radiated energy within a given band is zero, $f_i$ is also zeroed for that band, and the other $f_i$ values are renormalised to sum to one. Choosing the best stratified sampling approach is an important consideration for accurately calculating the ionization structure of the wind while minimizing the total number of photons generated in each cycle.

With weights specified, the remaining variables that initialise an $r$-packet are its direction ($\theta, \varphi$), position and frequency ($\nu$). The initial direction and location are chosen depending on the radiation source the $r$-packet is assigned to. For example, for a star, a random location on the stellar surface is generated, and the direction is chosen by sampling $p(\theta, \varphi) d\theta d\varphi = \eta(\theta) \sin \theta \cos \theta d\theta d\varphi$, where $\eta(\theta)$ represent a limb-darkening function obtained from the standard Eddington approximation. Alternatively, the central object could correspond to an isotropic point source. For the accretion disc, the direction is chosen in a similar manner to the stellar case, while the initial radius is chosen by sampling the radial emissivity profile of the disc. The latter is achieved for each frequency band by breaking the disc up into $3000$ annuli, with boundaries adjusted so that each generates the same luminosity. We then choose a random annulus to emit from. 

We choose the frequency of $r$-packets by first assigning them a source, such that the number of packets for each source in band $i$, $N_{\gamma,j,i}$ satisfies $\sum_j {\cal N}_{\gamma,j,i} = f_i {\cal N}_\gamma$ and ${\cal N}_{\gamma,j,i} \epsilon_i = \int_{\nu_{i}}^{\nu_{i + 1}} L_{\nu, j} ~ d\nu$. This ensures that the number of packets generated correctly produces the required band-limited luminosity of that source and matches the desired stratified sampling strategy. The frequency of a given packet from source $j$ is then chosen from a probability distribution $p(\nu) \propto L_{\nu,j}$. The exact procedure here depends on the source in question, but typically involves exploiting cumulative distribution functions (CDFs). These CDFs are calculated either numerically or analytically, as appropriate for the form of the spectrum (see \citetalias{long02}). In classic mode, $r$-packets frequencies are effectively fixed (aside from Doppler shifts and recoil effects in Compton scattering) throughout their transport, but their weights can be reduced. In hybrid macro-atom mode, $r$-packets have fixed weights, but their frequencies can shift during interactions.

The choice of ${\cal N}_\gamma$, the number of $r$-packets generated per cycle, is important. The key point is that a good representation of the local radiation field is required in any wind region that might contribute significantly to the emergent spectrum. The estimators required to obtain this representation are constructed from photon passages through the relevant wind region. The number of $r$-packets therefore has to be large enough to yield sufficiently accurate and precise estimators over the entire relevant frequency range. 
\footnote{If one is simulating the spectrum of a fixed wind structure in radiative equilibrium, cells which receive few photons from external sources do not contribute significantly to the observed spectrum of the object, and it is less important that the radiation field of such cells be well defined.  As a result one can tolerate some cells with few photon passages.  If however, one wants to use \xcode\ as the radiation part of a hydro simulation, then the temperature and pressure in all cells needs to be determined accurately and to do this, one must ensure that the radiation field in all cells is well-defined by sufficient numbers of photon passages.}  In addition, there will ideally be enough $r$-packets that other important Monte Carlo estimators (see section~\ref{sec:estimators}) are not overly noisy. The number of photons one requires to improve a solution from its current state depends on how far one is from the final solution. Therefore options are provided to vary the number of photons per cycle logarithmically (see section \ref{sec:converge_cv}). Finally, the total number of photons (${\cal N}_\gamma \times N_s$) must provide the user with a reasonable signal to noise ratio in the final output spectrum.

\subsection{Photon transport and special relativity}
\label{sec:photon_transport}
Photon transport in \xcode\ takes place in 3D, allowing fully for the effects of special relativity. Whenever an $r$-packet has been created or undergone an interaction, a uniform random number $p$ between $0$ and $1$ is used to generate the optical depth to the next interaction. This ``interaction'' optical depth is given by
\begin{equation}
\tau_{\rm int}= -  \ln (p).
\end{equation}
Each $r$-packet is transported through the wind (on a cell-by-cell basis) until it reaches an optical depth $\tau_{\rm int}$, hits a surface or escapes from the system. Continuum opacities (Compton scattering, free-free and bound-free) are built up gradually throughout the cell, whereas bound-bound opacities are incremented at individual Sobolev surfaces, following the accumulation procedure described by, e.g., \cite{mazzali93} and \cite{noebauer19}.  Once $\tau \geq \tau_{\rm int}$, a decision is made to process the packet depending on the relative contributions of each opacity source. 

For much of its life, \xcode\  made the assumption that outflow velocities were small compared to the speed of light. Doppler shifts were therefore treated to linear order in $v/c$, and no other corrections associated with relativistic effects were considered. However, \xcode\ now fully incorporates special relativistic effects, and it explicitly distinguishes (and correctly transforms) between the co-moving (local/fluid) and laboratory frames. Energy is conserved in the co-moving frame, and the relevant special relativistic transformations are applied when converting between frames. \xcode\ therefore accounts for angle aberration, as well as for the frame-dependence of photon frequencies and weights \citep[see, e.g.,][]{castor72, Mihalas_Mihalas1984}. 

We take a {\em mixed-frame} approach to radiative transfer, in that we make convenient choices as to which frame to use for a given task, following, e.g., \cite{lucy2005}. The simulation grid mesh is defined in the laboratory frame, and so is photon transport between wind cells. Opacities are generally calculated first in the co-moving frame (since internally we store model densities in the local frame) and translated to the laboratory frame. This is done by using either the fact that $\tau$ is a Lorentz invariant or by direct translation of (continuum) opacities from the co-moving to laboratory frame, viz. \cite[Eq. 2,][]{castor72}
\begin{equation}
\kappa_{\nu} = \frac{\nu_o}{\nu} \kappa_{\nu,o}.
\end{equation}
If a $r$-packet scatters, the photon is transformed to the local frame, a new scattering direction and in some cases frequency and weight is selected depending on process. Then, the photon is converted back to the laboratory frame and a new $\tau_{\rm scat}$ generated. This process therefore accounts for relativistic aberration through the transformation of the $r$-packet direction. In general, estimators of the radiation field as well as all the atomic and heating/cooling processes, are calculated in the co-moving frame of the fluid.

\subsection{Ionization cycles: determining the plasma state}
\label{sec:ionization_cycles}

\subsubsection{Monte Carlo estimators}
\label{sec:estimators}
During photon propagation, Monte Carlo estimators are recorded which are used to characterise the mean intensity of the radiation field, $J_\nu$, in the cell and estimate the rate of any process which depends on this (such as photoionization or radiative heating processes). Following \cite{lucy_computing_1999}, we generally construct volume-based estimators. For example, the Monte Carlo estimator for the mean intensity has the form 
\begin{equation}
    \bar{J}_\nu~\Delta \nu = \frac{1}{4\pi}\frac{1}{\Delta V} \sum_{\Delta \nu} \epsilon_i \Delta s,
\label{eq:jnu}    
\end{equation}
where the sum is over all $r$-packets within the co-moving frame frequency range $\nu,\nu+\Delta \nu$. In practice, within \xcode, we do not calculate this estimator across a large number of frequency bins, because this is both noisy and often unnecessary. Instead, we increment individual estimators for macro-atom processes (i.e., for specific bound-bound and bound-free transitions) in the forms given by \cite{lucy_monte_2003}. 

For the ionization states of simple atoms, a model for the mean intensity is required. We always record the total mean intensity, $\bar{J}$, from equation~\ref{eq:jnu}, with $\Delta \nu$ covering the full frequency range. We also record the mean frequency of $r$-packets passing through the cell, weighted by the path length, given by  
\begin{equation}
\overline{\nu}=\frac{\sum_{\rm paths}{\epsilon_{\rm ave}\nu \Delta s}}{\sum_{\rm paths}{\epsilon_{\rm ave} \Delta s}},
\end{equation}
where $\epsilon_{\rm ave}$ is the mean energy carried by an $r$-packet as it travels a distance $\Delta s$ in the cell.The radiation temperature, $T_r$, and dilution factor, $W$, can then be estimated as $T_r = (h \bar{\nu})/(3.832 k_B)$ and $W = \pi \bar{J} / (\sigma T_r^4)$
where $\sigma$ is the Stefan-Boltzmann constant. These quantities are also used for assessing convergence (section~\ref{sec:converge}) and calculating level populations (section~\ref{sec:ionization}).  The user then has the option to model the mean intensity either as a dilute blackbody ($J_\nu = W B_\nu (T_r)$) or following 
the procedure described by \cite{higginbottom13}. In the latter case, we construct a model for the mean intensity by splitting the entire frequency range into a series of coarse bands, where the bands are usually explicitly tied to the photon generation (but with a minimum of seven frequency bands for the spectral model). In each band, we make use of the band-limited versions of $\bar{\nu}$, $\bar{J}$ and the standard deviation of the frequency,
\begin{equation}
\sigma_{\nu}=\sqrt{\frac{\sum{\nu^2 \epsilon_{\rm ave}\Delta s}}{\sum{\epsilon_{\rm ave}\Delta s}}-\overline{\nu}^2} .
\end{equation}
These three moments of the mean intensity are collected in a series of frequency bands, so that the specific mean intensity $j_{\nu}$ can be modeled in each cell of the wind. We model the specific mean intensity in each band using either a power law or exponential representation, which we choose between based which most closely matches the standard deviation of packet frequencies. The full spectral model is then used in the ionization calculation described in section~\ref{sec:ionization}.

In addition to rate estimators and estimators characterising the radiation field, we also record heating rate estimators for each radiative heating process. Typically, this involves incrementing an estimator by the heating contribution $\epsilon_i \kappa(\nu) \Delta s$ for each $r$-packet as it travels a distance $\Delta s$ in a cell. In hybrid macro-atom mode, there is then also the possibility that a $k$-packet will be created, while in classic mode the $r$-packet weight is reduced to account for the energy transfer to the thermal pool. Both modes should asymptotically produce the same results. The full set of macro-atom heating estimators are given by \cite{matthews16}. 

\subsubsection{Thermal, ionization and excitation state: the wind update step}
\label{sec:ionization}
The purpose of the ionization cycles in \xcode\ is to establish the thermal, ionization and excitation state of the wind plasma. This process requires multiple iterations because it involves the solution of a coupled radiative transfer problem: the radiation field and ionization state are non-linearly and non-locally dependent on each other. We initialize the ion abundances in a model using the Saha equation with a user-defined initial electron temperature for the wind, $T_{e,0}$.  At the end of each ionization cycle, we update the ionization and thermal structure of the wind based on the Monte Carlo estimators that were constructed during the cycle. 

During each cycle, once all the $r$-packets have been transported through the wind, we update the temperature structure, defining the current temperature $T_{e,n}$. The first step is to calculate, for each cell, a new guess at the electron temperature, $T_{e,{\rm opt}}$: this is the temperature within the range $0.7 T_{e,n} < T_{e,{\rm opt}} < 1.3 T_{e,n}$  that results in the cooling most closely matching the heating rate obtained from the relevant MC estimators. The improved estimate of the electron temperature in the cell for the next cycle, $T_{e,n+1}$, is then obtained from the formula 
\begin{equation}
    T_{e,n+1} = (1 - {\cal G}) T_{e,n} + {\cal G} T_{e,{\rm opt}},
\end{equation}
where ${\cal G}$ is a ``gain'' factor between $0.1$ and $0.999$ that varies strategically from cycle to cycle depending mostly on the difference between $T_{e,n}$ and $T_{e,{\rm opt}}$. 
We adopt this procedure so as to prevent large changes in temperature and dramatic changes in, e.g., the wind emissivity from cycle to cycle. The goal here is to iterate smoothly and gradually towards a converged solution in which heating and cooling rates balance. At this point, the thermal structure of the wind should no longer change significantly from cycle to cycle.

In addition to updating the temperature, we also calculate the ion abundances at this stage. The exact treatment of ionization can differ between macro-atoms and simple atoms, but our principle method involves a rate matrix calculation. For simple atoms, the photoionization rate for each transition is obtained by numerically integrating the modelled mean intensity obtained from estimators as described in section~\ref{sec:estimators}. We typically use total recombination rates from the literature including dielectronic recombination, although using the Milne relation is also an option for deriving the recombination rates in the absence of this data. Given a set of photoionization and recombination rates, we then populate a square rate matrix $\boldsymbol{\textsf R}$ with size determined by the total number of ions. This allows us to set up the system $\boldsymbol{\textsf R} \boldsymbol{N} = {\boldsymbol 0}$, where $\boldsymbol{N}$ is a vector of the ion populations. To close this system of equations, we finally impose that the ion abundances in each element must sum to the total abundance of that element and solve  the matrix equation using the GNU scientific libraries (GSL) implementation of LU decomposition \citep{turing} to obtain the solution vector $\boldsymbol{N}$. We repeat the calculation until we have converged on $n_e$ (that is, the value of $n_e$ from one iteration to the next changes by less than $0.03$). This iterative process is required because $n_e$ depends on the ionization state, but the ionization state depends on $n_e$ via the recombination rates. 

In addition to the rate matrix method, we also provide various other options for calculating simple ion abundances. There are two diagnostic modes in which we assume LTE (based on either $T_e$ or $T_r$). It is also possible to use an on-the-spot approximation and a modified version of the on-the-spot approximation \citep{mazzali93}, as described in detail by \citetalias{long02}. Finally, one can impose a fixed ionization state, uniform in space, by manually specifying the fractional abundances of individual ions.  

The level populations of simple atoms are also needed in order to calculate bound-bound and bound-free opacities. If the model for the mean intensity is that of a dilute blackbody radiation field then the level populations are approximated using the dilute Boltzmann approximation. That is, the population of level $j$ relative to the ground state $n_1$ is then given by
\begin{equation}
\frac{n_{j}}{n_{1}} = W \frac{g_j}{g_{1}} \exp(-E_j/kT_r).
\label{eq:dilute_boltzmann}
\end{equation}
where $g_j$ is the statistical weight. In the case of a more complex local radiation field, we make the assumption that all two-level ions are in the ground state.

Historically, and by default, macro-atom ionization and excitation states are obtained simultaneously by solving the equations of statistical equilibrium based on the rate estimators from \cite{lucy_monte_2003}. 
The procedure is similar to the ion calculation above, except that size of the square matrix ${\boldsymbol{\textsf R}}_m$ is now set by the number of macro-atom levels. The resulting system of rate equations is ${\boldsymbol{\textsf R}}_m \boldsymbol{n} = \boldsymbol {0}$, where $\boldsymbol{n}$ is a vector of the fractional level populations. Again, we make use of a closure relation -- that the fractional level populations across all ions in an element must sum to one -- and solve the resulting matrix equation as above to obtain $\boldsymbol{n}$. Following \cite{lucy_monte_2002} and \cite{sim05}, the photoionization and radiative excitation rates obtained from MC estimators and collisional processes are included. Recombination rates are obtained from the Milne relation in this case for each possible bound-free transition in the macro-atom. 

\subsubsection{Checking whether the calculation has converged} 
\label{sec:converge}
Each  \xcode\  run begins with a series of cycles designed to determine the ionization and thermal structure of the wind.  Hopefully, each ionization cycle brings one to a converged solution. Since  \xcode\  is a Monte Carlo code, the degree to which even a converged solution stays constant from cycle to cycle is limited by counting statistics.  

To check convergence in  \xcode\ , we monitor the radiation temperature T$_r$, the electron temperature T$_e$, and the total heating ${\cal H}_{\rm tot}$ and cooling ${\cal C}_{\rm tot}$ in each cell as a function of the ionization cycle, $n$. A cell is said to be converged if it satisfies the following tests:
\begin{align}
\left | \frac{T_e^n-T_e^{n-1}}{T_e^n+T_e^{n-1}} \right | &< \delta \\
\left | \frac{T_r^n-T_r^{n-1}}{T_r^n+T_r^{n-1}} \right | &< \delta \\
\left | 
\frac{{\cal H}_{\rm tot}^n- {\cal C}_{\rm tot}^{n}}
{{\cal H}_{\rm tot}^n + {\cal C}_{\rm tot}^{n}} 
\right | &< \delta
\end{align}
where  we set $\delta = 0.05$. It is relatively rare that all of the cells in a model will satisfy all of these criteria.  That is partly because the number of $r$-packets that pass through a cell can vary considerably, and the statistical noise depends on the number of photons. It is important to note that, often, the cells that contribute most to the spectra of an object will be those which have the most $r$-packets passing through them. However, this does not have to be the case at all viewing angles. 
Typically, we consider a model to be well converged if $\approx 90\%$ of the cells meet these criteria. 
However, the convergence criteria are only advisory; the number of ionization cycles in a calculation is user-determined, and if the user decides the calculation has not converged sufficiently, they can simply continue the calculation with more cycles and/or more $r$-packets.
 
 \subsection{Creation of the final spectrum}
 \label{sec:spectrum}

The final stage of a \xcode\ simulation is the calculation of the detailed spectrum at a series of user-requested viewing angles. During these spectral cycles, we assume that one is interested in a limited spectral range, and \xcode\ generates photons that all carry the same weight. The relative number of $r$-packets generated for the various sources is determined by the luminosity and spectrum of the source within the desired band. Photons from all spectral cycles contribute to the final spectrum; the only reason multiple cycles are needed at all is to limit memory requirements (by not having too many photons in existence at any one time). That said, the use of multiple spectral cycles is also useful from a user perspective. Since it is not always obvious in advance how many photons are required in order to achieve the required S/N in the output spectra, it is often helpful to monitor the evolution of the synthetic spectra across spectral cycles. In general, we recommended choosing the best ${\cal N}_\gamma$ for the ionization state calculation and adjusting $N_s$, the number of spectral cycles, to achieve the desired signal to noise. 

In classic mode, the generation of the $r$-packets from the wind during spectral cycles more-or-less mirrors that of the ionization cycles: for continuum processes, $r$-packets are apportioned based on the band-limited luminosity of each process.  For line radiation, one only need consider lines within (or near, within a typical Doppler shift) the desired wavelength band, and no particular preparation before the spectral cycles is required. However, in hybrid macro-atom mode, the process is more complex. In the normal macro-atom formulation, the wavelengths of $r$-packets can change by large amounts as a result of interactions. Thus, one cannot simply generate all of the photons from external sources. Instead, using the current state of the plasma and the known broad-band radiation field, we undertake a macro-atom emissivity calculation just before the spectral cycles begin (see Fig.~\ref{fig:flowchart}). Here, we calculate the band-limited emissivities of all processes within the band. We use these emissivities to generate $r$-packets similarly to the classic mode, which accounts for the radiation that would have been generated in the desired band from macro-atom interactions. During photon transport, all $r$-packets that interact with macro-atoms or $k$-packets are destroyed and do not contribute to the spectrum. The end result is that both radiative equilibrium and co-moving frame energy conservation are still enforced, but less directly than in the ionization calculation (that is, not strictly at the point of interaction).
    
\xcode\ normally uses a "peel-off" variance reduction method \citep{zadeh84} -- also described by \cite{knigge95} as the ``viewpoint'' technique -- to construct detailed spectra in a specific number of user-defined directions. In the peel-off method, every time a photon is created or scattered, \xcode\  calculates the relative probability of the photon travelling in the each of the desired directions. The weight and frequency of the $r$-packet is adjusted to reflect this probability and then ``extracted'' from the system. Special relativistic effects are fully accounted for and anisotropy of scattering or emission is also taken into consideration.  During the extraction, all continuum opacities are treated as purely absorptive processes that reduce the photon weight. If the extracted photon encounters  the disc or the primary star, the packet is processed according to different modes described in section~\ref{sec:surfaces}. If the packet encounters the (optional) secondary star, it is destroyed, which allows one to construct, for example, the in-eclipse spectrum of an accreting binary seen at high inclination. We have carried out extensive tests with \xcode\ to show that the extract method agrees with what we term the `live-or-die' method, where one simply counts (correcting for solid-angle) $r$-packets escaping the system within  $\pm 0.2^\circ$ of defined directions; the former method produces a much less noisy spectrum for a given number of $r$-packets transported.

\subsection{Logistical details}

\subsubsection{Parallelization}

One of the main computational challenges of MCRT is that it can be expensive. Since each photon packet is independent, MCRT simulations -- or, at least portions of them -- fall into the class of ``embarrassingly parallelizable'' algorithms. In \xcode , parallel computing is exploited to decrease the time spent on photon transport and ionization calculations by splitting the workload across multiple processors using the \texttt{Message Passing Interface} (MPI). MPI follows the \textit{distributed-memory} paradigm, hence each parallel process requires its own copy of global variables, such as the simulation grid, atomic data and Monte Carlo estimators. This can unfortunately translate into a large memory requirement when using a large number of processors, or a very high resolution grid. For example, \xcode\ can use about $1$ GB of memory per process for a simulation with a grid resolution of $150\times150$ and $10^{8}$ photons. Some models with larger grids and complex data sets, or extremely  large ${\cal N_\gamma}$, can approach 8 GB per process.  

The process of generating $r$-packets is divided between MPI processes, such that for $N_p$ processes, each process transports ${\cal N}_\gamma/ N_p$ $r$-packets, sampled from the same frequency distribution, through the wind. Each process records  its own estimators and diagnostic quantities, which are communicated and normalized between all the processes during the wind update step. The wind update step and macro atom emissivity calculations are split such that each process works on a subset of the grid independently. Each process then communicates their subset back to the main process, which communicates the updated grid, in full, to each process for use during the next photon transport cycle. 

\begin{figure}
    \centering
    \includegraphics[width=\linewidth]{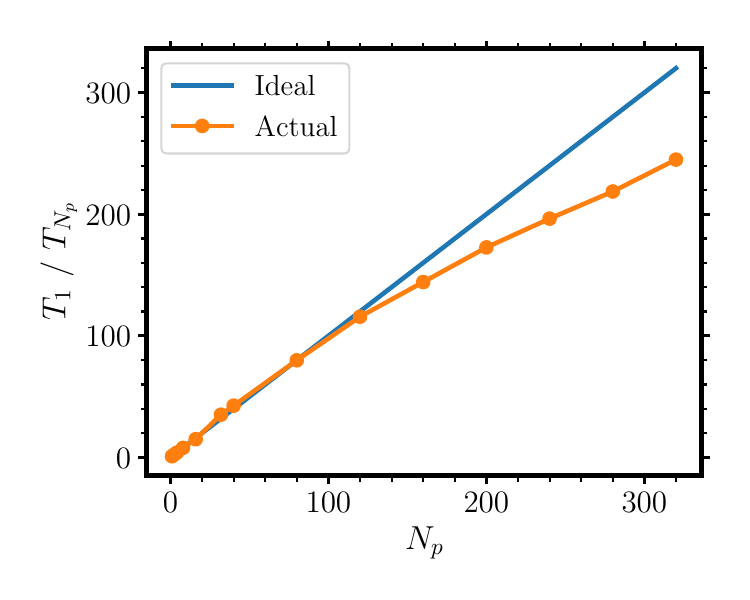}
    \caption[Parallel speed up]{The parallel speed up of a quasar model as a function of number of MPI processes $N_{p}$. Near-optimal speed-up is obtained up to $N_p \approx 120$, and substantial speed-up can still be achieved at higher $N_p$.}
    \label{fig: parallel_speedup}
\end{figure}

The exact performance increase from running in parallel depends on the model used and how dominant each step in the calculation is, in terms of cumulative run time. Generally speaking, the speedup for the photon transport cycles is close to optimal, whilst the cumulative run time drops off as the number of processes increases. Shown in Fig.~\ref{fig: parallel_speedup} is the speedup for a quasar model from \cite{matthews16}. We have defined the speedup factor as the ratio of the multi-process run time, $T_{N_p}$, to the single processor one, $T_1$. For $N_p \lesssim 120$ the speed up is nearly optimal. The curve then gradually flattens off, but still shows significant improvements even for hundreds of processors. The sub-optimal speed up at very large $N_p$ is due to, in-part, increased communication time, but also because parts of the code are executed in serial and there can be uneven workloads between processes.
    
\subsubsection{User interface}
Inputs to a \xcode\ simulation are provided using a parameter file with a keyword value approach. There are two principal ways to build a model. One option is for the user to run \xcode\ in an interactive mode in which one is presented with a series of choices of inputs. From this, the code creates a parameter file, which one can then  edit for future runs. Alternatively, a user can start with one of the provided example parameter files included within the code repository and adapt it accordingly. Many of the typical input parameters are given in Table~\ref{tab:demo_model_parameters}, and they are fully described within the code documentation. 

\subsubsection{Atomic data}

\xcode\ uses atomic data assembled from a variety of sources, which is read in from a series of ascii text files at the beginning of each run. The data are accessed in a hierarchical fashion so one can choose which atoms, ions, etc. to incorporate into a particular \xcode\ run. This makes it possible to modify atomic data for testing or for a specific scientific purpose. The quality of the output spectra is directly related to the quality of the atomic data for a particular purpose. Typically, we include the elements H, He, C, N, O, Ne, Na, Mg, Al, Si, S, Ar, Ca and Fe, although this can be easily adapted. Given the complexity of the atomic data needed for running \xcode, we have collated four recommended data-sets for the user, which are described in the documentation. The documentation also provides more detail on our atomic data sources and the limitations of each data-set. We also provide an ancillary program, \textsc{atomix}, for exploring atomic datasets within a graphical user interface. Here, we provide only a brief overview of the basic data sources and references. 

The photoionization cross-sections  are  obtained either from TopBase \citep{cunto93} or \cite{verner96}, with the latter source supplemented by cross-sections from \cite{verner95} at high energies or when there are missing species. For TopBase, we have extrapolated the cross-sections out to 100 keV to prevent unphysical discontinuities. We have also adjusted some of the threshold energies using data from the {\sl NIST} Atomic Spectra Database\footnote{\url{https://www.nist.gov/pml/atomic-spectra-database}}. We obtain electron yield data (for Auger ionization) from \cite{kaastra1993}. 
Basic data on ions, such as ionization potentials and ionization states come from \cite{verner96_lines}, as do the majority of the 6000 lines in our standard set of atomic data.
Alternatively, one can use energy levels and lines selected from the line lists assembled by \cite{kurucz1995}. However, \xcode\ cannot handle the entire Kurucz line lists (see section~\ref{sec:limitations}), so in this case we include about 54,000 lines selected on the basis of atomic abundances and oscillator strengths. The level information for simple atoms is constructed  from the line lists using the technique described by \cite{lucy_improved_1999}. We make extensive use of the Chianti database \citep{dere97,delzanna21}, from which we obtain radiative recombination rate coefficients, line collision strengths (following the fit coefficients given by \cite{burgess1992}) and dielectronic recombination rates. We use free-free Gaunt factors from \cite{sutherland_accurate_1998}. Ground state recombination rates from \cite{badnell_radiative_2006} are adopted where available, otherwise the code defaults to calculating recombination rates from the Milne relation. We use charge exchange rate coefficients from \cite{kingdon1996} and energy defects from \cite{kingdon1999}.

\subsubsection{Outputs}
The primary outputs of \xcode\ are spectra, which are written to simple ASCII files. 
The actual structure of the wind grid, including temperatures and ion abundances, is stored as a binary file, with ancillary programs provided to convert these into other machine-readable table formats. 

\section{Wind models}
\label{sec:wind}
\xcode\ requires an input model with a specified density and velocity field. This can be achieved either by importing a model grid or by making use of one of the built-in kinematic prescriptions.
    
\subsection{Imported models}
\label{sec:imported}
\xcode\ is designed to be flexible.  It can accept imported models in 1D spherical coordinates or 2D cylindrical or spherical polar coordinates.  Models which are 2D are assumed to be both azimuthally and reflection symmetric, that is they are intended to model a bi-conical wind, emerging from both side of an axially symmetric disc.  Models can fill all of space, but do not necessarily have to do so. 
     
We have used imported model grids to study snapshots of hydrodynamic simulations in, so far, AGN \citep{higginbottom14} and TDEs \citep{parkinson2024}. We also use this machinery when running full radiation-hydrodynamics simulations of XRBs, AGN, and CVs (see section~\ref{sec:rad-hydro}). We are currently engaged in an effort to import self-similar MHD wind models from the so-called jet emitting disc (JED) framework \citep[e.g.][]{jacquemin2019} and have successfully applied \xcode\ to self-similar MHD solutions in the context of FU Ori systems \citep{milliner19}.

\subsection{Parameterized models}
\begin{figure}
    \centering
    \includegraphics[width=\linewidth]{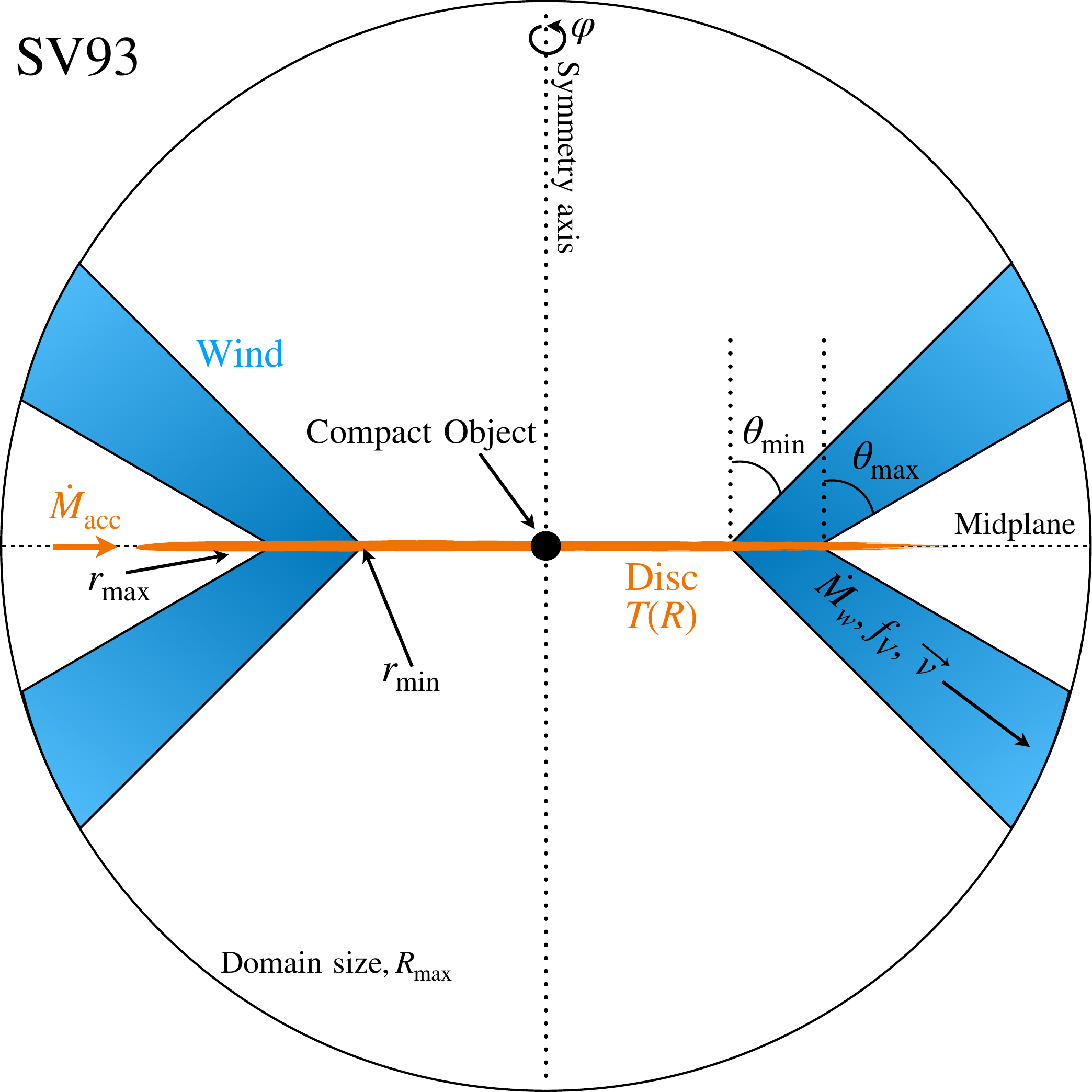}
    \caption[Wind Models]{
    Schematic representation of the SV93 wind parameterizations. Key wind parameters (radii, angles, etc.) are labelled. {\em Blue} represents wind plasma through which radiative transfer occurs and within which thermal balance takes place. Orange represents surfaces that are sources of radiation (quantised as $r$-packets) and can absorb or reflect $r$-packets through the adopted boundary condition. In models such as this, photon transport takes place in 3D but the model has azimuthal symmetry around the central axis and reflective symmetry about the disc midplane. 
   }
    \label{fig:sv_kwd}
\end{figure}

\xcode\ can generate several types of parameterized models.  These include the 2D  wind parameterizations developed by \cite[][hereafter \citetalias{shlosman93}]{shlosman93} for characterizing the spectra of CVs and AGN, and by \cite[][hereafter \citetalias{knigge95}]{knigge95} for studying CVs, as well as several spherical outflows, which have been used to model stellar winds and SNe. We describe each of these briefly below. A schematic diagram for the \citetalias{shlosman93} prescription is shown in Fig.~\ref{fig:sv_kwd}. We describe each of these briefly below.
    
\subsubsection{The Shlosman \& Vitello parameterization}
In the SV parameterization, shown in the left panel of Fig.\ \ref{fig:sv_kwd}, the wind is a biconical flow arising from the disc between cylindrical radii $r_{\rm min}$ and $r_{\rm max}$, with the streamlines varying in angle as
\begin{equation}
    \theta = \theta_{\rm min} + \left ( \theta_{\rm max}-\theta_{\rm min} \right ) x^{\gamma} \, .
\end{equation}
Here $x=(r_0-r_{\rm max})/(r_{\rm max}-r_{\rm min})$ is the fractional distance of the streamline along the wind base.  The velocity is separated into a poloidal ($v_l = \sqrt {v_{\rho}^2+v_z^2}$) and azimuthal ($v_\varphi$) component. The streamlines are straight in the $r_{\rm cyl}-z$ plane, and the poloidal velocity along the streamlines in this plane obeys
\begin{equation}
    v_l = v_0 + (v_{\infty}-v_0) \left ( \frac{(l/R_v)^{\alpha_s}}{(l/R_v)^{\alpha_s}+1}  \right) \,  .
\end{equation}
Here, $l$ is the poloidal distance along the streamline from its footpoint, $R_v$ sets the length scale for the velocity to grow and $\alpha_s$ defines the sharpness of the transition in poloidal velocity. The minimum poloidal velocity, $v_0$ can either be set to a fixed value or a multiple of the sound speed at that point in the disc. The maximum poloidal velocity along a streamline is not constant, but is instead set to a multiple of the escape velocity at the footpoint of the streamline. Angular momentum is assumed to be conserved from an initially Keplerian value, $v_{\varphi, 0} (r_0)$, so that the azimuthal velocity is given by $v_{\varphi}r=v_{\varphi, 0}r_0$. Finally, the differential mass loss per unit area varies as   
\begin{equation}
    \frac{d\dot{M}_w}{d A} \propto \dot{M}_{w}  r_0^{\lambda_s} 
    \cos \theta(r_o) \, ,
\end{equation}
where $\dot{M}_{w}$ is the total mass loss rate. The parameter $\lambda_s$ determines how concentrated the mass loss is toward the inner edge of the wind.

\subsubsection{The Knigge, Woods \& Drew parameterization}
The \citetalias{knigge95} parameterization is similar to the \citetalias{shlosman93} wind in many respects. For example, the poloidal streamlines are straight and the azimuthal velocity is again determined by angular momentum conservation.  However in the \citetalias{knigge95} framework, the poloidal velocity, much like the law often used to parameterize stellar winds, is given by
\begin{equation}
    v(l)=v_0 + (v_{\infty} -v_0) \left (1 - \frac{R_s}{l-R_s}   \right )^{\beta_k} \, ,
\end{equation}  
where the initial velocity, $v_0$, is set to a fixed multiple of the sound speed of the disc and $R_s$ is an acceleration length. Additionally, the direction of the streamlines are determined by a focal point at a distance $d$ measured below the origin (a ``split-monopole'' geometry). As defined by \citetalias{knigge95}, the inner and outer streamlines are set by the point at which a streamline grazes the central object and the outer edge of the disc.  However, as implemented in  \xcode, the inner and outer streamlines are defined through positions in the disc for greater flexibility. Finally, the local mass loss rate is given
\begin{equation}
    \frac{d\dot{M}_w}{d A} \propto \dot{M}_{w} T^{4\alpha_k} \, .
\end{equation}
Thus for $\alpha_k=0$ the mass loss per unit area is constant, whereas for $\alpha_k$ = 1, it is proportional to the luminosity per unit area. 

\subsubsection{Models with spherical symmetry}
The primary use case for \xcode\ is a 2D azimuthally symmetric geometry, but we also include a few spherically symmetric models. These are particularly useful for comparing to state-of-the art 1D radiative transfer codes -- such as \cmfgen\ \citep{hillier98}, \textsc{powr} \citep{grafener2002,sander2015} or \tardis\ \citep{kerzendorf_spectral_2014} 
-- designed to reproduce the spectra of stars and/or SNe. We include a model describing a homologous outflows
(for SNe; see also section~\ref{sec:tardis}), and a simple parameterization for stellar winds (typically used to model O-stars). In the latter case, we use a standard \cite{castor79} velocity law, given by
\begin{equation}
    v(r)=v_0 + (v_{\infty} -v_0) \left(1- \frac{r_0}{r} \right)^{\beta}.
    \label{eq:stellar_wind_vlaw}
\end{equation}
Here, $v_0$ is the velocity at the base $r_0$ (normally the stellar surface) of the wind, $v_{\infty}$ is the terminal velocity, and $\beta$ is the acceleration exponent. If $\beta=1$ , then assuming $v_{\infty} \gg v_0$, the wind will reach half the terminal velocity at $r_0/2$; if $\beta>1$, the wind will accelerate more rapidly.

\subsection{Coordinate grids}
Although many of the models described above are continuous, \xcode\ is grid-based. As part of initialization, the model is discretised onto a 2D coordinate grid with user-specified dimensions. The coordinates and velocities of each cell are calculated at the cell edges, whereas other quantities, such as the density or various estimators of the radiation field, are cell-centered. The grid is usually in cylindrical ($r_{\rm cyl}, z$) or polar ($r,\theta$) coordinates, depending on which is more appropriate for the model. The cylindrical grid is logarithmically spaced in both $r_{\rm cyl}$ and $z$, while the polar grid is logarithmically spaced only in the $r$ direction. Spherical grids are also available, usually used for simulating stellar winds, SNe or for testing purposes (see section~\ref{sec:tests}). 

For the imported models described in section~\ref{sec:imported}, the cell coordinates are defined by what is read in, as are the velocities, densities, and, optionally, the radiation and electron temperatures. In the inbuilt parametrized models,  grid cells are spaced monotonically, and typically uniform in logarithmic space. However, this is not a requirement for an imported grid. \xcode\ is able to read in grids for simulations which use different cell spacing for different regions of the grid, such as those using adaptive mesh refinement. Unstructured grids, however, are not supported.
    
 \subsection{Additional Options}
 
 \subsubsection{Micro-clumping}
\xcode\ was originally developed to model smooth outflows, such that the density at any given point is determined by only the kinematic parameters, the geometry of the outflow, and the mass loss rate of the wind. In reality, various instabilities are likely to break up a smooth flow into clumps; for example, the line-deshadowing instability  \citep[LDI][]{owocki_solar_1983} and thermal instabilities \citep{mccourt_characteristic_2018, waters_cloud_2019, dannen_clumpy_2019} are both important clumping mechanisms in outflows. Clumping also offers a potential solution to the long-standing challenge of ``over-ionization'' in quasar line-driven winds. Quasar winds can easily become overionized when exposed to the intense (E)UV and X-ray radiation field near the central engine \citep{proga02, higginbottom13}. For line-driven winds, if the wind is too highly ionized it can be impossible to sustain the wind \citep{proga02,higginbottom_line-driven_2014,higginbottom24}. Irrespective of driving mechanism, overionization also can prevent the formation of absorption and emission features.

From both a computational and astrophysical standpoint, addressing the problem of clumping is difficult. First and foremost, the physical scale length and density contrast of clumps are not usually well-constrained from either observations or theory. Nor is it simple to know where to place clumps in a kinematic model, or how their shape, size and density could change as a function of, for example, radius. Secondly, not only does clumping introduce additional complexity into already complex models, the spatial resolution of the computational grid is required to be fine enough to resolve each individual clump. In \xcode, this is not yet feasible due to current memory restrictions limiting the size of the simulation grid. Hence, as a first step, we have implemented a necessarily simple approximation known as \textit{micro-clumping} \citep{hamann_spectrum_1998, hillier_constraints_1999, oskinova_x-raying_2008}, a parameterization commonly used in stellar wind modelling.
    
One of the key assumptions in the micro-clumping framework is that the clumps are optically thin and smaller than all relevant length scales. In \xcode, the smallest such length scale is often the so-called Sobolev length, $l_{s}$, which defines the size of the narrow resonance spatial region within which an $r$-packet can interact with a bound-bound transition. For example, in a simulation where $l_{s} \sim 10^{10}-10^{12}$ cm and where the optical depth for a resonance can reach upwards of $\tau \sim 10^{6}$, clumps would have to be smaller than $l_{c} \ll l_{s} / \tau \sim 10^{4}$ cm in order to remain optically thin. Moreover, the micro-clumping treatment also assumes that clumps are embedded within a vacuum. If these assumptions are satisfied, clumps can be treated simply in terms of a volume filling factor, $f_V$, which we take to be independent of position in the grid. 

Since the clumps are not resolved by the grid, this framework can  easily be applied to a wide range of wind models without any modification to the grid itself. The densities of clumps are enhanced by a factor $D = 1 / f_V$ relative to the equivalent smooth flow. The opacities, $\kappa$, and emissivities, $j$, are then given by, $\kappa = f_V \kappa_{c} (D)$ and $j = f_V j_{c} (D)$, where the subscript $c$ denotes that the quantity is calculated using the enhanced density of the clump. At a fixed temperature and ionization state, processes which scale linearly with density (e.g. electron scattering) remain unchanged, since the effects on density and emitting volume cancel exactly ($D / f_V = 1$). However, processes which scale with the square of the density (e.g. collisional excitation or recombination) are enhanced by a factor $D$. Additionally, the increased density of clumps increases the rate of recombination and moderates the ionization state. 
 
 \subsubsection{Complex `multi-domain' models}
    
In a typical \xcode\ simulation, only one wind model will be included. However, it is possible to simulate more complex geometries that include multiple wind models. One can create a simulation which includes a fast polar wind but, also a slow and dense flow close to the disc plane. In such a simulation, each wind is contained within its own ``domain'', which $r$-packets are able to freely traverse between. Each domain is non-overlapping, and has its own grid and coordinate system. This means one could create a simulation using both a \citetalias{shlosman93} type wind and an imported grid, or by using multiple \citetalias{shlosman93} winds.

\section{Code Testing and Benchmarking}
\label{sec:tests}
In order to validate the code and better understand its limitations, we have carried out tests against several other publicly available radiative transfer codes. Some of these comparisons are available elsewhere: we refer the reader to \cite{long02} for tests of simpler ionization calculations with the code; to \citetalias{matthews15} for tests of level populations solvers for H and He macro-atoms; to \cite{higginbottom13} for tests of ionization balance; and to \cite{matthews_phd_2016} for a summary of macro-atom code validation exercises with earlier code versions. Finally, we note that \cite[][see their appendix A]{parkinson2024} have recently presented a comparison between \xcode\ and {\sc Sedona} for 1D models with high velocities and optical depths. These provide a good test of \xcode 's ability to describe continuum bulk reprocessing effects due to many Compton scatters. 

We have conducted many more tests of the code that we do not present in this paper, and a small suite of unit tests for more basic code operation are now regularly conducted within a continuous integration framework. Below, we focus on up-to-date tests of \xcode 's ionization, thermal and spectral synthesis calculations. For these tests, we compare the results obtained with  \xcode\ to those from  the photoionization code \cloudy\ \citep{ferland17}, the stellar atmospheres/wind code \cmfgen\ \citep{hillier98}, and the SN code \tardis\ \citep{kerzendorf_spectral_2014}.

\subsection{{Ionization and thermal balance comparisons to \cloudy}}
\label{sec:cloudy}
\cloudy\ \citep{ferland17} is a photoionization and spectral synthesis code that computes, among other things, the ion abundances and thermal balance of a cloud, as well as its transmitted spectrum, when irradiated by an external source of ionizing photons. \cloudy\ is the most widely-used program of its type. It is often employed in modelling of, e.g., nebulae, AGN broad-line regions and the intracluster medium. \cloudy\ represents the state-of-the-art in 1D photoionization calculations. As a result, \cloudy\ is an ideal code for us to compare to check our ionization balance and heating and cooling rates. 

We compare to  \cloudy\ by considering an optically and geometrically thin shell of plasma illuminated by a central power-law spectrum, broadly repeating the same test presented by \cite{higginbottom13} for Carbon. We compute the ionization state at a fixed temperature of $T_e=10,000~{\rm K}$ and Hydrogen number density of $n_H = 10^7~{\rm cm}^{-3}$. We then compare the ion abundances obtained as a function of the ionization parameter typically used in \cloudy, $U$, defined as the ratio of the number density of ionizing photons to number density of Hydrogen, given by (in the optically thin limit)
\begin{equation}
    U = \frac{Q_H}{4 \pi r^2 n_H c}.
\end{equation}
Here, $Q_H$ is the number of Hydrogen ionizing photons emitted by the central source per unit time. We adjust this parameter by varying the total luminosity for a fixed ionizing continuum shape and shell radius. To facilitate the comparison, we introduce the ``ionization degree'', $\chi$ for each element with atomic number $Z$, defined as 
\begin{equation}
    \chi = \frac{1}{Z} \left[ \left(\sum^Z_i N_i / N_{\rm elem} \right) - 1 \right].
\end{equation}
Here, $N_i$ is the number density of ion $i$ and $N_{\rm elem}$ is the total number density of the element, so that $\chi=0$ for a neutral ion and $\chi=1$ for an fully-stripped ion. 

 \begin{figure*}
    \centering
	\includegraphics[width=\linewidth]{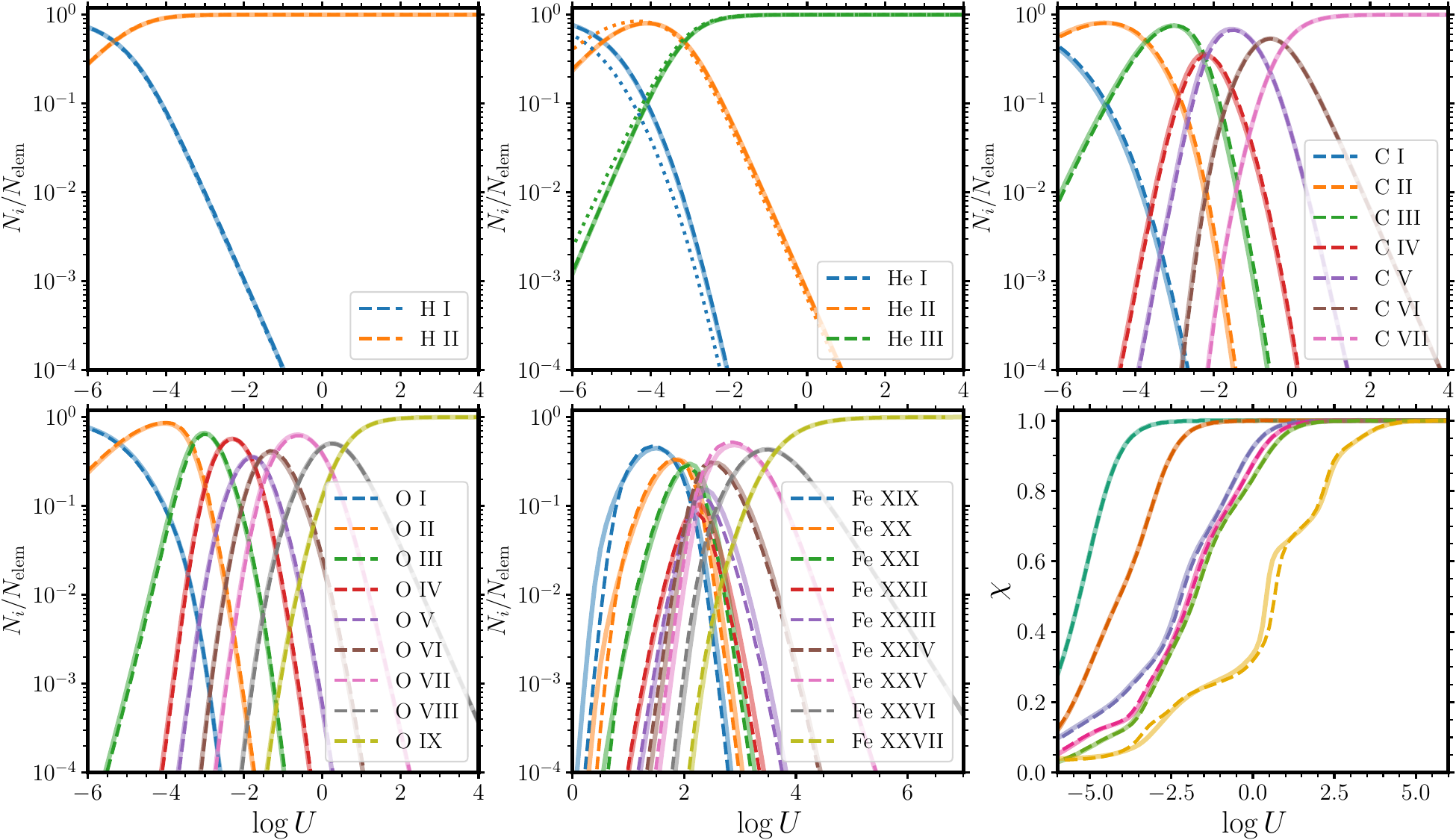}
    \caption[\cloudy\ versus \xcode\ ion abundances test]{
    \cloudy\ versus \xcode\ ion abundances test. The first five panels show ion adundances $N_i / N_{\rm elem}$ for all ions of H, He, C and O, and the upper nine ionization stages of Fe. The bottom right panel shows the ionization degree, $\chi$, for the same elements plus N. $\chi$ is defined such that  $\chi=0$ for a neutral ion and $\chi=1$ for an fully-stripped ion. In each case \cloudy\ results are shown in solid translucent lines and \xcode\ in dashed lines, where the \xcode\ results use the spectral modelling plus recombination rates approach. Additionally, for H and He, we show, with dotted lines, ion abundances calculated with the macro-atom estimator rate matrix. These agree almost perfectly for H and for He when $\log U \gtrsim -3$, but there are some discrepancies at low $U$ for He driven by incomplete recombination rates in the macro-atom framework. 
    }
    \label{fig:cloudy}
\end{figure*}

In Fig.~\ref{fig:cloudy}, we show ion adundances $N_i / N_{\rm elem}$ for all ions of H, He, C, O and Fe as well as $\chi$ for the same elements plus N. \cloudy\ abundances are shown as solid translucent lines, while \xcode\ abundances are shown either as  dashed lines (for ionization states calculated with the spectral model plus recombination rates approach; mode A) and as dotted lines (for macro-atom ionization and excitation states calculation from Lucy-style estimators; mode B). In mode A, we agree extremely well with \cloudy\ for all ions apart from the lower and middle stages of Fe, and the comparison for Carbon now looks somewhat improved compared to that presented by \cite{higginbottom13}. In mode B, we agree perfectly for H, but start to disagree for low ionization stages of He. The comparisons of C,N,O and Fe for mode B are not shown here for brevity, and because these datasets are still in development, but we find similar results to He for C, N, O, Fe when they are treated as full macro-atoms -- that is, there is an offset in the ion abundances such that \xcode\ is slightly overionized compared to \cloudy\, likely due to the neglect of dielectronic recombination rates in the macro-atom rate matrix. Overall, the agreement between \xcode\ and \cloudy\ is extremely encouraging. 

\begin{figure}
    \centering
	\includegraphics[width=\linewidth]{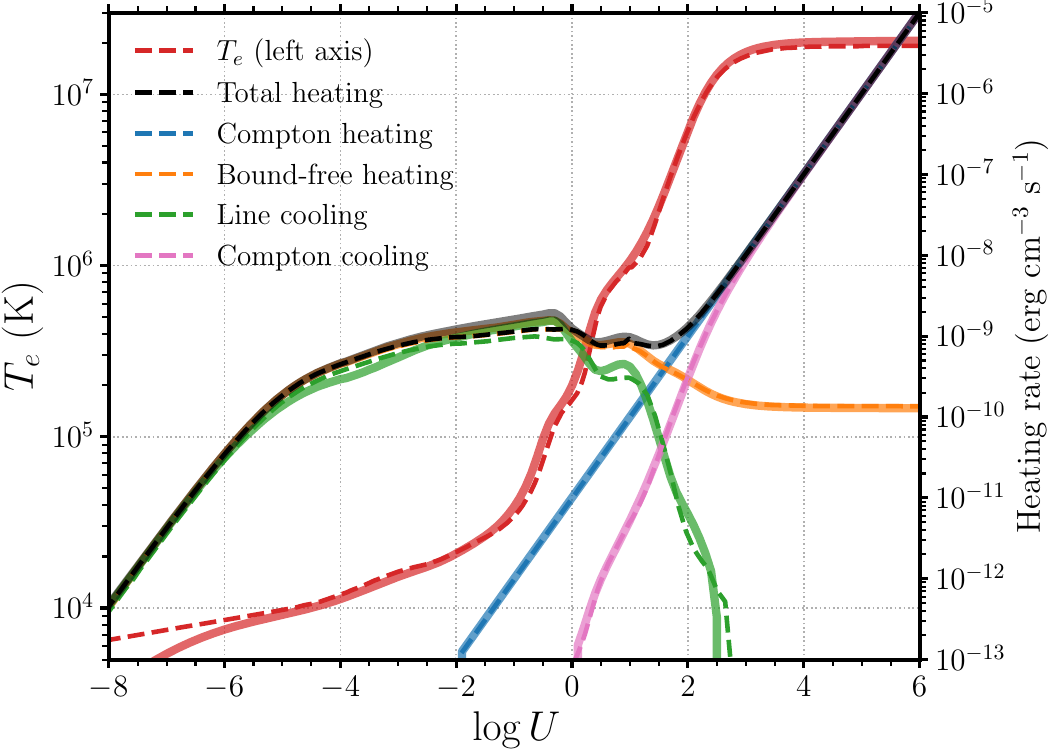}
    \caption[\cloudy\ versus \xcode\ heating and cooling test]{
    \cloudy\ versus \xcode\ heating and cooling test. The electron temperature corresponds to the left axis and heating and cooling rates to the right. \cloudy\ values are shown in solid translucent lines, and \xcode\ in dashed lines. The rates and electron temperature generally agree very well apart from at very low $U$. 
    }
    \label{fig:cloudy_heat}
\end{figure}

We also present tests of the heating and cooling rates in \xcode\ and the calculation of an equilibrium temperature. For this purpose, we take the same thin shell models, but relax the fixed temperature assumption. Instead, we now also solve for the temperature in both codes. We simulate the same range of $U$ and record $T_e$, as well as the important heating and cooling rates. The results are shown in Fig.~\ref{fig:cloudy_heat}. At low temperatures, the thermal balance is a competition between primarily bound-free heating and line cooling, whereas at high temperatures Compton processes become important, eventually balancing each other at the Compton temperature. We find very good agreement between \xcode\ and \cloudy\ in this test for classic mode, with all the important heating and cooling rates matching closely with only very minor differences in line cooling. The temperature tracks closely apart from at very low $U$, where we do not include all of the heating and cooling processes present in \cloudy. 

\subsection{1D Supernova comparison to \tardis} 
\label{sec:tardis}
\tardis\ \citep{kerzendorf_spectral_2014} is a 1D MCRT code intended for modelling the spectra of SNe that makes full use of the macro-atom and indivisible packet frameworks.  $r$-packets are injected at an inner spherical boundary and pass through an atmosphere that is assumed to be in homologous expansion (that is, $v \propto r$) with a prescribed density profile, $\rho (r)$.  \tardis\ uses a nebular approximation for determining ionic abundance, following \cite{mazzali93}. This is very similar to the ``on the spot''  option that exists in \xcode. \tardis\ uses Monte Carlo estimators to obtain values for the radiation temperature and dilution factor throughout the model, in a very similar manner to \xcode. However, in its standard mode, \tardis\ follows \cite{mazzali93} in assuming that $T_e = 0.9 T_r$, thus avoiding the need for a thermal equilibrium calculation.
 
 For our comparison, we use a simple supernova model, following \cite{kerzendorf_spectral_2014}. Specifically, the model considers a computational domain extending from $r_{\rm min} = 1.2 \times 10^{15}~{\rm cm}$ to $r_{\rm max} = 2.2 \times 10^{15}~{\rm cm}$ in which the mass density
 \begin{equation}
 \rho = 8 \times 10^{-14} (r/r_{\rm min})^{-7} {\rm g~cm}^{-3} 
 \end{equation}
 is assumed along with a spatially uniform  composition (by mass) of intermediate mass elements O:Mg:Si:S:Ca:Ar of 19:3:52:19:3:4. The velocity law is homologous, extending from $v = 1.1 \times 10^{4}~{\rm km}~{\rm s}^{-1}$ at $r_{\rm min}$ to $v = 2 \times 10^{4}~{\rm km}~{\rm s}^{-1}$ at $r_{\rm max}$. We adopt a temperature of $10^4~{\rm K}$ for the radiation field injected at the inner boundary. To mimic \tardis, we impose $T_e=0.9~T_r$ for this test in \xcode\ and use the \cite{mazzali93} modified Saha equation (with their $\delta$ parameter set to one in both codes). 
 
\begin{figure}
    \centering
	\includegraphics[width=0.9\linewidth]{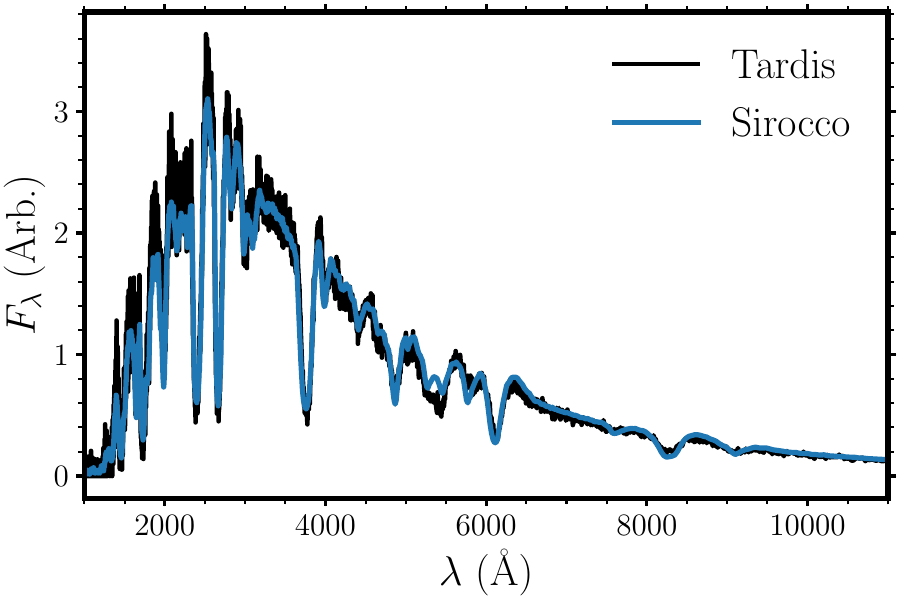}
    \caption[\tardis\ versus \xcode\ comparison for 1D supernova model]{
    \tardis\ versus \xcode\ comparison for 1D supernova model. This example uses a simple homologous model with simplified ionization balance, and shows that the two codes produce a very similar UV-optical spectrum.
    }
    \label{fig:tardispython}
\end{figure}

A comparison between the spectra created with \xcode\ and \tardis\ is  shown in Fig.~\ref{fig:tardispython}.  Although there are some small differences between the two spectra, all the main atomic line features seen in the \tardis\ spectrum are also seen in \xcode. The shapes and depths of the absorption profiles are generally very similar and the overall agreement between the two codes is very good. 
 
 \subsection{1D stellar wind comparison to \cmfgen}
 
 \cmfgen\ \citep{hillier98} is a 1D, non-LTE radiative transfer code originally written for calculating the ionization structure and emission for stars with stellar winds.  As in \xcode, \cmfgen\ assumes the wind's velocity field can be described as a $\beta$ velocity law, and fully accounts for co-moving frame effects.  Unlike \xcode, it is based on more traditional radiative transfer techniques, using $\lambda$-iteration.

We use a stellar wind model with a velocity law given by equation~\ref{eq:stellar_wind_vlaw}. The central star has a surface temperature of $34,000~{\rm K}$, a mass of $38~M_\odot$ and a radius of $R_*=4\times10^{12}~{\rm cm}$. The wind is launched from $R_*$ with $v_0=0.1~{\rm km~s}^{-1}$, $v_\infty = 900~{\rm km~s}^{-1}$ and $\beta=2$. We use the hybrid macro-atom scheme with H and He macro-atoms. We use as an input spectrum at the stellar surface the same continuum that emerges from the \cmfgen\ stellar atmosphere. The overall model resembles an O star wind. Out comparison is in a similar spirit to that presented by \cite{kurosawa_three-dimensional_2009}, except that we do not import the local ionizing spectrum from \cmfgen\ or fix the \ion{He}{ii} continuum as in that test, but allow \xcode\ to self-consistently solve for the radiation field throughout the wind. 

\begin{figure}
    \centering
	\includegraphics[width=\linewidth]{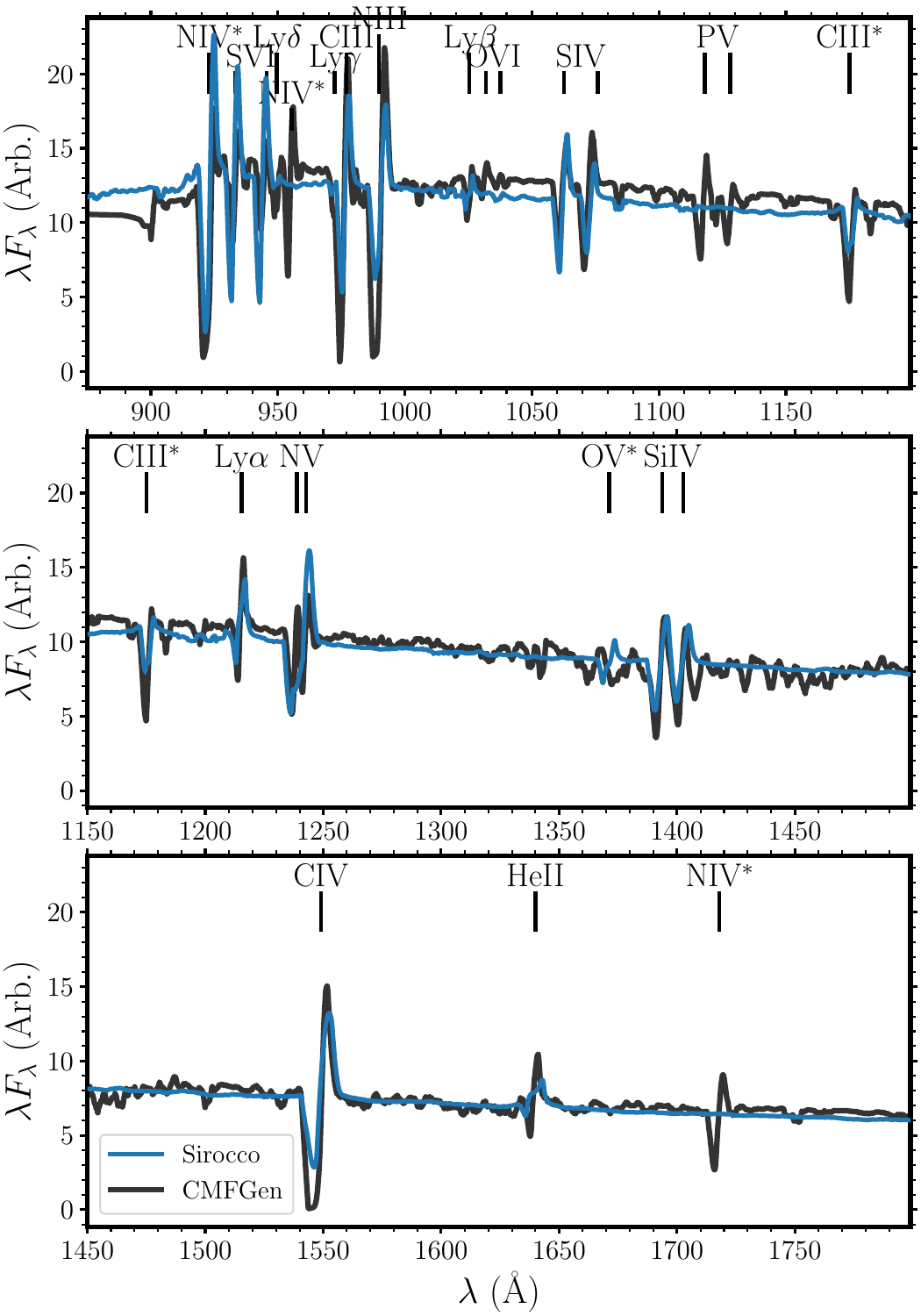}
    \caption[CMFGEN Comparison]
    {
    Comparison between the spectrum obtained from \xcode\ and \cmfgen\ calculations using a one dimensional O-star wind model. 
    }
    \label{fig:cmfgen}
\end{figure}

The UV spectrum from our calculation is shown in Fig.~\ref{fig:cmfgen}, compared to the output from \cmfgen. We find that all the lines present in our model produce comparable P-Cygni profiles, in species such as \civ, \nv, \ion{Si}{iv}, \ion{He}{ii} and \ion{S}{iv}. Some profiles agree better than others, and in general we see some discrepancies between the depths of the absorption features in particular. The continuum matches fairly well across the UV range, with some small differences in level particular in the $1000-1200$\AA\ and $1000-1200$\AA\ regions of the spectrum. We do not include all the semi-forbidden lines present in \cmfgen, or Phosphorous, so these lines are missing the spectrum.  

\begin{figure}
    \centering
	\includegraphics[width=0.9\linewidth]{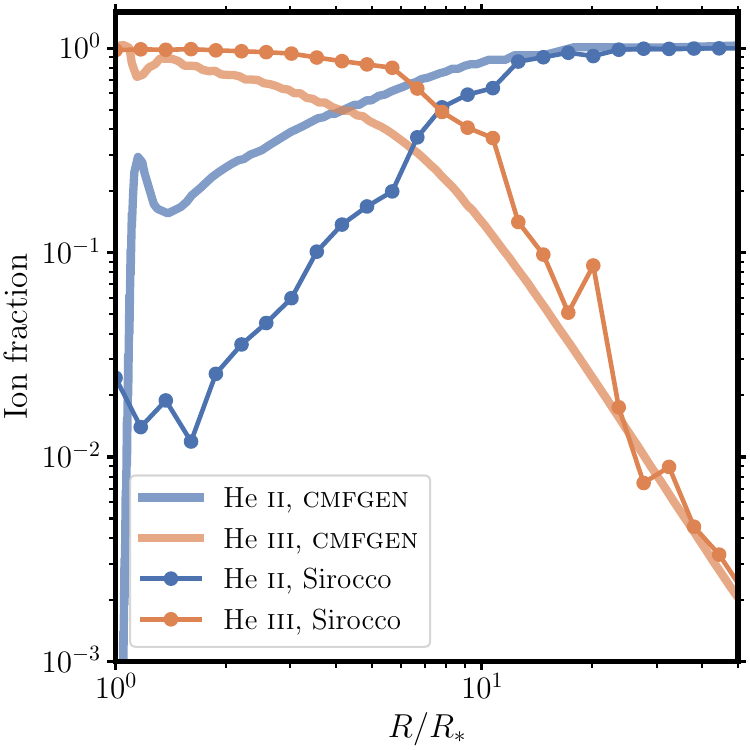}
    \caption[CMFGEN Ions]
    {
    Comparison between the He ionization state obtained from \xcode\ and \cmfgen\ calculations using a one dimensional O-star wind model. 
    }
    \label{fig:cmfgen2}
\end{figure}

In comparing to \cmfgen, we found that relatively good agreement in the spectrum could sometimes hide differences in ionization state. In Fig.~\ref{fig:cmfgen2}, we show a comparison between the He ion fractions as a function of radius in both models. \xcode\ clearly reproduces the qualitative trends in \cmfgen, in that there is a gradual evolution from \ion{He}{iii} being the dominant ion at the inner edge, transitioning to a dominant \ion{He}{ii} state further out in the wind. While agreement in the outer regions is rather good, there is a disagreement in the inner regions where \xcode\ obtains a more highly ionized solution than \cmfgen. The dominant ion transition happens at $R/R_* \approx 8$ in \xcode\ but at $R/R_* \approx 4$ in \cmfgen. In addition, there is a spike in the \ion{He}{ii} ion fractions in the inner region that is not well-produced by \xcode.

\begin{table}
    \centering
    \resizebox{\columnwidth}{!}{
        \begin{tabular}{lcccc}
            \hline
            Parameter & CV & Quasar & XRB & TDE  \\ 
            \hline
            $M_{\rm co}~(M_{\odot})$     & $0.8$ & $10^9$ & $10$ & $3\times10^{6}$  \\
            $R_*$~(cm)     & $0.8$ & $8.85\times 10^{14}$ & $10^{7}$ & $3\times10^{6}$  \\
            $r_{\rm min}$~(cm)           & $2.8\times 10^9$ & $4.4\times 10^{16}$ & $4.82\times 10^{10}$ & $10^{13}$  \\
            $r_{\rm max}$~(cm)           & $8.4\times 10^9$ & $8.9\times 10^{16}$ & $9.64\times 10^{11}$ & $10^{15}$  \\
            $R_{\rm disc}$~(cm)          & $2.4\times 10^{10}$ & $10^{20}$ & $10^{11}$ & $10^{15}$  \\
            $R_{v}$~(cm)                 & $7\times 10^{10}$ & $10^{19}$ & $5\times 10^{12}$ & $5 \times 10^{16}$ \\
            $\theta_{\rm min~(^{\circ})}$      & $20$ & $45$ & $60$ & $20$  \\
            $\theta_{\rm max}~(^{\circ})$      & $65$ & $60$ & $89.9$ & $65$  \\
            $v_0$      & $10~{\rm km~s}^{-1}$ & $c_s(r_0)$ & $10~{\rm km~s}^{-1}$ & $c_s(r_0)$  \\
            $v_\infty/v_{\rm esc}(r_0)$ & $3$ & $1$ & $1$ & $0.3$  \\
            $\alpha$                & $1.5$ & $1$ & $2$ & $4$ \\
            $\gamma$                & $1$ & $1$ & $1$ & $1$  \\
            $\lambda_s$               & $0$ & $0$  & $-1$ & $2$  \\
            $f_V$                 & $1$ & $0.01$ & $1$ & $0.1$  \\
            $\dot{M}_{\rm acc}~(M_{\odot}~\text{yr}^{-1})$ &  $10^{-8}$ & $5$ & -- & $10^{-2}$ \\
            $L_{\rm bol}~(\text{erg~s}^{-1})$ & $4.5\times10^{34}$ & $2.3\times10^{46}$ & $3.2\times10^{37}$ & $4.7\times10^{43}$  \\
            $L_{\rm x}~(\text{erg~s}^{-1})$ & -- & $10^{43}$ & $1.5\times10^{37}$ & --  \\
            $\dot{M}_{\rm w}~(M_{\odot}~\text{yr}^{-1})$   & $10^{-9}$ & $5$ & $1.4 \times 10^{-8}$ & $10^{-2}$ \\
            ${\cal N}_{\gamma}$ & $10^7$ & $1.5\times10^7$ & $10^8$ & $5\times10^7$ \\
            ${\cal N}_{i}$ & $20$ & $30$ & $20$ & $25$ \\
            ${\cal N}_{s}$ & $10$ & $20$ & $1$ &  $5$\\
            Grid~size & $30\times30$ & $100\times100$ & $30\times30$ &  $100\times100$ \\
            Time (core hr) & $\approx7-23$ & $\approx85$ & $\approx22$ & $\approx360$ \\
            Mode & Varied & \multicolumn{3}{c}{Hybrid Macro} \\
            Macro-atoms & Varied & H,He & H,Fe & H,He \\
            Convergence & $100\%$ & $81.4\%$ & $91.7\%$ & $98.4\%$ \\
            Section   & \ref{sec:cv} & \ref{sec:quasar} & \ref{sec:xrb} & \ref{sec:tde} \\
            \hline
        \end{tabular} 
    }
    \caption{Wind parameters for models in section \ref{sec: demo_models} using the \citetalias{shlosman93} parameterization. Each input parameter file is available as part of the code release. $L_{\rm x}$ is the $2-10$~keV X-ray luminosity, and all other symbols are described in the text.}
    \label{tab:demo_model_parameters}
\end{table}

\section{Illustrative Models}
\label{sec: demo_models}

\xcode\  was originally intended to model the UV and optical spectra of CVs, but our goal since has been to create a much more versatile, general purpose program that can be used to model a wide variety of systems. We thus start by considering a simple example of the use of \xcode\ to model a CV, before moving on to models of a quasar, XRB and TDE. Each of the four models correspond to those shown in Fig.~\ref{fig:full_demo}, and the model parameters are given in Table~\ref{tab:demo_model_parameters}.

\subsection{A simple example of the code: A cataclysmic variable wind}
\label{sec:cv}

\begin{figure*}  
    \centering 
    \includegraphics[width=1.0\linewidth]{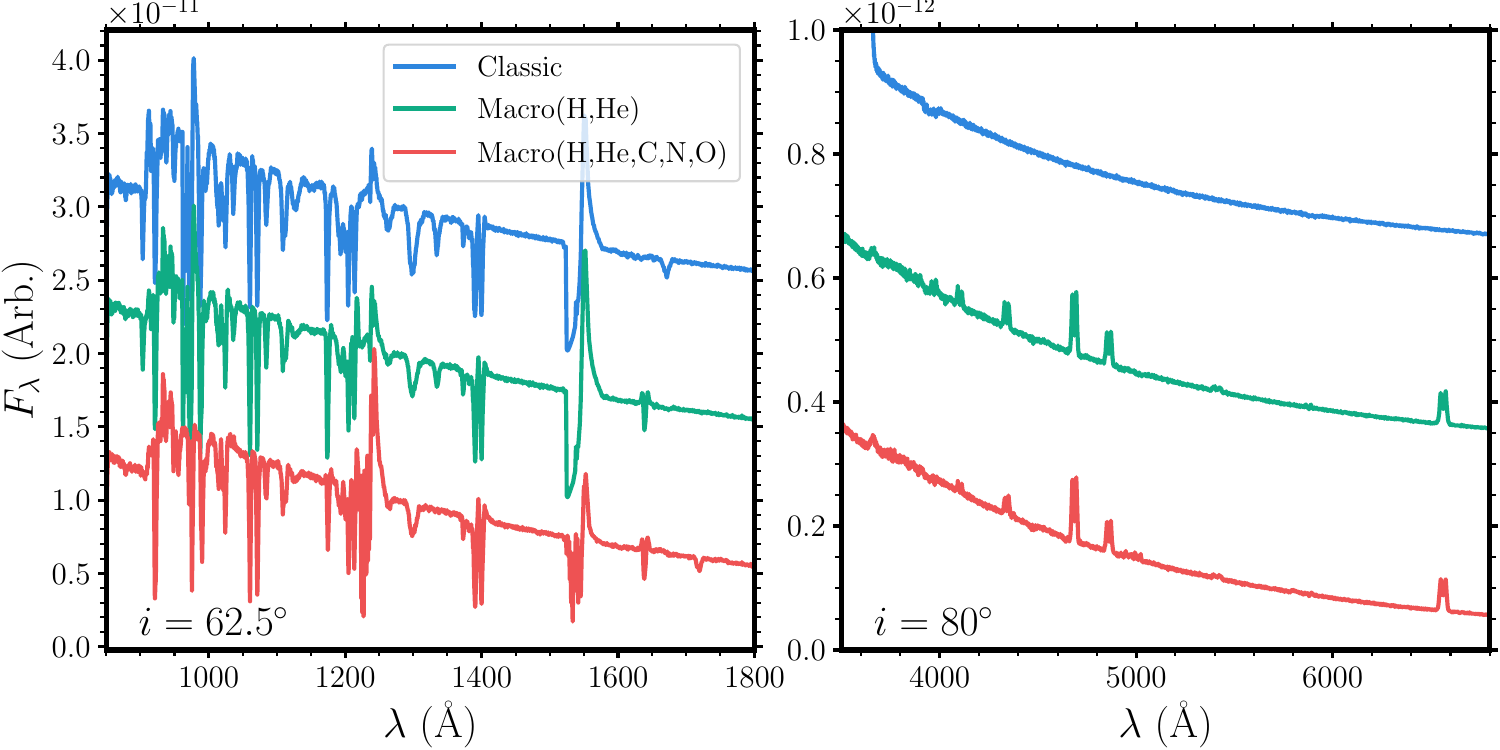}
        \caption{
        Spectra from the CV model in the UV (left) and optical (right) wavelength ranges. The spectra are computed with three different radiative transfer modes: classic mode, a macro atom calculation in which detailed models are used for H and He, and a macro-atom calculation in which C, N and O are also included as full macro-atoms. The UV spectrum is computed at $62.5^\circ$, looking into the wind cone, so that blue shifted absorption lines are produced in addition to emission. The optical spectrum is shown at high inclination ($80^\circ$) and the lines are purely in emission. 
        }
    \label{fig:cv_example}
\end{figure*}

\begin{figure}  
    \centering 
    \includegraphics[width=1.0\linewidth]{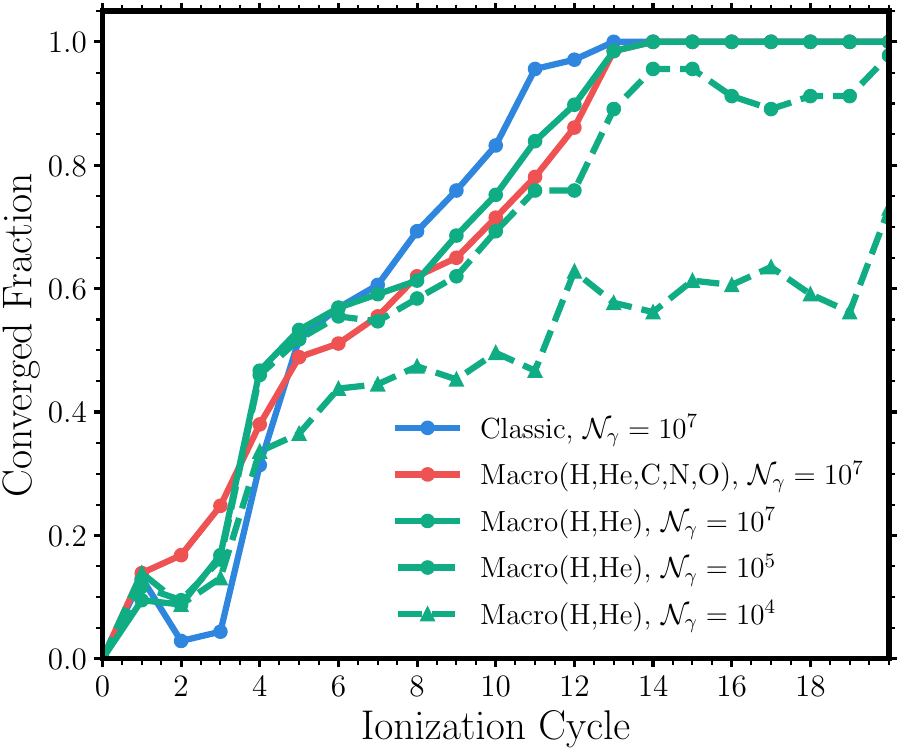}
        \caption[Convergence CV model]{
        The fraction of cells that are converged in a simple model for a CV as a function of ionization cycle assuming that varying numbers of photons per cycle ranging from 10$^5$ to 10$^7$.  
        }
    \label{fig:converge}
\end{figure}

Cataclysmic variables (CVs) are interacting binary systems in which a white dwarf (WD) primary accretes material from a Roche-lobe-filling secondary, usually via a disc. In the so-called nova-like CVs (NLs), the mass-transfer rate through the disc is relatively high, and the accretion process takes place in a steady state. By contrast, in CVs belonging to the dwarf nova (DN) class, the disc cycles between a low-$\dot{M}$ quiescent state and a high-$\dot{M}$ outburst state on time-scales ranging from weeks to years. Clear evidence for outflows is seen whenever the accretion rate is sufficiently high, i.e. in both NLs and erupting DNe. For example, the UV resonance lines in these systems exhibit blue-shifted absorption troughs \citep[when viewed face-on][]{cordova82,hartley02} or weak eclipses \citep[when viewed edge on][]{cordova85,knigge98}. Similarly, the evolution of line profiles during eclipses strongly suggests that line formation takes place in a rotating bipolar outflow \citep{cordova85,drew88, knigge_eclipse_1997}. P~Cygni features have also been observed in the optical H and He lines \citep{kafka04,cuneo23}, and the recombination continuum produced by the wind may also make a significant contribution to the overall UV-optical SED (\citealt{hassall1985,knigge98}; \citetalias{matthews15}). 

\xcode\  has been applied to CVs several times. The formation of UV absorption and emission line signatures was studied by \cite{noebauer10}, while the impact of these outflows on optical lines and continua was explored by \citetalias{matthews15}. More recently, the code has been used to model the optical spectrum of a DN in outburst \citep{tampo2024}. To illustrate the application of \xcode\  to CVs, here we focus on one of the models from \citetalias{matthews15}. Briefly, this is an SV-type model describing a biconical outflow from an accretion disc surrounding a WD. The disc extends from the WD surface to an outer radius of $R_{d} = 34.3~R_{WD}$ and radiates as a multi-temperature blackbody. The wind is launched from the disc between $4~R_{WD}$ and $12~R_{WD}$ and the inner and outer opening angles of the wind cone are $20^\circ$ and $65^\circ$, respectively. Additional model parameters are listed in Table~\ref{tab:demo_model_parameters}. The simulation presented here used $N_i=20$ ionization cycles and $N_s=10$ spectral cycles, with ${\cal N}_\gamma=10^7$. The run time for such a simulation on a typical laptop depends on the mode; the classic mode model takes $\approx 7$~core hr, whereas a hybrid macro-atom model using H, He, C, N and O macro-atoms is more expensive, taking  $\approx 23$~core hr.  

Fig.~\ref{fig:cv_example} shows the spectra obtained from this model, with three different radiative transfer modes: classic mode, a macro atom calculation in which detailed models are used for H and He, and a macro-atom calculation in which C, N and O are also included as full macro-atoms. The left panel shows the three UV spectra at an inclination of $i=62.5^\circ$, looking into the wind cone. Key features here are the UV resonance line wind features. At visible wavelengths, the model shows very little other than continuum emission, and so the visible spectra are shown for a higher inclination of $i=80^\circ$, where the disc is relatively fainter. In the UV, we observe a range of absorption and emission features, particularly in resonance lines such as \civline\ and \nvline. Indeed, the UV spectra in all three modes are rather similar, except for some differences caused by changes in ionization state when C, N and O are treated as full macro-atoms. Although we do not present the trends with inclination here, the UV lines evolve from pure blue-shifted absorption features to P~Cygni profiles to pure emission lines with increasing inclination, as shown by \citetalias{matthews15}. Overall, the properties of the computed UV spectra are similar to those seen in observations of high-state CVs (nova-likes and DNe in outburst). 

In the optical, classic mode produces no features apart from an unrealistically large Balmer jump in emission. By contrast, the macro-atom models produces recombination lines in features such as H$\alpha$, H$\beta$ and \ion{He}{ii}~4686\AA. The reason for this difference, as discussed by \citetalias{matthews15}, is that the macro-atom formalism allows us to accurately model the recombination cascade responsible for producing these lines. Conversely, when line formation is modelled in the two-level approximation then the population of the upper level by recombination (and cascades from higher levels) is not accounted for. The spectra thus serve as a reminder of the advantages of the hybrid macro-atom mode, particularly for bound-free processes. The optical emission lines are double-peaked, because they originate in the dense base of the outflow, which is rotationally dominated. The lines in the optical from this particular wind prescription are rather weak compared to those observed, but as we have found in other studies, modifying, for example, the velocity law of the outflow, or its clumpiness, can significantly enhance the optical features and produce a good match to observed optical spectra of CV. 
 
 \subsubsection{Convergence} \label{sec:converge_cv}
Fig.~\ref{fig:converge} illustrates how the level of convergence (see section~\ref{sec:converge}) changes as a function of ionization cycles for the same three calculations of the UV spectrum of a CV presented above. We also show two additional calculations with fewer photons (${\cal N}_\gamma = 10^5,10^4$) to illustrate how low numbers of packets can cause issues. In the models with large ${\cal N}_\gamma$, the convergence reaches $100\%$ after 13 cycles by our definition. As is typical, the convergence rises rapidly in the early cycles and then levels out; the cells that converge first tend to be those with larger numbers of photon passages and those in regions where the spectrum of photons passing through the cell is not changing from cycle to cycle. In the models with few $r$-packets, a significant number of cells are still converged, and in fact the spectrum from these models is rather similar. However, there are clearly problems, and in the ${\cal N}_\gamma = 10^4$ run the convergence asymptotes around $60\%$. 
   
\subsection{Emission and absorption lines in quasars}
\label{sec:quasar}
\begin{figure}
    \centering
    \includegraphics[width=\linewidth]{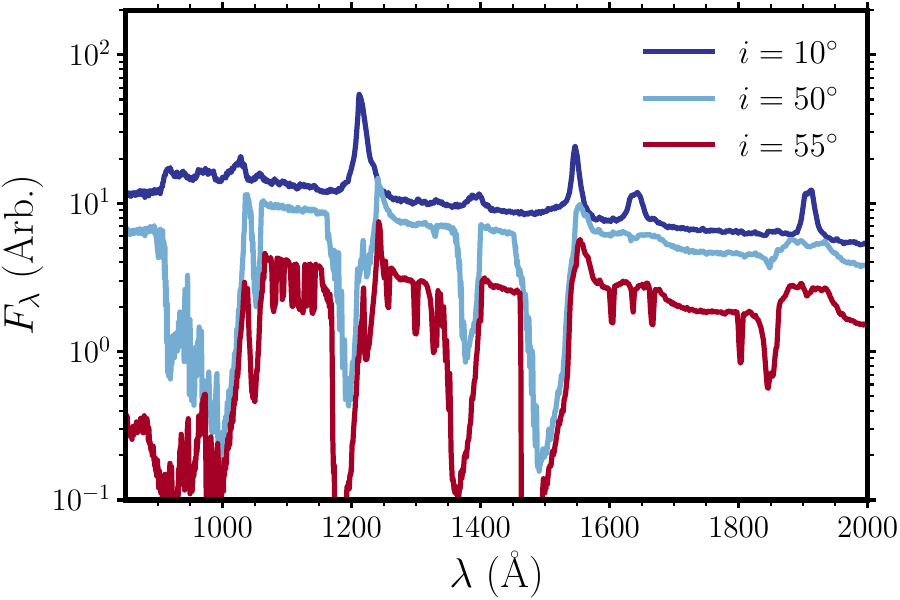}
    \caption[Quasar spectra]{The synthetic spectra at three inclinations for a quasar model, described by the parameters in Table \ref{tab:demo_model_parameters}.}
    \label{fig:demo_quasar}
\end{figure}

AGN are powered by accreting super-massive black holes (SMBHs), and quasars are the highest-luminosity (i.e. highest $M_{\rm BH}$ and/or $\dot{M}_{\rm acc}$) objects among these systems.
Approximately $\simeq20\%$ of quasars \citep{trump_catalog_2006,dai2008,knigge_intrinsic_2008,allen_strong_2011} exhibit blueshifted broad absorption lines (BALs) associated with high-ionization UV resonance transitions (e.g. \ion{N}{v}, \ion{Si}{iv}, \ion{C}{iv}) and sometimes in low-ionization (e.g. \ion{Mg}{ii}, \ion{Al}{iii}) species \citep{voit1993,reichard_continuum_2003}. A few systems even show BALs in optical lines 
\citep[e.g. the Balmer series and \ion{He}{i};][]{hall2007,leighly2011}. Other evidence for outflows from AGN includes the warm absorbers and highly blueshifted ``ultrafast outflow'' features \citep{reeves_massive_2003,pounds_high-velocity_2003,gofford_suzaku_2013}. These X-ray signatures have been modelled using MCRT techniques in the past \citep{sim_multidimensional_2008,sim_multidimensional_2010,sim_synthetic_2012}, and can be modelled using \xcode, although accurate Fe atomic data sets are still in development. X-ray photoabsorption is already treated relatively accurately in \xcode. For a discussion of the X-ray properties of quasar models, we refer the reader to \cite{matthews16,matthews23}.

The relatively low incidence of BALs among quasars is likely to be an orientation effect: BALs are only seen for sightlines that view the disc through the outflow, so the BAL fraction provides an immediate estimate of the solid angle subtended by the flow \citep[e.g.][]{turnshek1984,morris1988,weymann_comparisons_1991,krolik_what_1998}. 
If the outflows giving rise to BALs are actually present in {\em all} AGN and quasars, they may also be responsible for their distinctive broad  emission lines. The broad line region (BLR) in these systems could also be the dense base of the same accretion disc wind that produces the blue-shifted BALs \citep[e.g.][]{murray_accretion_1995,elvis_structure_2000,matthews16,matthews20}. Alternatively, strong wind components to the broad emission lines might only be present at high mass and high Eddington fraction, forming part of a multi-component BLR with kinematics and constituents that vary with BH mass and accretion rate \citep[e.g.][]{giustini2019,temple2023}. 

Testing these ideas, particularly the simpler geometric interpretations, has been the main thrust of applications of \xcode\  to AGN and quasars to date. More specifically, \cite{higginbottom13} and \cite{matthews16,matthews20} showed that disc winds can naturally produce both absorption and emission line signatures similar to those seen in observations of quasars. Below, we illustrate these applications of \xcode\  to accreting SMBHs with the help of a kinematic quasar model. We use the SV93 prescription and adopt the parameters corresponding to Model B from \cite{matthews23}. The system has a black hole mass of $M_{\rm BH} = 10^9~M_{\odot}$ and accretion rate  of $\dot{M}_{\rm acc} = 5~M_{\odot}~\mathrm{yr}^{-1} \simeq 0.2 \dot{M}_{\rm Edd}$. The wind mass loss rate is set to the accretion rate, and the wind is clumpy with $f_V=0.01$. Other key parameters are listed in Table \ref{tab:demo_model_parameters}, and the input file is available as part of the code documentation.

Fig.~\ref{fig:demo_quasar} shows synthetic UV spectra generated by \xcode\ for this quasar model. Pure emission line spectra are produced for face-on orientations, where the sightline to the inner disc does not pass through the outflow. However, for more edge-on orientations, looking into the wind, the quasar model spectra exhibit strong blue-shifted broad absorption lines. At $i=50^\circ$, the spectrum broadly resembles that of a BAL quasar, and at $i=55^\circ$ the sightline intersects more low ionization material and starts to produce strong low-ionization BAL features. In addition, at low inclination, the \ion{C}{iv}~1550\AA\ line has a blueshifted asymmetry (``blueshift'') as is typical of quasars with comparable bolometric luminosities and masses to BAL quasars \citep[e.g.][]{richards2011,rankine2020,temple2023}, although the detailed shape and skew of the line does not match observations. An opaque midplane is essential for forming this blueshift. 

As discussed in our previous works on the topic, there are various issues with a simple biconical wind model for the BLR. For example, the wind must be clumpy and/or highly mass-loaded to avoid over-ionization. Also, if the wind is equatorial, then the angular dependence of a disc-like continuum makes it hard to achieve the observed similarity in emission line properties in BAL and non-BAL quasars within a simple unification scenario \citep{matthews2017}. However, these problems are either fundamental in an astrophysical sense, or associated with the specific model used; they are not associated with the modelling capabilities of \xcode\ itself. In fact, we are able to model the complex gradients in temperature/ionization structure and account fully for self-shielding and multiple scattering. This allows us to compute a emergent spectra in a more self-consistent manner than when using 1D or quasi-1D approaches.

\subsubsection{Reverberation Mapping}
\label{sec:reverb}
\xcode\ can also be used to predict the emission line reverberation signatures associated with any given model outflow. \cite{mangham17} implemented the tracking of photon light-travel times through the flow, which allows the construction of velocity-delay maps that account for the responsivity of the gas within a linearised framework. Overall, they found velocity-delay maps which were fairly symmetric, with an outer envelope that follows the ``virial'' relation $v \propto \tau^{-1/2}$. In analyses of observational data, such maps are typically interpreted as Keplerian disc signatures, since inflow- or outflow-dominated kinematics are usually expected to produce maps with diagonal ``blue-leads-red'' or ``red-leads-blue'' asymmetries, respectively \citep[e.g.][]{welsh1991}. A key conclusion of \cite{mangham17} was that real disc winds may not give rise to obvious "outflow signatures" like this, since line formation tends to take place mainly in the dense, rotation-dominated wind base. In a follow-up work, \cite{mangham19} used \xcode\ to generate mock reverberation mapping data and then conducted a blind test of two commonly used RM modelling tools (\textsc{memecho} [\citealt{memecho}] and \textsc{caramel} [\citealt{pancoast2011,pancoast2014}]). This work is a demonstration of how MCRT modelling can be used to test the robustness of inference techniques. For brevity, we do not present a reverberation mapping example here, but instead refer the reader to these studies. Instructions for constructing velocity-delay maps can be found in the code documentation.

\subsection{A parameterized X-ray binary model}
\label{sec:xrb}
 \begin{figure}
\centering
\includegraphics[width=\linewidth]{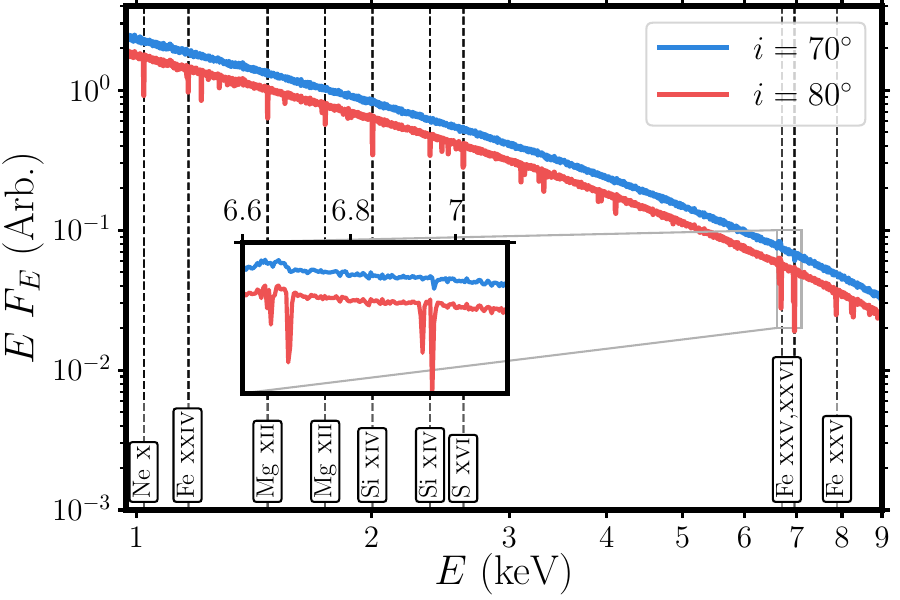}
\caption[X-ray spectrum, from 1-9 keV, from a parameterised XRB wind model]{
X-ray spectrum, from 1-9  keV, from a parameterised XRB wind model. The inset shows a zoom-in of the 6.6-7.1 keV region to show the \ion{Fe}{xxv} 6.7 keV and \ion{Fe}{xxvi} doublet. A series of absorption lines are produced by the wind, with prominent lines from highly ionized (hydrogenic) species of Ne, Mg, Si and S labelled. 
}
\label{fig:xrb}
\end{figure}

Low-mass X-ray binaries (LMXRBs) are systems consisting of a compact object primary (a BH or neutron star) with a main-sequence or sub-giant secondary. In LMXRBS, the secondary loses mass via Roche-lobe overflow, which is then accreted by the primary via a disc \citep[see ][for a review]{Bahramian2023}. These systems are characterised by strong X-ray emission originating, in a state-dependent manner, from the central parts of the accretion disc and a Comptonised "corona". 

LMXRBs are variable, typically spending most of the time in a quiescent state with a low luminosity and a relatively hard SED, but occasionally undergoing an outburst. During these outbursts, the source will typically brighten quickly, maintaining a hard spectrum (hard state) before transitioning, typically at a high luminosity around $\approx 0.2-0.5~L_{\rm Edd}$, to a soft state. The path from the outburst back to quiescence is often complex with changes in SED and luminosity.

There have been several observations of blue shifted absorption lines in XRBs during outbursts -- the smoking-gun signature of outflowing gas. Initially, wind signatures were observed only in the soft state and only in X-ray transitions associated with highly ionized species of Fe. However, since then, absorption features in optical \ion{He}{i}, \ion{He}{ii} and Balmer lines \citep{munoz2016,munoz2018,jimenez2019,charles2019} and in UV resonance lines \citep{castro2022,fijma2023} have also been observed \citep{munoz2016,munoz2018,jimenez2019,charles2019}, and the optical lines have also been seen in the hard state \citep{munoz2019}. The X-ray, UV and optical lines are thought to be associated with a disc wind, but the driving mechanism is an area of active research \citep[e.g.][]{diaztrigo2016,tomaru2023}, with the winds likely to be either thermal, Compton-heated flows or driven by MHD processes. There have been a number of efforts to model the X-ray features, often using quasi-1D treatments and/or simplified treatments of ionization and radiative transfer \citep{chakravorty2016,tomaru_thermal-radiative_2019,fukumura2021}. 

\xcode\ has been used as part of an effort to model thermally driven disc winds in XRBs \citep{higginbottom17,higginbottom18,higginbottom19,higginbottom20}. In such outflows, the central X-ray source Compton-heats the surface of the outer accretion disc to a temperature where the gas becomes unbound and escapes from the system. All of these simulations were hydrodynamic, with \xcode\ being used as the radiation module (see section~\ref{sec:rad-hydro}). We present here results from a parametrized Sv93 disc-wind  -- designed to mimic the sorts of thermal winds that emerge from radiation-hydrodynamic simulations -- to demonstrate the X-ray behaviour of \xcode. We refer the reader to \cite{koljonen2023} for a study of optical emission features, using similar techniques, from XRB winds and/or disc atmospheres. 

The full wind parameters are listed in  Table \ref{tab:demo_model_parameters}. The outflow is roughly equatorial, with a mass-loss rate of $1.4\times 10^{-8}~M_\odot~{\rm yr}^{-1}$, comparable to the expected accretion rate, and is illuminated by a $4.8$ keV Bremsstrahlung spectrum with a 2-10 keV X-ray luminosity of $1.5\times10^{37}$. 
We use a partial Fe macro-atom dataset that includes fluorescence and Auger processes, with Fe ions from \ion{Fe}{xxvii} upwards, and a 10 level H macro-atom, with other elements treated as simple atoms. 

In Fig.~\ref{fig:xrb}, we show the X-ray spectrum obtained from this simulation at two relatively edge-on viewing angles. At $i=80^\circ$, we see a series of soft X-ray absorption lines in ions such as \ion{Ne}{x}, \ion{Mg}{xii}, \ion{Si}{xiv} and \ion{S}{xvi}, with the most strongly absorbed features labelled in the figure. The inset shows the prominent \ion{Fe}{xxv} 6.7 keV and \ion{Fe}{xxvi} doublet features. At $i=70^\circ$, we only see the \ion{Fe}{xxv} and \ion{Fe}{xxv} high energy lines. 

\begin{figure*}
    \centering
    \includegraphics[width=\linewidth]{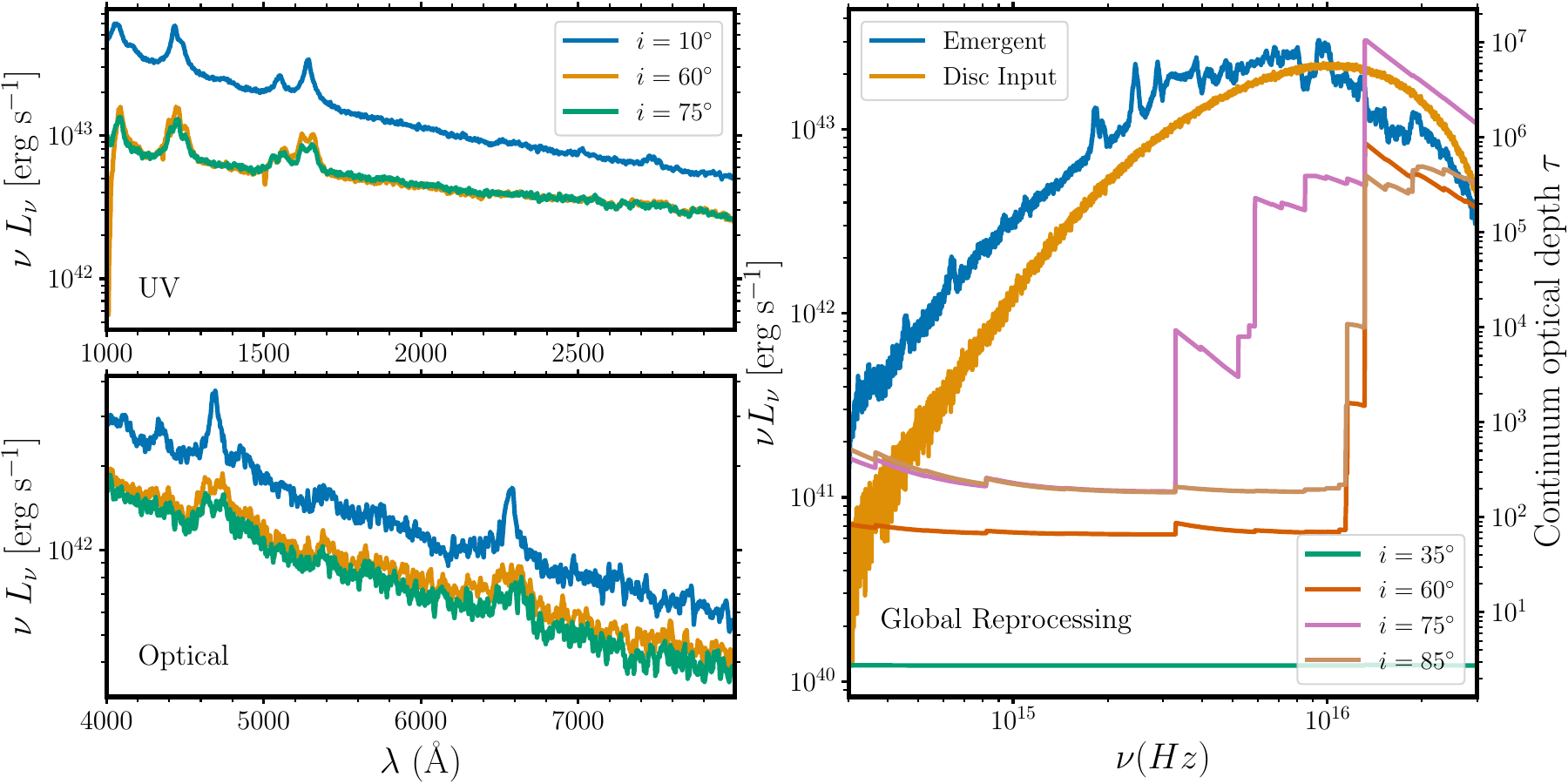}
    \caption[TDE Spectra]{The synthetic spectra and optical depths for our benchmark TDE model, described by the parameters in Table \ref{tab:demo_model_parameters}. The top and bottom left panels show the rest-frame UV and optical spectra, respectively, for several sight lines. The right panel shows the continuum optical depth, again, for several sight lines as well as the angle-integrated input (disc) and output (emergent) SEDs.}
    \label{fig: demo_tde_uv_optical}
\end{figure*}

\subsection{Tidal disruption events: broad-band reprocessing and UV/optical lines}
\label{sec:tde}
Tidal disruption events occur when an unlucky star passes too close to a SMBH and is ripped apart by tidal forces. The fallback of the stellar debris onto the SMBH is expected to take place via an accretion disc. Early in their evolution, the mass-transfer rate through a TDE disc is often comparable to the Eddington-limited rate \citep{Strubbe2009, wu2018, roth2020}. Under these conditions, powerful radiation-pressure driven outflows are expected to form; direct evidence for them comes from the observation of broad blue-shifted absorption and P Cygni lines in the UV spectra of several TDE \citep{Brown2017, Blagorodnova19, Hung2019}. The outflows are expected to be optically thick to electron scattering and can therefore reprocess much of the luminosity emitted by the disc. As a result, outflows may be the key to understanding the shape and orientation dependence of the observed SEDs; which exhibit a wide range of X-ray to optical ratios and are much redder than expected for directly observed disc emission.

\xcode\ can be used to model both the line formation and broad-band reprocessing in TDE accretion disc winds. For example, \citet{parkinson2020} show that the UV absorption and emission lines can be used as orientation indicators, while \citet{parkinson2021} demonstrate that disc wind reprocessing could also account for the observed optical lines and continua. In order to demonstrate \xcode's capabilities in the context of TDEs, the wind model we adopt here is one of the SV-type models explored by \citet{parkinson2021}. It describes an outflow from an accretion disc around a $M_{\rm BH} = 3 \times 10^6~M_{\odot}$ SMBH, with a mass-transfer rate of $\dot{M}_{\rm acc} = 0.15~\dot{M}_{\rm Edd}$. The mass-loss rate of the wind is $\dot{M}_{w} = 0.3~\dot{M}_{\rm acc}$. The inner edge of the disc and wind are located at roughly $20~R_g$, and the inner and outer opening angles are 20$^\circ$ and 65$^\circ$, respectively. Hydrogen and Helium are treated as macro-atoms, ensuring that reprocessing and line formation via photoionization and recombination, in the dominant atomic species, are treated accurately. The remaining model parameters are listed in Table \ref{tab:demo_model_parameters}, and the full input parameter file is available as part of the documentation of the code.

The left two panels of Fig.~\ref{fig: demo_tde_uv_optical} show the UV and optical spectra produced for several sight lines. The optical emission line spectra, featuring Balmer and He {\sc ii} emission, are reminiscent of those observed in the TDE-Bowen class \citep{vanVelzen2020}. The strong, high-ionization UV lines in the model are also commonly seen in the spectra of TDEs \citep{Blagorodnova19, Hung2019}. Since the outflow is moderately optically thick ($\tau_{es} \simeq 1$, depending on viewing angle), it reprocesses much of the disc's radiation field toward longer wavelengths. This happens partly via thermal processes (absorption followed by reemission) and partly via ``adiabatic reprocessing'' associated with repeated scatterings in an outflowing medium \citep{roth2018}. The overall effect is illustrated in the right panel of Fig.~\ref{fig: demo_tde_uv_optical}, showing the angle-averaged input (disc) and output (emergent) SEDs of the model. Reprocessing dominates, and strongly enhances, the radiation field at optical wavelengths. 

Running TDE models with \xcode\ is computationally challenging. The main issue is that the average number of interactions per photon scales as $\tau^{2}$, so transporting large numbers of photons through the outflow is time-consuming. Using $10^{8}$ photons on a typical desktop computer, it takes, on average, 360 core hr for this type of model to reach a converged state after 15 ionization cycles. Spectra can, however, be constructed with fewer photons, since the wavelength range is usually reduced for spectral cycles; we typically use $10^{6}$ photons over 10-15 cycles to create sufficiently high-S/N UV and optical spectra. However, TDE models with larger mass-loss rates, such as those shown by \citet{parkinson2021}, can take upwards of 250 hours on the same desktop machine to reach a converged state because of larger optical depths.

 \begin{figure*}
\centering
\includegraphics[width=1.0\linewidth]{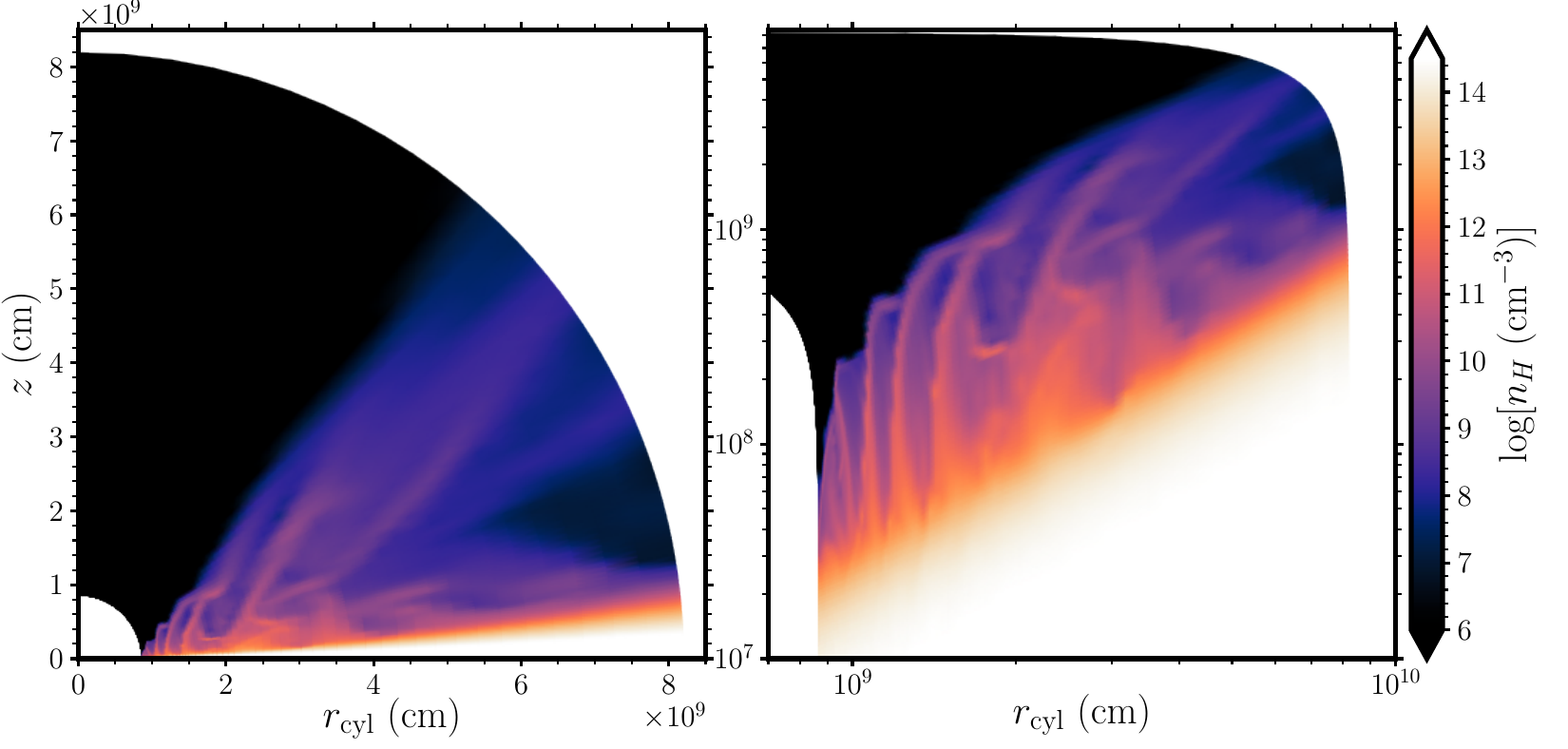}
\caption[Rad-hydro figure]{
The Hydrogen number density from a simulation snapshot (HK22D) of a CV line-driven wind with an isothermal equation of state, described in full by \cite{higginbottom24}. \xcode\ is used to calculate the ionization state and fluxes in a full MCRT-hydrodynamics framework, and {\sc Pluto} is used for the hydrodynamics steps. A successful disc wind is produced, but it is significantly weaker (in terms of power and mass-loss rate) than that obtained with an idealized treatment of ionization and radiative transfer. 
}
\label{fig:rad-hydro}
\end{figure*}

\subsection{Monte Carlo Radiation-hydrodynamics}
\label{sec:rad-hydro}
\xcode\ has been used as the radiation module in radiation-hydrodynamics simulations of accretion disc winds, originally coupled to {\sc zeus} \citep{zeus} and more recently to {\sc pluto} \citep{mignone_pluto:_2007}. Radiation transport is important in determining the radiative heating of the plasma, the ionization state, and the momentum transferred to the flow through the radiation force. We have used \xcode\ in a hydrodynamics context to study thermal or Compton-heated winds in XRBs \citep{higginbottom17, higginbottom18, higginbottom19, higginbottom20}, radiation line-driven winds in CVs \citep{higginbottom24} and radiation line-driven winds in AGN (Scepi et al., {\sl in prep.}). Briefly, the basic approach is to use an ``operator splitting'' style approach, where \xcode\ is called frequently, but not after every hydrodynamic time step. In each call, a small number of ionization cycles are carried out 
and used to update the ionization and temperature structure of the flow. The way this feeds back into the hydrodynamic calculation depends on the application, but one can choose to update the heating/cooling rates and/or radiation forces. In the latter case, we use the stand-alone code of \cite{parkin2013} to calculate the line force using a more complete line list. The aim of these methods is to model the frequency-dependent, multidimensional radiative transfer as accurately as possible and fully simulate the dynamic system, rather than relying on parameterised models with limited predictive power and/or cruder treatments of the plasma-radiation interaction. 

Our aim presently is not to provide details of the method, which are given by the above references. Rather, we present an example snapshot from a simulation that is representative of the current \xcode-{\sc Pluto} capabilities
and gives the reader a feel for what is possible when using \xcode\ for MCRT-hydrodynamics. The left-hand panel of  Fig.~\ref{fig:rad-hydro} shows a snapshot from the simulation of a line-driven CV wind presented by \cite[][run HK22D]{higginbottom24}, while the right-hand panel shows the same quantity on logarithmic axes. In this simulation, there are significant differences compared to using hydrodynamics without without a detailed treatment of ionization and radiative transfer; the simulated wind has a mass-loss rate a factor of $\approx 350$ times lower than when the idealised ``$k$-$\alpha$'' formalism is used. These results highlight the importance of using accurate radiative transfer methods whenever radiative heating and photoionization are important for the gas dynamics; this is particularly the case when multiple scattering and frequency-dependent opacities play a significant role in determining the plasma state. In the simulation here, we assumed an isothermal equation of state (EOS) with $T=40,000~{\rm K}$, but our simulations of XRB thermal winds necessarily include a full treatment of heating and cooling with an ideal EOS. We are currently in the process of improving our line-driven wind simulations to include full heating and cooling (Mosallanezhad et al, {\sl in prep}). We are also applying the same  techniques to model line-driven outflows in AGN (Scepi et al, {\sl in prep}).

\section{Limitations and Caveats}
\label{sec:limitations}

\xcode\ is a powerful code that includes a wide range of physical processes. It is capable of synthesizing the electromagnetic signatures of a diverse set of astrophysical systems on all scales. However, there are, of course, important limitations of which the user should be aware. Perhaps most importantly, \xcode\ is not a stellar atmosphere code: it is not designed for extremely optically thick and/or static media. Below, we explore these and some other constraints in more detail.

\subsection{Extreme optical depths}

Since \xcode\ is a Monte Carlo code that tracks the discrete interactions of simulated photon packets, the computational cost of a wind model scales with the characteristic optical depth it presents to these packets. In most environments of interest, at least hydrogen is nearly fully ionized, so it is convenient to define $\tau_{\rm ch}$ as the characteristic electron scattering optical depth. More specifically, the computational cost depends on the mean number of scatterings, $\overline{N}_{\rm scat}$, packets undergo before escaping. In some of our TDE models $\overline{N}_{\rm scat}$ approaches $100 - 1000$, corresponding to $\tau_{\rm ch} \sim 10 - 100$. Such models take $\mathcal{O}(1000)$ core hr to converge and create spectra.

Optically thick media can give rise to a variety of symptoms. Moreover, geometry matters. In 2D models, optical depths can be highly direction dependent, with $\tau_{\rm min} \ll \tau_{\rm max}$ (as demonstrated in Fig.~\ref{fig: demo_tde_uv_optical}). In this case, $r$-packets can and will escape preferentially along the most transparent direction(s), i.e. $\tau_{\rm ch} \sim \tau_{\rm min}$. This is both good and bad. Good, because $\tau_{\rm ch} \ll \tau_{\rm max}$ implies that even models with very high $\tau_{\rm max}$ are, in principle, computationally feasible. Bad, because such models often encounter serious convergence problems, as there are not enough $r$-packets crossing cells embedded deep within (or located behind) high-$\tau$ structures. In practice, unconverged cells in such regions can often be safely ignored, as they contribute little to the synthesized spectra. However, this decision has to be made by the user.

In 1D (spherical) models, $r$-packets can more easily get trapped, as there are no preferred directions along which they can escape. In practice, computational resources and/or convergence issues tend to be the limiting factor for optically thick 1D models. However, in classic mode, there is an additional failure mode for such models. In the limit where most photons packets are (re)absorbed and (re)emitted many times in a single cell, it becomes extremely difficult for the code to maintain energy conservation. The issue is that the radiation field locally becomes nearly isotropic. The {\em net (outward) flux} can then be a tiny fraction of the {\em mean (angle-averaged) intensity}. In Monte Carlo terms, the luminosity (number of $r$-packets) that has to be created within a cell is far greater than the net number of packets that actually leaves the cell. The luminosity carried by the escaping population can then become highly uncertain simply due to Poisson scatter. Yet global energy conservation across a simulation relies on the correct flow rates of $r$-packets across cells. This is not a concern when \xcode\ is run in macro-atom mode, as this locally enforces radiative equilibrium and hence guarantees global energy conservation.

\subsection{The Sobolev and related approximations}

\xcode\ is designed to synthesize spectra formed in {\em moving} media and treats line transfer in the Sobolev approximation. This means that an $r$-packet is only allowed to interact with a bound-bound transition at the exact location where its co-moving frequency coincides with the rest frequency of the transition. In reality, interactions take place across a spatial scale defined by the {\em Sobolev length}, $\Delta s \simeq v_{\rm th} / \left(dv_{s}/ds\right)$, where $v_{\rm th}$ is the local thermal speed, $s$ is distance along the $r$-packet's line of flight, and $v_{s}$ is the projected velocity along this direction. The Sobolev approximation therefore amounts to the assumption that the physical properties of the flow do not change significantly on scales as small as $\Delta s$, which is not always a valid assumption. In line with this treatment, \xcode\ neglects any thermal or microturbulent broadening associated with bound-bound transitions. It is the responsibility of the user to ensure that the Sobolev approximation is valid for the models they run. Static media and flows in which thermal speeds are comparable to bulk velocities should not be simulated with \xcode\ if focusing on line interactions, although such a simulation is appropriate if the focus is on continuum processes or ionization state. 

\subsection{The size of the line list}

An important challenge for \xcode\ is to identify spectral lines that might be in resonance with a particular $r$-packet as it passes through a given cell. The computational cost of this task scales directly with the size of the line list. It is therefore not feasible to simply import the latest Kurucz line list\footnote{http://kurucz.harvard.edu/linelists/gfnew/}, for example, which contains $\simeq 10^9$ transitions. 
Our default line list contains $\simeq 10^4$ transitions, which is sufficient for many applications. However, for radiation-hydrodynamics simulations where line-driving forces are important \citep{higginbottom20, higginbottom24}, is is important to have as complete a line list as possible. Our approach here is to use \xcode\ to estimate the ionization state and SED in each cell, then pass this information  to an external code that estimates the total line forces by summing over the complete Kurucz line list.

\subsection{General Relativity}

\xcode\ self-consistently carries out the special relativistic frame transformations required as $r$-packets travel through the grid and interact with the moving material in each cell. However, it does not account for any purely {\em general} relativistic effects. In particular, $r$-packets in \xcode\ always travel in straight lines, rather than along geodesics. This will primarily affect the angular distributions of photons that are emitted or travel within, say, $\simeq 10$ gravitational radii of a compact object. This caveat should be kept in mind when modelling AGN and XRBs with \xcode . However, we console ourselves with the thought that the physical and radiative properties of accretion flows in this regime remain quite uncertain. 

\subsection{Polarization}

In principle, polarization can be included quite naturally in Monte-Carlo radiative transfer \citep[e.g.][]{kasen2006,bulla2015,peest2017}, but no polarization treatment is implemented in the current version of \xcode. 

\subsection{Thermal and statistical equilibrium}

\xcode\ assumes that the flow is always and everywhere in (local) thermal and statistical equilibrium. That is, the code iterates towards a temperature and ionization state for each cell in which the heating and cooling rates in each cell balance and the net transition rate {\em into} any given atomic/ionic level matches the net transition rate {\em out of} that level. This implies that there is no concept of time in \xcode\ -- the code is not designed to deal with non-equilibrium and/or time-dependent conditions.\footnote{We currently do use \xcode\ in two time-dependent settings: in the context of radiation-hydrodynamics (section~\ref{sec:rad-hydro}) and reverberation mapping (section~\ref{sec:reverb}). However, in both cases, the time dependence is dealt with externally, and it is still assumed that thermal and/or statistical equilibrium are established instantaneously.}

This limitation can be important even if the input radiation field is steady. For example, if the flow velocity in a grid cell with characteristic size $\Delta x$ is given by $v$, matter will flow through the cell on a time-scale $t_{\rm flow} \sim \Delta x / v$. However, ionization equilibrium is established on a time-scale of $t_{\rm rec} \sim (\alpha n_e)^{-1}$, where $\alpha$ is the relevant recombination coefficient, and $n_e$ is the local electron density. Thus if $t_{\rm flow} < t_{\rm rec}$, the cell cannot be in ionization equilibrium. In sufficiently fast-moving flows, the ionization state can then become ``frozen-in'', i.e. fixed to approximately the state at the last point where equilibrium could be established. \xcode\ currently does not check for or deal with such non-equilibrium conditions. It is up to the user to carry out the relevant consistency checks on their models. There are also related issues when a dynamical time is short compared to the radiation `diffusion' time, since we implicitly assume that the conditions in the outflow (e.g. density, velocity, temperature, ionization) do not evolve on timescales comparable to photon residence times within the model.

\subsection{Dimensionality and resolution limits}
At present, \xcode\ is (at most) a 2.5-dimensional code. That is, the coordinate grid is restricted to 2D and assumed to be symmetric about the $z$-axis. However, photon transport takes place in 3D and allows for a rotational ($v_\varphi$) component of motion around the $y$-axis. In principle, upgrading \xcode\  to ``full'' 3D is fairly straightforward, at least conceptually, but running such models would require significantly more computing resources. Memory usage, in particular, scales fairly directly with the number of grid cells used in a simulation, and it is already a limitation for high-resolution models in 2D (or 2.5-D). This is the reason we have set an upper limit of $500\times500$ cells as a default. \xcode's memory requirements for computationally demanding simulations are partly driven by the \texttt{MPI} approach to parallelization, which requires the computational grid to be copied in full to each \texttt{MPI} process. This situation could be improved by adopting a hybrid \texttt{OpenMP} and \texttt{MPI} approach.
    
\section{Summary and Conclusions}
\label{sec:conclusions}

\xcode\ (Simulating Ionization and Radiation in Outflows Created by Compact Objects -- or ``the code formerly known as \emph{Python}'') is a 2.5D grid-based Monte Carlo radiative transfer and (photo)ionization code. It is intended for modelling either spherical or biconical outflows in a variety of systems with a range of masses and sizes. In particular, \xcode\ can be used to synthesize spectra from winds in CVs, AGN, XRBs, TDEs, YSOs and stars across a range of wavelengths from infra-red to X-ray. The code self-consistently computes the radiation field and ionization state of the plasma in an iterative manner and deals well with multiple scattering as well as frequency-dependent opacities and reprocessing. It is also flexible in terms of how the illuminating spectrum and wind structure is specified. Not all radiative mechanisms are included -- for example, \xcode\ is not currently suitable for computing spectra dominated by synchrotron, dust or molecular features -- but \xcode\ is capable of modelling accurately most of the important emission lines and physical processes in the rest-frame optical, near infra-red, X-ray and UV spectra, particularly for accreting compact objects in a thermally dominated state. The ability to import models means that user-defined wind geometries (for example, from self-similar MHD wind models) can be used. In addition, the code can be coupled to hydrodynamics codes such as {\sc Pluto} to conduct Monte Carlo radiation-hydrodynamics simulations. Overall, \xcode\ represents a powerful modelling tool that can help us better understand the physics of -- and observable signatures produced by -- winds and outflows in accreting systems. 
    
Here, we have provided an overview of \xcode\ and described its operation at the time of code release. We have outlined the key assumptions underlying the code, its capabilities, and the treatment of radiation transfer and ionization within it. We have also presented comparisons to other radiative transfer codes, tests of our methods, and some illustrative models of biconical winds in a CV, XRB, TDE and Quasar. Together with the documentation, we hope this makes it clear how others could make use of \xcode, if they wish. More generally, our work highlights the power of Monte Carlo radiative transfer as a technique for studying outflows in accreting systems. 

\section*{Data availability}
\xcode\ is freely available via \href{https://github.com/sirocco-rt/sirocco}{GitHub}.  
Documentation can be found on \href{https://sirocco-rt.readthedocs.io/en/latest/}{ReadTheDocs} or generated locally from the source code.  
The specific code release and documentation associated with this paper can be accessed via Zenodo (\href{https://doi.org/10.5281/zenodo.14243841}{DOI:10.5281/zenodo.14243841}). 
In addition, the data and scripts needed to recreate the majority of the figures can be found on 
\href{https://github.com/sirocco-rt/release-models}{GitHub} or via the associated Zenodo DOI (\href{https://doi.org/10.5281/zenodo.13993033}{DOI:10.5281/zenodo.13993033}). 

\section*{Acknowledgements}
We would like to thank Noel Castro Segura, Karri Koljonen, Ulrich Noebauer, Yusuke Tampo, Andreas Sander, Tim Harries, Daniel Proga, Randall Dannen, D. John Hillier, Wolfgang Kerzendorf, Ryota Tomaru, Chris Done, Shane Davis, Sergei Dyda, Matt Stepney, Keara Carter, Daniel Baig, Alessandra Ambrifi, James Gillanders and many others for helpful discussions and collaborations. We gratefully acknowledge the use of the following software packages both for the writing of this paper and for the development of \xcode: astropy \citep{astropy-collaboration13,astropy-collaboration18},  \cloudy\ \citep{ferland17}, matplotlib \citep{hunter07}, GNU Science Library \citep{Galassi2018_gsl}, \tardis\ \citep{kerzendorf_spectral_2014, tardis_zenodo}, \cmfgen\ \citep{hillier98}. JHM acknowledges a Royal Society University Research Fellowship. The work by CK, NSH, NS and AM was supported by STFC via grant ST/V001000/1.  AW's contribution was supported by a PhD studentship funded by STFC. EJP acknowledges financial support from the EPSRC Centre for Doctoral Training in Next Generation Computational Modelling grant EP/L015382/1. This software was developed with assistance from the Southampton Research Software Group.  Partial support for KSL's effort on the project was provided by NASA through grant numbers HST-GO-16489 and HST-GO-16659 and from the Space Telescope Science Institute, which is operated by AURA, Inc., under NASA contract NAS 5-26555.

\input{sirocco.bbl}


\begin{thebibliography}{}
\makeatletter
\relax
\def\mn@urlcharsother{\let\do\@makeother \do\$\do\&\do\#\do\^\do\_\do\%\do\~}
\def\mn@doi{\begingroup\mn@urlcharsother \@ifnextchar [ {\mn@doi@} {\mn@doi@[]}}
\def\mn@doi@[#1]#2{\def\@tempa{#1}\ifx\@tempa\@empty \href {http://dx.doi.org/#2} {doi:#2}\else \href {http://dx.doi.org/#2} {#1}\fi \endgroup}
\def\mn@eprint#1#2{\mn@eprint@#1:#2::\@nil}
\def\mn@eprint@arXiv#1{\href {http://arxiv.org/abs/#1} {{\tt arXiv:#1}}}
\def\mn@eprint@dblp#1{\href {http://dblp.uni-trier.de/rec/bibtex/#1.xml} {dblp:#1}}
\def\mn@eprint@#1:#2:#3:#4\@nil{\def\@tempa {#1}\def\@tempb {#2}\def\@tempc {#3}\ifx \@tempc \@empty \let \@tempc \@tempb \let \@tempb \@tempa \fi \ifx \@tempb \@empty \def\@tempb {arXiv}\fi \@ifundefined {mn@eprint@\@tempb}{\@tempb:\@tempc}{\expandafter \expandafter \csname mn@eprint@\@tempb\endcsname \expandafter{\@tempc}}}

\bibitem[\protect\citeauthoryear{{Abbott} \& {Lucy}}{{Abbott} \& {Lucy}}{1985}]{abbott85}
{Abbott} D.~C.,  {Lucy} L.~B.,  1985, \mn@doi [\apj] {10.1086/162834}, \href {https://ui.adsabs.harvard.edu/abs/1985ApJ...288..679A} {288, 679}

\bibitem[\protect\citeauthoryear{Allen, Hewett, Maddox, Richards  \& Belokurov}{Allen et~al.}{2011}]{allen_strong_2011}
Allen J.~T.,  Hewett P.~C.,  Maddox N.,  Richards G.~T.,   Belokurov V.,  2011, \mn@doi [\mnras] {10.1111/j.1365-2966.2010.17489.x}, 410, 860

\bibitem[\protect\citeauthoryear{{Astropy Collaboration} et~al.,}{{Astropy Collaboration} et~al.}{2013}]{astropy-collaboration13}
{Astropy Collaboration} et~al., 2013, \mn@doi [\aap] {10.1051/0004-6361/201322068}, \href {https://ui.adsabs.harvard.edu/abs/2013A&A...558A..33A} {558, A33}

\bibitem[\protect\citeauthoryear{{Astropy Collaboration} et~al.,}{{Astropy Collaboration} et~al.}{2018}]{astropy-collaboration18}
{Astropy Collaboration} et~al., 2018, \mn@doi [\aj] {10.3847/1538-3881/aabc4f}, \href {https://ui.adsabs.harvard.edu/abs/2018AJ....156..123A} {156, 123}

\bibitem[\protect\citeauthoryear{Badnell}{Badnell}{2006}]{badnell_radiative_2006}
Badnell N.~R.,  2006, \mn@doi [\apjs] {10.1086/508465}, 167, 334

\bibitem[\protect\citeauthoryear{{Bahramian} \& {Degenaar}}{{Bahramian} \& {Degenaar}}{2023}]{Bahramian2023}
{Bahramian} A.,  {Degenaar} N.,  2023, in , Handbook of X-ray and Gamma-ray Astrophysics.
p.~120, \mn@doi{10.1007/978-981-16-4544-0_94-1}

\bibitem[\protect\citeauthoryear{Blagorodnova et~al.,}{Blagorodnova et~al.}{2019}]{Blagorodnova19}
Blagorodnova N.,  et~al., 2019, \mn@doi [ApJ] {10.3847/1538-4357/ab04b0}, 873, 92

\bibitem[\protect\citeauthoryear{Brown et~al.,}{Brown et~al.}{2017}]{Brown2017}
Brown J.~S.,  et~al., 2017, \mn@doi [MNRAS] {10.1093/mnras/stx2372}, 473, 1130

\bibitem[\protect\citeauthoryear{{Bu}, {Qiao}  \& {Yang}}{{Bu} et~al.}{2023}]{fu2023}
{Bu} D.-F.,  {Qiao} E.,   {Yang} X.-H.,  2023, \mn@doi [\mnras] {10.1093/mnras/stad1696}, \href {https://ui.adsabs.harvard.edu/abs/2023MNRAS.523.4136B} {523, 4136}

\bibitem[\protect\citeauthoryear{{Bulla}, {Sim}  \& {Kromer}}{{Bulla} et~al.}{2015}]{bulla2015}
{Bulla} M.,  {Sim} S.~A.,   {Kromer} M.,  2015, \mn@doi [\mnras] {10.1093/mnras/stv657}, \href {https://ui.adsabs.harvard.edu/abs/2015MNRAS.450..967B} {450, 967}

\bibitem[\protect\citeauthoryear{{Burgess} \& {Tully}}{{Burgess} \& {Tully}}{1992}]{burgess1992}
{Burgess} A.,  {Tully} J.~A.,  1992, \aap, \href {https://ui.adsabs.harvard.edu/abs/1992A&A...254..436B} {254, 436}

\bibitem[\protect\citeauthoryear{Camps \& Baes}{Camps \& Baes}{2015}]{camps_skirt_2015}
Camps P.,  Baes M.,  2015, \mn@doi [Astronomy and Computing] {10.1016/j.ascom.2014.10.004}, 9, 20

\bibitem[\protect\citeauthoryear{{Castor}}{{Castor}}{1972}]{castor72}
{Castor} J.~I.,  1972, \mn@doi [\apj] {10.1086/151834}, \href {https://ui.adsabs.harvard.edu/abs/1972ApJ...178..779C} {178, 779}

\bibitem[\protect\citeauthoryear{{Castor} \& {Lamers}}{{Castor} \& {Lamers}}{1979}]{castor79}
{Castor} J.~I.,  {Lamers} H.~J.~G.~L.~M.,  1979, \mn@doi [\apjs] {10.1086/190583}, \href {https://ui.adsabs.harvard.edu/abs/1979ApJS...39..481C} {39, 481}

\bibitem[\protect\citeauthoryear{{Castro Segura} et~al.,}{{Castro Segura} et~al.}{2022}]{castro2022}
{Castro Segura} N.,  et~al., 2022, \mn@doi [\nat] {10.1038/s41586-021-04324-2}, \href {https://ui.adsabs.harvard.edu/abs/2022Natur.603...52C} {603, 52}

\bibitem[\protect\citeauthoryear{{Chakravorty} et~al.,}{{Chakravorty} et~al.}{2016}]{chakravorty2016}
{Chakravorty} S.,  et~al., 2016, \mn@doi [\aap] {10.1051/0004-6361/201527163}, \href {https://ui.adsabs.harvard.edu/abs/2016A&A...589A.119C} {589, A119}

\bibitem[\protect\citeauthoryear{{Charles}, {Matthews}, {Buckley}, {Gandhi}, {Kotze}  \& {Paice}}{{Charles} et~al.}{2019}]{charles2019}
{Charles} P.,  {Matthews} J.~H.,  {Buckley} D. A.~H.,  {Gandhi} P.,  {Kotze} E.,   {Paice} J.,  2019, \mn@doi [\mnras] {10.1093/mnrasl/slz120}, \href {https://ui.adsabs.harvard.edu/abs/2019MNRAS.489L..47C} {489, L47}

\bibitem[\protect\citeauthoryear{{Chatzikos} et~al.,}{{Chatzikos} et~al.}{2023}]{cloudy}
{Chatzikos} M.,  et~al., 2023, \mn@doi [\rmxaa] {10.22201/ia.01851101p.2023.59.02.12}, \href {https://ui.adsabs.harvard.edu/abs/2023RMxAA..59..327C} {59, 327}

\bibitem[\protect\citeauthoryear{{Cordova} \& {Mason}}{{Cordova} \& {Mason}}{1982}]{cordova82}
{Cordova} F.~A.,  {Mason} K.~O.,  1982, \mn@doi [\apj] {10.1086/160291}, \href {https://ui.adsabs.harvard.edu/abs/1982ApJ...260..716C} {260, 716}

\bibitem[\protect\citeauthoryear{{Cordova} \& {Mason}}{{Cordova} \& {Mason}}{1985}]{cordova85}
{Cordova} F.~A.,  {Mason} K.~O.,  1985, \mn@doi [\apj] {10.1086/163024}, \href {https://ui.adsabs.harvard.edu/\#abs/1985ApJ...290..671C} {290, 671}

\bibitem[\protect\citeauthoryear{{C{\'u}neo} et~al.,}{{C{\'u}neo} et~al.}{2023}]{cuneo23}
{C{\'u}neo} V.~A.,  et~al., 2023, \mn@doi [\aap] {10.1051/0004-6361/202347265}, \href {https://ui.adsabs.harvard.edu/abs/2023A&A...679A..85C} {679, A85}

\bibitem[\protect\citeauthoryear{{Cunto}, {Mendoza}, {Ochsenbein}  \& {Zeippen}}{{Cunto} et~al.}{1993}]{cunto93}
{Cunto} W.,  {Mendoza} C.,  {Ochsenbein} F.,   {Zeippen} C.~J.,  1993, \aap, \href {https://ui.adsabs.harvard.edu/abs/1993A&A...275L...5C} {275, L5}

\bibitem[\protect\citeauthoryear{{Dai}, {Shankar}  \& {Sivakoff}}{{Dai} et~al.}{2008}]{dai2008}
{Dai} X.,  {Shankar} F.,   {Sivakoff} G.~R.,  2008, \mn@doi [\apj] {10.1086/523688}, \href {https://ui.adsabs.harvard.edu/abs/2008ApJ...672..108D} {672, 108}

\bibitem[\protect\citeauthoryear{Dannen, Proga, Waters  \& Dyda}{Dannen et~al.}{2019}]{dannen_clumpy_2019}
Dannen R.,  Proga D.,  Waters T.,   Dyda S.,  2019, arXiv:2001.00133 [astro-ph]

\bibitem[\protect\citeauthoryear{{Del Zanna}, {Dere}, {Young}  \& {Landi}}{{Del Zanna} et~al.}{2021}]{delzanna21}
{Del Zanna} G.,  {Dere} K.~P.,  {Young} P.~R.,   {Landi} E.,  2021, \mn@doi [\apj] {10.3847/1538-4357/abd8ce}, \href {https://ui.adsabs.harvard.edu/abs/2021ApJ...909...38D} {909, 38}

\bibitem[\protect\citeauthoryear{{Dere}, {Landi}, {Mason}, {Monsignori Fossi}  \& {Young}}{{Dere} et~al.}{1997}]{dere97}
{Dere} K.~P.,  {Landi} E.,  {Mason} H.~E.,  {Monsignori Fossi} B.~C.,   {Young} P.~R.,  1997, \mn@doi [\aaps] {10.1051/aas:1997368}, \href {https://ui.adsabs.harvard.edu/abs/1997A&AS..125..149D} {125, 149}

\bibitem[\protect\citeauthoryear{{Dere}, {Del Zanna}, {Young}  \& {Landi}}{{Dere} et~al.}{2023}]{dere23}
{Dere} K.~P.,  {Del Zanna} G.,  {Young} P.~R.,   {Landi} E.,  2023, \mn@doi [\apjs] {10.3847/1538-4365/acec79}, \href {https://ui.adsabs.harvard.edu/abs/2023ApJS..268...52D} {268, 52}

\bibitem[\protect\citeauthoryear{{D{\'\i}az Trigo} \& {Boirin}}{{D{\'\i}az Trigo} \& {Boirin}}{2016}]{diaztrigo2016}
{D{\'\i}az Trigo} M.,  {Boirin} L.,  2016, \mn@doi [Astronomische Nachrichten] {10.1002/asna.201612315}, \href {https://ui.adsabs.harvard.edu/abs/2016AN....337..368D} {337, 368}

\bibitem[\protect\citeauthoryear{Dolence, Gammie, Mościbrodzka  \& Leung}{Dolence et~al.}{2009}]{dolence_grmonty_2009}
Dolence J.~C.,  Gammie C.~F.,  Mościbrodzka M.,   Leung P.~K.,  2009, \mn@doi [\apjs] {10.1088/0067-0049/184/2/387}, 184, 387

\bibitem[\protect\citeauthoryear{Done, Davis, Jin, Blaes  \& Ward}{Done et~al.}{2012}]{done_intrinsic_2012}
Done C.,  Davis S.~W.,  Jin C.,  Blaes O.,   Ward M.,  2012, \mn@doi [\mnras] {10.1111/j.1365-2966.2011.19779.x}, 420, 1848

\bibitem[\protect\citeauthoryear{{Drew} \& {Verbunt}}{{Drew} \& {Verbunt}}{1988}]{drew88}
{Drew} J.~E.,  {Verbunt} F.,  1988, \mn@doi [\mnras] {10.1093/mnras/234.2.341}, \href {https://ui.adsabs.harvard.edu/abs/1988MNRAS.234..341D} {234, 341}

\bibitem[\protect\citeauthoryear{Elvis}{Elvis}{2000}]{elvis_structure_2000}
Elvis M.,  2000, \mn@doi [\apj] {10.1086/317778}, 545, 63

\bibitem[\protect\citeauthoryear{Ercolano, Barlow, Storey  \& Liu}{Ercolano et~al.}{2003}]{ercolano_mocassin_2003}
Ercolano B.,  Barlow M.~J.,  Storey P.~J.,   Liu X.~W.,  2003, \mn@doi [\mnras] {10.1046/j.1365-8711.2003.06371.x}, 340, 1136

\bibitem[\protect\citeauthoryear{Ercolano, Barlow  \& Storey}{Ercolano et~al.}{2005}]{ercolano_dusty_2005}
Ercolano B.,  Barlow M.~J.,   Storey P.~J.,  2005, \mn@doi [\mnras] {10.1111/j.1365-2966.2005.09381.x}, 362, 1038

\bibitem[\protect\citeauthoryear{Ergon, Fransson, Jerkstrand, Kozma, Kromer  \& Spricer}{Ergon et~al.}{2018}]{ergon_monte-carlo_2018}
Ergon M.,  Fransson C.,  Jerkstrand A.,  Kozma C.,  Kromer M.,   Spricer K.,  2018, \mn@doi [\aap] {10.1051/0004-6361/201833043}, 620, A156

\bibitem[\protect\citeauthoryear{Fabian}{Fabian}{2012}]{fabian_observational_2012}
Fabian A.~C.,  2012, \mn@doi [\araa] {10.1146/annurev-astro-081811-125521}, 50, 455

\bibitem[\protect\citeauthoryear{{Fabrika}, {Ueda}, {Vinokurov}, {Sholukhova}  \& {Shidatsu}}{{Fabrika} et~al.}{2015}]{fabrika2015}
{Fabrika} S.,  {Ueda} Y.,  {Vinokurov} A.,  {Sholukhova} O.,   {Shidatsu} M.,  2015, \mn@doi [Nature Physics] {10.1038/nphys3348}, \href {https://ui.adsabs.harvard.edu/abs/2015NatPh..11..551F} {11, 551}

\bibitem[\protect\citeauthoryear{Ferland et~al.,}{Ferland et~al.}{2013}]{ferland_2013_2013}
Ferland G.~J.,  et~al., 2013, \rmxaa, 49, 137

\bibitem[\protect\citeauthoryear{{Ferland} et~al.,}{{Ferland} et~al.}{2017}]{ferland17}
{Ferland} G.~J.,  et~al., 2017, \rmxaa, \href {https://ui.adsabs.harvard.edu/abs/2017RMxAA..53..385F} {53, 385}

\bibitem[\protect\citeauthoryear{{Fijma}, {Castro Segura}, {Degenaar}, {Knigge}, {Higginbottom}, {Hern{\'a}ndez Santisteban}  \& {Maccarone}}{{Fijma} et~al.}{2023}]{fijma2023}
{Fijma} S.,  {Castro Segura} N.,  {Degenaar} N.,  {Knigge} C.,  {Higginbottom} N.,  {Hern{\'a}ndez Santisteban} J.~V.,   {Maccarone} T.~J.,  2023, \mn@doi [\mnras] {10.1093/mnrasl/slad125}, \href {https://ui.adsabs.harvard.edu/abs/2023MNRAS.526L.149F} {526, L149}

\bibitem[\protect\citeauthoryear{{Fukumura}, {Kazanas}, {Shrader}, {Tombesi}, {Kalapotharakos}  \& {Behar}}{{Fukumura} et~al.}{2021}]{fukumura2021}
{Fukumura} K.,  {Kazanas} D.,  {Shrader} C.,  {Tombesi} F.,  {Kalapotharakos} C.,   {Behar} E.,  2021, \mn@doi [\apj] {10.3847/1538-4357/abedaf}, \href {https://ui.adsabs.harvard.edu/abs/2021ApJ...912...86F} {912, 86}

\bibitem[\protect\citeauthoryear{Galassi}{Galassi}{2018}]{Galassi2018_gsl}
Galassi M. e.~a.,  2018, GNU Scientific Library Reference Manual, \url {https://www.gnu.org/software/gsl/}

\bibitem[\protect\citeauthoryear{{Gillanders}, {Smartt}, {Sim}, {Bauswein}  \& {Goriely}}{{Gillanders} et~al.}{2022}]{gillanders2022}
{Gillanders} J.~H.,  {Smartt} S.~J.,  {Sim} S.~A.,  {Bauswein} A.,   {Goriely} S.,  2022, \mn@doi [\mnras] {10.1093/mnras/stac1258}, \href {https://ui.adsabs.harvard.edu/abs/2022MNRAS.515..631G} {515, 631}

\bibitem[\protect\citeauthoryear{{Giustini} \& {Proga}}{{Giustini} \& {Proga}}{2019}]{giustini2019}
{Giustini} M.,  {Proga} D.,  2019, \mn@doi [\aap] {10.1051/0004-6361/201833810}, \href {https://ui.adsabs.harvard.edu/abs/2019A&A...630A..94G} {630, A94}

\bibitem[\protect\citeauthoryear{Gofford, Reeves, Tombesi, Braito, Turner, Miller  \& Cappi}{Gofford et~al.}{2013}]{gofford_suzaku_2013}
Gofford J.,  Reeves J.~N.,  Tombesi F.,  Braito V.,  Turner T.~J.,  Miller L.,   Cappi M.,  2013, \mn@doi [\mnras] {10.1093/mnras/sts481}, 430, 60

\bibitem[\protect\citeauthoryear{{Gr{\"a}fener}, {Koesterke}  \& {Hamann}}{{Gr{\"a}fener} et~al.}{2002}]{grafener2002}
{Gr{\"a}fener} G.,  {Koesterke} L.,   {Hamann} W.~R.,  2002, \mn@doi [\aap] {10.1051/0004-6361:20020269}, \href {https://ui.adsabs.harvard.edu/abs/2002A&A...387..244G} {387, 244}

\bibitem[\protect\citeauthoryear{Greenstein \& Oke}{Greenstein \& Oke}{1982}]{greenstein_rw_1982}
Greenstein J.~L.,  Oke J.~B.,  1982, \mn@doi [\apj] {10.1086/160069}, 258, 209

\bibitem[\protect\citeauthoryear{Hagino, Odaka, Done, Gandhi, Watanabe, Sako  \& Takahashi}{Hagino et~al.}{2015}]{hagino_origin_2015}
Hagino K.,  Odaka H.,  Done C.,  Gandhi P.,  Watanabe S.,  Sako M.,   Takahashi T.,  2015, \mn@doi [\mnras] {10.1093/mnras/stu2095}, 446, 663

\bibitem[\protect\citeauthoryear{{Hall}}{{Hall}}{2007}]{hall2007}
{Hall} P.~B.,  2007, \mn@doi [\aj] {10.1086/511272}, \href {https://ui.adsabs.harvard.edu/abs/2007AJ....133.1271H} {133, 1271}

\bibitem[\protect\citeauthoryear{Hamann \& Koesterke}{Hamann \& Koesterke}{1998}]{hamann_spectrum_1998}
Hamann W.-R.,  Koesterke L.,  1998, \aap, 335, 1003

\bibitem[\protect\citeauthoryear{{Harries}, {Haworth}, {Acreman}, {Ali}  \& {Douglas}}{{Harries} et~al.}{2019}]{harries2019}
{Harries} T.~J.,  {Haworth} T.~J.,  {Acreman} D.,  {Ali} A.,   {Douglas} T.,  2019, \mn@doi [Astronomy and Computing] {10.1016/j.ascom.2019.03.002}, \href {https://ui.adsabs.harvard.edu/abs/2019A&C....27...63H} {27, 63}

\bibitem[\protect\citeauthoryear{{Harrison}, {Costa}, {Tadhunter}, {Fl{\"u}tsch}, {Kakkad}, {Perna}  \& {Vietri}}{{Harrison} et~al.}{2018}]{harrison2018}
{Harrison} C.~M.,  {Costa} T.,  {Tadhunter} C.~N.,  {Fl{\"u}tsch} A.,  {Kakkad} D.,  {Perna} M.,   {Vietri} G.,  2018, \mn@doi [Nature Astronomy] {10.1038/s41550-018-0403-6}, \href {https://ui.adsabs.harvard.edu/abs/2018NatAs...2..198H} {2, 198}

\bibitem[\protect\citeauthoryear{{Hartley}, {Drew}, {Long}, {Knigge}  \& {Proga}}{{Hartley} et~al.}{2002}]{hartley02}
{Hartley} L.~E.,  {Drew} J.~E.,  {Long} K.~S.,  {Knigge} C.,   {Proga} D.,  2002, \mn@doi [\mnras] {10.1046/j.1365-8711.2002.05277.x}, \href {http://adsabs.harvard.edu/abs/2002MNRAS.332..127H} {332, 127}

\bibitem[\protect\citeauthoryear{{Hassall}}{{Hassall}}{1985}]{hassall1985}
{Hassall} B.~J.~M.,  1985, \mn@doi [\mnras] {10.1093/mnras/216.2.335}, \href {https://ui.adsabs.harvard.edu/abs/1985MNRAS.216..335H} {216, 335}

\bibitem[\protect\citeauthoryear{Hewett \& Foltz}{Hewett \& Foltz}{2003}]{hewett_frequency_2003}
Hewett P.~C.,  Foltz C.~B.,  2003, \mn@doi [\aj] {10.1086/368392}, 125, 1784

\bibitem[\protect\citeauthoryear{{Higginbottom}, {Knigge}, {Long}, {Sim}  \& {Matthews}}{{Higginbottom} et~al.}{2013}]{higginbottom13}
{Higginbottom} N.,  {Knigge} C.,  {Long} K.~S.,  {Sim} S.~A.,   {Matthews} J.~H.,  2013, \mn@doi [\mnras] {10.1093/mnras/stt1658}, \href {https://ui.adsabs.harvard.edu/abs/2013MNRAS.436.1390H} {436, 1390}

\bibitem[\protect\citeauthoryear{Higginbottom, Proga, Knigge, Long, Matthews  \& Sim}{Higginbottom et~al.}{2014a}]{higginbottom_line-driven_2014}
Higginbottom N.,  Proga D.,  Knigge C.,  Long K.~S.,  Matthews J.~H.,   Sim S.~A.,  2014a, \mn@doi [\apj] {10.1088/0004-637X/789/1/19}, 789, 19

\bibitem[\protect\citeauthoryear{{Higginbottom}, {Proga}, {Knigge}, {Long}, {Matthews}  \& {Sim}}{{Higginbottom} et~al.}{2014b}]{higginbottom14}
{Higginbottom} N.,  {Proga} D.,  {Knigge} C.,  {Long} K.~S.,  {Matthews} J.~H.,   {Sim} S.~A.,  2014b, \mn@doi [\apj] {10.1088/0004-637X/789/1/19}, \href {https://ui.adsabs.harvard.edu/abs/2014ApJ...789...19H} {789, 19}

\bibitem[\protect\citeauthoryear{{Higginbottom}, {Proga}, {Knigge}  \& {Long}}{{Higginbottom} et~al.}{2017}]{higginbottom17}
{Higginbottom} N.,  {Proga} D.,  {Knigge} C.,   {Long} K.~S.,  2017, \mn@doi [\apj] {10.3847/1538-4357/836/1/42}, \href {https://ui.adsabs.harvard.edu/abs/2017ApJ...836...42H} {836, 42}

\bibitem[\protect\citeauthoryear{{Higginbottom}, {Knigge}, {Long}, {Matthews}, {Sim}  \& {Hewitt}}{{Higginbottom} et~al.}{2018}]{higginbottom18}
{Higginbottom} N.,  {Knigge} C.,  {Long} K.~S.,  {Matthews} J.~H.,  {Sim} S.~A.,   {Hewitt} H.~A.,  2018, \mn@doi [\mnras] {10.1093/mnras/sty1599}, \href {https://ui.adsabs.harvard.edu/abs/2018MNRAS.479.3651H} {479, 3651}

\bibitem[\protect\citeauthoryear{{Higginbottom}, {Knigge}, {Long}, {Matthews}  \& {Parkinson}}{{Higginbottom} et~al.}{2019}]{higginbottom19}
{Higginbottom} N.,  {Knigge} C.,  {Long} K.~S.,  {Matthews} J.~H.,   {Parkinson} E.~J.,  2019, \mn@doi [\mnras] {10.1093/mnras/stz310}, \href {https://ui.adsabs.harvard.edu/abs/2019MNRAS.484.4635H} {484, 4635}

\bibitem[\protect\citeauthoryear{{Higginbottom}, {Knigge}, {Sim}, {Long}, {Matthews}, {Hewitt}, {Parkinson}  \& {Mangham}}{{Higginbottom} et~al.}{2020}]{higginbottom20}
{Higginbottom} N.,  {Knigge} C.,  {Sim} S.~A.,  {Long} K.~S.,  {Matthews} J.~H.,  {Hewitt} H.~A.,  {Parkinson} E.~J.,   {Mangham} S.~W.,  2020, \mn@doi [\mnras] {10.1093/mnras/staa209}, \href {https://ui.adsabs.harvard.edu/abs/2020MNRAS.492.5271H} {492, 5271}

\bibitem[\protect\citeauthoryear{{Higginbottom}, {Scepi}, {Knigge}, {Long}, {Matthews}  \& {Sim}}{{Higginbottom} et~al.}{2024}]{higginbottom24}
{Higginbottom} N.,  {Scepi} N.,  {Knigge} C.,  {Long} K.~S.,  {Matthews} J.~H.,   {Sim} S.~A.,  2024, \mn@doi [\mnras] {10.1093/mnras/stad3830}, \href {https://ui.adsabs.harvard.edu/abs/2024MNRAS.527.9236H} {527, 9236}

\bibitem[\protect\citeauthoryear{{Hillier} \& {Miller}}{{Hillier} \& {Miller}}{1998}]{hillier98}
{Hillier} D.~J.,  {Miller} D.~L.,  1998, \mn@doi [\apj] {10.1086/305350}, \href {https://ui.adsabs.harvard.edu/abs/1998ApJ...496..407H} {496, 407}

\bibitem[\protect\citeauthoryear{Hillier \& Miller}{Hillier \& Miller}{1999}]{hillier_constraints_1999}
Hillier D.~J.,  Miller D.~L.,  1999, \mn@doi [\apj] {10.1086/307339}, 519, 354

\bibitem[\protect\citeauthoryear{{Horne}}{{Horne}}{1994}]{memecho}
{Horne} K.,  1994, in {Gondhalekar} P.~M.,  {Horne} K.,   {Peterson} B.~M.,  eds,  Astronomical Society of the Pacific Conference Series Vol. 69, Reverberation Mapping of the Broad-Line Region in Active Galactic Nuclei. p.~23

\bibitem[\protect\citeauthoryear{Hung et~al.,}{Hung et~al.}{2019}]{Hung2019}
Hung T.,  et~al., 2019, \mn@doi [ApJ] {10.3847/1538-4357/ab24de}, 879, 119

\bibitem[\protect\citeauthoryear{{Hunter}}{{Hunter}}{2007}]{hunter07}
{Hunter} J.~D.,  2007, \mn@doi [Computing in Science and Engineering] {10.1109/MCSE.2007.55}, \href {https://ui.adsabs.harvard.edu/abs/2007CSE.....9...90H} {9, 90}

\bibitem[\protect\citeauthoryear{{Jacquemin-Ide}, {Ferreira}  \& {Lesur}}{{Jacquemin-Ide} et~al.}{2019}]{jacquemin2019}
{Jacquemin-Ide} J.,  {Ferreira} J.,   {Lesur} G.,  2019, \mn@doi [\mnras] {10.1093/mnras/stz2749}, \href {https://ui.adsabs.harvard.edu/abs/2019MNRAS.490.3112J} {490, 3112}

\bibitem[\protect\citeauthoryear{{Jim{\'e}nez-Ibarra}, {Mu{\~n}oz-Darias}, {Casares}, {Armas Padilla}  \& {Corral-Santana}}{{Jim{\'e}nez-Ibarra} et~al.}{2019}]{jimenez2019}
{Jim{\'e}nez-Ibarra} F.,  {Mu{\~n}oz-Darias} T.,  {Casares} J.,  {Armas Padilla} M.,   {Corral-Santana} J.~M.,  2019, \mn@doi [\mnras] {10.1093/mnras/stz2393}, \href {https://ui.adsabs.harvard.edu/abs/2019MNRAS.489.3420J} {489, 3420}

\bibitem[\protect\citeauthoryear{{Kaastra} \& {Mewe}}{{Kaastra} \& {Mewe}}{1993}]{kaastra1993}
{Kaastra} J.~S.,  {Mewe} R.,  1993, \aaps, \href {https://ui.adsabs.harvard.edu/abs/1993A&AS...97..443K} {97, 443}

\bibitem[\protect\citeauthoryear{{Kafka} \& {Honeycutt}}{{Kafka} \& {Honeycutt}}{2004}]{kafka04}
{Kafka} S.,  {Honeycutt} R.~K.,  2004, \mn@doi [\aj] {10.1086/424618}, \href {https://ui.adsabs.harvard.edu/abs/2004AJ....128.2420K} {128, 2420}

\bibitem[\protect\citeauthoryear{{Kallman} \& {Bautista}}{{Kallman} \& {Bautista}}{2001}]{kallman2001}
{Kallman} T.,  {Bautista} M.,  2001, \mn@doi [\apjs] {10.1086/319184}, \href {https://ui.adsabs.harvard.edu/abs/2001ApJS..133..221K} {133, 221}

\bibitem[\protect\citeauthoryear{{Kasen}, {Thomas}  \& {Nugent}}{{Kasen} et~al.}{2006}]{kasen2006}
{Kasen} D.,  {Thomas} R.~C.,   {Nugent} P.,  2006, \mn@doi [\apj] {10.1086/506190}, \href {https://ui.adsabs.harvard.edu/abs/2006ApJ...651..366K} {651, 366}

\bibitem[\protect\citeauthoryear{Kerzendorf \& Sim}{Kerzendorf \& Sim}{2014}]{kerzendorf_spectral_2014}
Kerzendorf W.~E.,  Sim S.~A.,  2014, \mn@doi [\mnras] {10.1093/mnras/stu055}, 440, 387

\bibitem[\protect\citeauthoryear{Kerzendorf et~al.,}{Kerzendorf et~al.}{2024}]{tardis_zenodo}
Kerzendorf W.,  et~al., 2024, tardis-sn/tardis: TARDIS v2024.08.25, \mn@doi{10.5281/zenodo.13370472}, \url {https://doi.org/10.5281/zenodo.13370472}

\bibitem[\protect\citeauthoryear{King}{King}{2003}]{king_black_2003}
King A.,  2003, \mn@doi [\apjl] {10.1086/379143}, 596, L27

\bibitem[\protect\citeauthoryear{{Kingdon} \& {Ferland}}{{Kingdon} \& {Ferland}}{1996}]{kingdon1996}
{Kingdon} J.~B.,  {Ferland} G.~J.,  1996, \mn@doi [\apjs] {10.1086/192335}, \href {https://ui.adsabs.harvard.edu/abs/1996ApJS..106..205K} {106, 205}

\bibitem[\protect\citeauthoryear{{Kingdon} \& {Ferland}}{{Kingdon} \& {Ferland}}{1999}]{kingdon1999}
{Kingdon} J.~B.,  {Ferland} G.~J.,  1999, \mn@doi [\apjl] {10.1086/312008}, \href {https://ui.adsabs.harvard.edu/abs/1999ApJ...516L.107K} {516, L107}

\bibitem[\protect\citeauthoryear{Knigge \& Drew}{Knigge \& Drew}{1997}]{knigge_eclipse_1997}
Knigge C.,  Drew J.~E.,  1997, \mn@doi [\apj] {10.1086/304519}, 486, 445

\bibitem[\protect\citeauthoryear{{Knigge}, {Woods}  \& {Drew}}{{Knigge} et~al.}{1995}]{knigge95}
{Knigge} C.,  {Woods} J.~A.,   {Drew} J.~E.,  1995, \mn@doi [\mnras] {10.1093/mnras/273.2.225}, \href {https://ui.adsabs.harvard.edu/abs/1995MNRAS.273..225K} {273, 225}

\bibitem[\protect\citeauthoryear{{Knigge}, {Long}, {Wade}, {Baptista}, {Horne}, {Hubeny}  \& {Rutten}}{{Knigge} et~al.}{1998}]{knigge98}
{Knigge} C.,  {Long} K.~S.,  {Wade} R.~A.,  {Baptista} R.,  {Horne} K.,  {Hubeny} I.,   {Rutten} R.~G.~M.,  1998, \mn@doi [\apj] {10.1086/305617}, \href {http://adsabs.harvard.edu/abs/1998ApJ...499..414K} {499, 414}

\bibitem[\protect\citeauthoryear{Knigge, Scaringi, Goad  \& Cottis}{Knigge et~al.}{2008}]{knigge_intrinsic_2008}
Knigge C.,  Scaringi S.,  Goad M.~R.,   Cottis C.~E.,  2008, \mn@doi [\mnras] {10.1111/j.1365-2966.2008.13081.x}, 386, 1426

\bibitem[\protect\citeauthoryear{{Koljonen}, {Long}, {Matthews}  \& {Knigge}}{{Koljonen} et~al.}{2023}]{koljonen2023}
{Koljonen} K.~I.~I.,  {Long} K.~S.,  {Matthews} J.~H.,   {Knigge} C.,  2023, \mn@doi [\mnras] {10.1093/mnras/stad809}, \href {https://ui.adsabs.harvard.edu/abs/2023MNRAS.521.4190K} {521, 4190}

\bibitem[\protect\citeauthoryear{{Konigl} \& {Pudritz}}{{Konigl} \& {Pudritz}}{2000}]{konigl2000}
{Konigl} A.,  {Pudritz} R.~E.,  2000, in {Mannings} V.,  {Boss} A.~P.,   {Russell} S.~S.,  eds, Protostars and Planets IV. p.~759 (\mn@eprint {arXiv} {astro-ph/9903168}), \mn@doi{10.48550/arXiv.astro-ph/9903168}

\bibitem[\protect\citeauthoryear{Krolik \& Voit}{Krolik \& Voit}{1998}]{krolik_what_1998}
Krolik J.~H.,  Voit G.~M.,  1998, \mn@doi [\apjl] {10.1086/311274}, 497, L5

\bibitem[\protect\citeauthoryear{{Kromer} \& {Sim}}{{Kromer} \& {Sim}}{2009}]{kromer09}
{Kromer} M.,  {Sim} S.~A.,  2009, \mn@doi [\mnras] {10.1111/j.1365-2966.2009.15256.x}, \href {https://ui.adsabs.harvard.edu/abs/2009MNRAS.398.1809K} {398, 1809}

\bibitem[\protect\citeauthoryear{Kurosawa \& Proga}{Kurosawa \& Proga}{2009}]{kurosawa_three-dimensional_2009}
Kurosawa R.,  Proga D.,  2009, \mn@doi [\apj, Volume 693, Issue 2, pp. 1929-1945 (2009).] {10.1088/0004-637X/693/2/1929}, 693, 1929

\bibitem[\protect\citeauthoryear{{Kurosawa}, {Romanova}  \& {Harries}}{{Kurosawa} et~al.}{2011}]{kurosawa2011}
{Kurosawa} R.,  {Romanova} M.~M.,   {Harries} T.~J.,  2011, \mn@doi [\mnras] {10.1111/j.1365-2966.2011.19216.x}, \href {https://ui.adsabs.harvard.edu/abs/2011MNRAS.416.2623K} {416, 2623}

\bibitem[\protect\citeauthoryear{{Kurucz} \& {Bell}}{{Kurucz} \& {Bell}}{1995}]{kurucz1995}
{Kurucz} R.~L.,  {Bell} B.,  1995, {Atomic line list}

\bibitem[\protect\citeauthoryear{{Kusterer}, {Nagel}, {Hartmann}, {Werner}  \& {Feldmeier}}{{Kusterer} et~al.}{2014}]{kusterer2014}
{Kusterer} D.~J.,  {Nagel} T.,  {Hartmann} S.,  {Werner} K.,   {Feldmeier} A.,  2014, \mn@doi [\aap] {10.1051/0004-6361/201321438}, \href {https://ui.adsabs.harvard.edu/abs/2014A&A...561A..14K} {561, A14}

\bibitem[\protect\citeauthoryear{{Lee}, {Reynolds}, {Remillard}, {Schulz}, {Blackman}  \& {Fabian}}{{Lee} et~al.}{2002}]{lee2002}
{Lee} J.~C.,  {Reynolds} C.~S.,  {Remillard} R.,  {Schulz} N.~S.,  {Blackman} E.~G.,   {Fabian} A.~C.,  2002, \mn@doi [\apj] {10.1086/338588}, \href {https://ui.adsabs.harvard.edu/abs/2002ApJ...567.1102L} {567, 1102}

\bibitem[\protect\citeauthoryear{{Leighly}, {Dietrich}  \& {Barber}}{{Leighly} et~al.}{2011}]{leighly2011}
{Leighly} K.~M.,  {Dietrich} M.,   {Barber} S.,  2011, \mn@doi [\apj] {10.1088/0004-637X/728/2/94}, \href {https://ui.adsabs.harvard.edu/abs/2011ApJ...728...94L} {728, 94}

\bibitem[\protect\citeauthoryear{{Long} \& {Knigge}}{{Long} \& {Knigge}}{2002}]{long02}
{Long} K.~S.,  {Knigge} C.,  2002, \mn@doi [\apj] {10.1086/342879}, \href {http://adsabs.harvard.edu/abs/2002ApJ...579..725L} {579, 725}

\bibitem[\protect\citeauthoryear{Lucy}{Lucy}{1999a}]{lucy_computing_1999}
Lucy L.~B.,  1999a, \aap, 344, 282

\bibitem[\protect\citeauthoryear{Lucy}{Lucy}{1999b}]{lucy_improved_1999}
Lucy L.~B.,  1999b, \aap, 345, 211

\bibitem[\protect\citeauthoryear{Lucy}{Lucy}{2002}]{lucy_monte_2002}
Lucy L.~B.,  2002, \mn@doi [\aap] {10.1051/0004-6361:20011756}, 384, 725

\bibitem[\protect\citeauthoryear{Lucy}{Lucy}{2003}]{lucy_monte_2003}
Lucy L.~B.,  2003, \mn@doi [\aap] {10.1051/0004-6361:20030357}, 403, 261

\bibitem[\protect\citeauthoryear{{Lucy}}{{Lucy}}{2005}]{lucy2005}
{Lucy} L.~B.,  2005, \mn@doi [\aap] {10.1051/0004-6361:20041656}, \href {https://ui.adsabs.harvard.edu/abs/2005A&A...429...19L} {429, 19}

\bibitem[\protect\citeauthoryear{{Mangham}, {Knigge}, {Matthews}, {Long}, {Sim}  \& {Higginbottom}}{{Mangham} et~al.}{2017}]{mangham17}
{Mangham} S.~W.,  {Knigge} C.,  {Matthews} J.~H.,  {Long} K.~S.,  {Sim} S.~A.,   {Higginbottom} N.,  2017, \mn@doi [\mnras] {10.1093/mnras/stx1863}, \href {https://ui.adsabs.harvard.edu/abs/2017MNRAS.471.4788M} {471, 4788}

\bibitem[\protect\citeauthoryear{{Mangham} et~al.,}{{Mangham} et~al.}{2019}]{mangham19}
{Mangham} S.~W.,  et~al., 2019, \mn@doi [\mnras] {10.1093/mnras/stz1713}, \href {https://ui.adsabs.harvard.edu/abs/2019MNRAS.tmp.1671M} {p.~1671}

\bibitem[\protect\citeauthoryear{{Matthews}}{{Matthews}}{2016}]{matthews_phd_2016}
{Matthews} J.~H.,  2016, PhD thesis, University of Southampton, UK

\bibitem[\protect\citeauthoryear{{Matthews}, {Knigge}, {Long}, {Sim}  \& {Higginbottom}}{{Matthews} et~al.}{2015}]{matthews15}
{Matthews} J.~H.,  {Knigge} C.,  {Long} K.~S.,  {Sim} S.~A.,   {Higginbottom} N.,  2015, \mn@doi [\mnras] {10.1093/mnras/stv867}, \href {http://adsabs.harvard.edu/abs/2015MNRAS.450.3331M} {450, 3331}

\bibitem[\protect\citeauthoryear{{Matthews}, {Knigge}, {Long}, {Sim}, {Higginbottom}  \& {Mangham}}{{Matthews} et~al.}{2016}]{matthews16}
{Matthews} J.~H.,  {Knigge} C.,  {Long} K.~S.,  {Sim} S.~A.,  {Higginbottom} N.,   {Mangham} S.~W.,  2016, \mn@doi [\mnras] {10.1093/mnras/stw323}, \href {https://ui.adsabs.harvard.edu/abs/2016MNRAS.458..293M} {458, 293}

\bibitem[\protect\citeauthoryear{{Matthews}, {Knigge}  \& {Long}}{{Matthews} et~al.}{2017}]{matthews2017}
{Matthews} J.~H.,  {Knigge} C.,   {Long} K.~S.,  2017, \mn@doi [\mnras] {10.1093/mnras/stx231}, \href {https://ui.adsabs.harvard.edu/abs/2017MNRAS.467.2571M} {467, 2571}

\bibitem[\protect\citeauthoryear{{Matthews}, {Knigge}, {Higginbottom}, {Long}, {Sim}, {Mangham}, {Parkinson}  \& {Hewitt}}{{Matthews} et~al.}{2020}]{matthews20}
{Matthews} J.~H.,  {Knigge} C.,  {Higginbottom} N.,  {Long} K.~S.,  {Sim} S.~A.,  {Mangham} S.~W.,  {Parkinson} E.~J.,   {Hewitt} H.~A.,  2020, \mn@doi [\mnras] {10.1093/mnras/staa136}, \href {https://ui.adsabs.harvard.edu/abs/2020MNRAS.492.5540M} {492, 5540}

\bibitem[\protect\citeauthoryear{{Matthews} et~al.,}{{Matthews} et~al.}{2023}]{matthews23}
{Matthews} J.~H.,  et~al., 2023, \mn@doi [\mnras] {10.1093/mnras/stad2895}, \href {https://ui.adsabs.harvard.edu/abs/2023MNRAS.526.3967M} {526, 3967}

\bibitem[\protect\citeauthoryear{{Mazzali} \& {Lucy}}{{Mazzali} \& {Lucy}}{1993}]{mazzali93}
{Mazzali} P.~A.,  {Lucy} L.~B.,  1993, \aap, \href {https://ui.adsabs.harvard.edu/abs/1993A&A...279..447M} {279, 447}

\bibitem[\protect\citeauthoryear{McCourt, Oh, O'Leary  \& Madigan}{McCourt et~al.}{2018}]{mccourt_characteristic_2018}
McCourt M.,  Oh S.~P.,  O'Leary R.,   Madigan A.-M.,  2018, \mn@doi [\mnras] {10.1093/mnras/stx2687}, 473, 5407

\bibitem[\protect\citeauthoryear{{Middleton}, {Walton}, {Roberts}  \& {Heil}}{{Middleton} et~al.}{2014}]{middleton2014}
{Middleton} M.~J.,  {Walton} D.~J.,  {Roberts} T.~P.,   {Heil} L.,  2014, \mn@doi [\mnras] {10.1093/mnrasl/slt157}, \href {https://ui.adsabs.harvard.edu/abs/2014MNRAS.438L..51M} {438, L51}

\bibitem[\protect\citeauthoryear{{Middleton}, {Higginbottom}, {Knigge}, {Khan}  \& {Wiktorowicz}}{{Middleton} et~al.}{2022}]{middleton2022}
{Middleton} M.~J.,  {Higginbottom} N.,  {Knigge} C.,  {Khan} N.,   {Wiktorowicz} G.,  2022, \mn@doi [\mnras] {10.1093/mnras/stab2991}, \href {https://ui.adsabs.harvard.edu/abs/2022MNRAS.509.1119M} {509, 1119}

\bibitem[\protect\citeauthoryear{Mignone, Bodo, Massaglia, Matsakos, Tesileanu, Zanni  \& Ferrari}{Mignone et~al.}{2007}]{mignone_pluto:_2007}
Mignone A.,  Bodo G.,  Massaglia S.,  Matsakos T.,  Tesileanu O.,  Zanni C.,   Ferrari A.,  2007, \mn@doi [\apj Supplement Series] {10.1086/513316}, 170, 228

\bibitem[\protect\citeauthoryear{{Mihalas} \& {Mihalas}}{{Mihalas} \& {Mihalas}}{1984}]{Mihalas_Mihalas1984}
{Mihalas} D.,  {Mihalas} B.~W.,  1984, {Foundations of radiation hydrodynamics}.
Oxford University Press

\bibitem[\protect\citeauthoryear{{Miller} et~al.,}{{Miller} et~al.}{2004}]{miller2004}
{Miller} J.~M.,  et~al., 2004, \mn@doi [\apj] {10.1086/380196}, \href {https://ui.adsabs.harvard.edu/abs/2004ApJ...601..450M} {601, 450}

\bibitem[\protect\citeauthoryear{{Miller}, {Raymond}, {Fabian}, {Steeghs}, {Homan}, {Reynolds}, {van der Klis}  \& {Wijnands}}{{Miller} et~al.}{2006a}]{miller2006a}
{Miller} J.~M.,  {Raymond} J.,  {Fabian} A.,  {Steeghs} D.,  {Homan} J.,  {Reynolds} C.,  {van der Klis} M.,   {Wijnands} R.,  2006a, \mn@doi [\nat] {10.1038/nature04912}, \href {https://ui.adsabs.harvard.edu/abs/2006Natur.441..953M} {441, 953}

\bibitem[\protect\citeauthoryear{{Miller} et~al.,}{{Miller} et~al.}{2006b}]{miller2006b}
{Miller} J.~M.,  et~al., 2006b, \mn@doi [\apj] {10.1086/504673}, \href {https://ui.adsabs.harvard.edu/abs/2006ApJ...646..394M} {646, 394}

\bibitem[\protect\citeauthoryear{{Milliner}, {Matthews}, {Long}  \& {Hartmann}}{{Milliner} et~al.}{2019}]{milliner19}
{Milliner} K.,  {Matthews} J.~H.,  {Long} K.~S.,   {Hartmann} L.,  2019, \mn@doi [\mnras] {10.1093/mnras/sty3197}, \href {https://ui.adsabs.harvard.edu/abs/2019MNRAS.483.1663M} {483, 1663}

\bibitem[\protect\citeauthoryear{{Morganti}}{{Morganti}}{2017}]{morganti2017}
{Morganti} R.,  2017, \mn@doi [Frontiers in Astronomy and Space Sciences] {10.3389/fspas.2017.00042}, \href {https://ui.adsabs.harvard.edu/abs/2017FrASS...4...42M} {4, 42}

\bibitem[\protect\citeauthoryear{{Morris}}{{Morris}}{1988}]{morris1988}
{Morris} S.~L.,  1988, \mn@doi [\apjl] {10.1086/185210}, \href {https://ui.adsabs.harvard.edu/abs/1988ApJ...330L..83M} {330, L83}

\bibitem[\protect\citeauthoryear{{Mu{\~n}oz-Darias} et~al.,}{{Mu{\~n}oz-Darias} et~al.}{2016}]{munoz2016}
{Mu{\~n}oz-Darias} T.,  et~al., 2016, \mn@doi [\nat] {10.1038/nature17446}, \href {https://ui.adsabs.harvard.edu/abs/2016Natur.534...75M} {534, 75}

\bibitem[\protect\citeauthoryear{{Mu{\~n}oz-Darias}, {Torres}  \& {Garcia}}{{Mu{\~n}oz-Darias} et~al.}{2018}]{munoz2018}
{Mu{\~n}oz-Darias} T.,  {Torres} M. A.~P.,   {Garcia} M.~R.,  2018, \mn@doi [\mnras] {10.1093/mnras/sty1711}, \href {https://ui.adsabs.harvard.edu/abs/2018MNRAS.479.3987M} {479, 3987}

\bibitem[\protect\citeauthoryear{{Mu{\~n}oz-Darias} et~al.,}{{Mu{\~n}oz-Darias} et~al.}{2019}]{munoz2019}
{Mu{\~n}oz-Darias} T.,  et~al., 2019, \mn@doi [\apjl] {10.3847/2041-8213/ab2768}, \href {https://ui.adsabs.harvard.edu/abs/2019ApJ...879L...4M} {879, L4}

\bibitem[\protect\citeauthoryear{Murray, Chiang, Grossman  \& Voit}{Murray et~al.}{1995}]{murray_accretion_1995}
Murray N.,  Chiang J.,  Grossman S.~A.,   Voit G.~M.,  1995, \mn@doi [\apj] {10.1086/176238}, 451, 498

\bibitem[\protect\citeauthoryear{{Noebauer} \& {Sim}}{{Noebauer} \& {Sim}}{2019}]{noebauer19}
{Noebauer} U.~M.,  {Sim} S.~A.,  2019, \mn@doi [Living Reviews in Computational Astrophysics] {10.1007/s41115-019-0004-9}, \href {https://ui.adsabs.harvard.edu/abs/2019LRCA....5....1N} {5, 1}

\bibitem[\protect\citeauthoryear{{Noebauer}, {Long}, {Sim}  \& {Knigge}}{{Noebauer} et~al.}{2010}]{noebauer10}
{Noebauer} U.~M.,  {Long} K.~S.,  {Sim} S.~A.,   {Knigge} C.,  2010, \mn@doi [\apj] {10.1088/0004-637X/719/2/1932}, \href {http://adsabs.harvard.edu/abs/2010ApJ...719.1932N} {719, 1932}

\bibitem[\protect\citeauthoryear{Oskinova, Hamann  \& Feldmeier}{Oskinova et~al.}{2008}]{oskinova_x-raying_2008}
Oskinova L.~M.,  Hamann W.-R.,   Feldmeier A.,  2008, arXiv:0806.2348 [astro-ph]

\bibitem[\protect\citeauthoryear{Owocki, Holzer  \& Hundhausen}{Owocki et~al.}{1983}]{owocki_solar_1983}
Owocki S.~P.,  Holzer T.~E.,   Hundhausen A.~J.,  1983, \mn@doi [\apj] {10.1086/161538}, 275, 354

\bibitem[\protect\citeauthoryear{{Pancoast}, {Brewer}  \& {Treu}}{{Pancoast} et~al.}{2011}]{pancoast2011}
{Pancoast} A.,  {Brewer} B.~J.,   {Treu} T.,  2011, \mn@doi [\apj] {10.1088/0004-637X/730/2/139}, \href {https://ui.adsabs.harvard.edu/abs/2011ApJ...730..139P} {730, 139}

\bibitem[\protect\citeauthoryear{{Pancoast}, {Brewer}  \& {Treu}}{{Pancoast} et~al.}{2014}]{pancoast2014}
{Pancoast} A.,  {Brewer} B.~J.,   {Treu} T.,  2014, \mn@doi [\mnras] {10.1093/mnras/stu1809}, \href {https://ui.adsabs.harvard.edu/abs/2014MNRAS.445.3055P} {445, 3055}

\bibitem[\protect\citeauthoryear{{Parkin} \& {Sim}}{{Parkin} \& {Sim}}{2013}]{parkin2013}
{Parkin} E.~R.,  {Sim} S.~A.,  2013, \mn@doi [\apj] {10.1088/0004-637X/767/2/114}, \href {https://ui.adsabs.harvard.edu/abs/2013ApJ...767..114P} {767, 114}

\bibitem[\protect\citeauthoryear{Parkinson, Knigge, Long, Matthews, Higginbottom, Sim  \& Hewitt}{Parkinson et~al.}{2020}]{parkinson2020}
Parkinson E.~J.,  Knigge C.,  Long K.~S.,  Matthews J.~H.,  Higginbottom N.,  Sim S.~A.,   Hewitt H.~A.,  2020, \mn@doi [\mnras] {10.1093/mnras/staa1060}, 494, 4914

\bibitem[\protect\citeauthoryear{Parkinson, Knigge, Matthews, Long, Higginbottom, Sim  \& Mangham}{Parkinson et~al.}{2021}]{parkinson2021}
Parkinson E.~J.,  Knigge C.,  Matthews J.~H.,  Long K.~S.,  Higginbottom N.,  Sim S.~A.,   Mangham S.~W.,  2021, \mnras, 1, 1

\bibitem[\protect\citeauthoryear{{Parkinson}, {Knigge}, {Dai}, {Thomsen}, {Matthews}  \& {Long}}{{Parkinson} et~al.}{2024}]{parkinson2024}
{Parkinson} E.~J.,  {Knigge} C.,  {Dai} L.,  {Thomsen} L.~L.,  {Matthews} J.~H.,   {Long} K.~S.,  2024, \mn@doi [arXiv e-prints] {10.48550/arXiv.2408.16371}, \href {https://ui.adsabs.harvard.edu/abs/2024arXiv240816371P} {p. arXiv:2408.16371}

\bibitem[\protect\citeauthoryear{{Peest}, {Camps}, {Stalevski}, {Baes}  \& {Siebenmorgen}}{{Peest} et~al.}{2017}]{peest2017}
{Peest} C.,  {Camps} P.,  {Stalevski} M.,  {Baes} M.,   {Siebenmorgen} R.,  2017, \mn@doi [\aap] {10.1051/0004-6361/201630157}, \href {https://ui.adsabs.harvard.edu/abs/2017A&A...601A..92P} {601, A92}

\bibitem[\protect\citeauthoryear{Ponti, Fender, Begelman, Dunn, Neilsen  \& Coriat}{Ponti et~al.}{2012}]{ponti_ubiquitous_2012}
Ponti G.,  Fender R.~P.,  Begelman M.~C.,  Dunn R. J.~H.,  Neilsen J.,   Coriat M.,  2012, \mn@doi [\mnras] {10.1111/j.1745-3933.2012.01224.x}, 422, L11

\bibitem[\protect\citeauthoryear{Pounds, Reeves, King, Page, O'Brien  \& Turner}{Pounds et~al.}{2003}]{pounds_high-velocity_2003}
Pounds K.~A.,  Reeves J.~N.,  King A.~R.,  Page K.~L.,  O'Brien P.~T.,   Turner M. J.~L.,  2003, \mn@doi [\mnras] {10.1046/j.1365-8711.2003.07006.x}, 345, 705

\bibitem[\protect\citeauthoryear{{Proga} \& {Kallman}}{{Proga} \& {Kallman}}{2002}]{proga02}
{Proga} D.,  {Kallman} T.~R.,  2002, \mn@doi [\apj] {10.1086/324534}, \href {http://adsabs.harvard.edu/abs/2002ApJ...565..455P} {565, 455}

\bibitem[\protect\citeauthoryear{{Puls}, {Vink}  \& {Najarro}}{{Puls} et~al.}{2008}]{puls2008}
{Puls} J.,  {Vink} J.~S.,   {Najarro} F.,  2008, \mn@doi [\aapr] {10.1007/s00159-008-0015-8}, \href {https://ui.adsabs.harvard.edu/abs/2008A&ARv..16..209P} {16, 209}

\bibitem[\protect\citeauthoryear{{Rankine}, {Hewett}, {Banerji}  \& {Richards}}{{Rankine} et~al.}{2020}]{rankine2020}
{Rankine} A.~L.,  {Hewett} P.~C.,  {Banerji} M.,   {Richards} G.~T.,  2020, \mn@doi [\mnras] {10.1093/mnras/staa130}, \href {https://ui.adsabs.harvard.edu/abs/2020MNRAS.492.4553R} {492, 4553}

\bibitem[\protect\citeauthoryear{Reeves, O'Brien  \& Ward}{Reeves et~al.}{2003}]{reeves_massive_2003}
Reeves J.~N.,  O'Brien P.~T.,   Ward M.~J.,  2003, \mn@doi [\apj, Volume 593, Issue 2, pp. L65-L68.] {10.1086/378218}, 593, L65

\bibitem[\protect\citeauthoryear{{\VAN{Regemorter}{Van}{van} Regemorter}}{{\VAN{Regemorter}{Van}{van} Regemorter}}{1962}]{vanregemorter62}
{\VAN{Regemorter}{Van}{van} Regemorter} H.,  1962, \mn@doi [\apj] {10.1086/147445}, \href {https://ui.adsabs.harvard.edu/abs/1962ApJ...136..906V} {136, 906}

\bibitem[\protect\citeauthoryear{Reichard et~al.,}{Reichard et~al.}{2003}]{reichard_continuum_2003}
Reichard T.~A.,  et~al., 2003, \mn@doi [The Astronomical Journal] {10.1086/379293}, 126, 2594

\bibitem[\protect\citeauthoryear{{Richards} et~al.,}{{Richards} et~al.}{2011}]{richards2011}
{Richards} G.~T.,  et~al., 2011, \mn@doi [\aj] {10.1088/0004-6256/141/5/167}, \href {https://ui.adsabs.harvard.edu/abs/2011AJ....141..167R} {141, 167}

\bibitem[\protect\citeauthoryear{Robitaille}{Robitaille}{2011}]{robitaille_hyperion_2011}
Robitaille T.~P.,  2011, \mn@doi [åp] {10.1051/0004-6361/201117150}, 536, A79

\bibitem[\protect\citeauthoryear{Roth \& Kasen}{Roth \& Kasen}{2018}]{roth2018}
Roth N.,  Kasen D.,  2018, \mn@doi [ApJ] {10.3847/1538-4357/aaaec6}, 855, 54

\bibitem[\protect\citeauthoryear{Roth, Rossi, Krolik, Piran, Mockler  \& Kasen}{Roth et~al.}{2020}]{roth2020}
Roth N.,  Rossi E.~M.,  Krolik J.,  Piran T.,  Mockler B.,   Kasen D.,  2020, \mn@doi [Space Sci Rev] {10.1007/s11214-020-00735-1}, 216, 114

\bibitem[\protect\citeauthoryear{{Rybicki} \& {Hummer}}{{Rybicki} \& {Hummer}}{1983}]{rybicki83}
{Rybicki} G.~B.,  {Hummer} D.~G.,  1983, \mn@doi [\apj] {10.1086/161454}, \href {https://ui.adsabs.harvard.edu/abs/1983ApJ...274..380R} {274, 380}

\bibitem[\protect\citeauthoryear{Sander}{Sander}{2024}]{sander2024}
Sander A. A.~C.,  2024, Hot Star Winds (\mn@eprint {arXiv} {2410.12484}), \url {https://arxiv.org/abs/2410.12484}

\bibitem[\protect\citeauthoryear{{Sander}, {Shenar}, {Hainich}, {G{\'\i}menez-Garc{\'\i}a}, {Todt}  \& {Hamann}}{{Sander} et~al.}{2015}]{sander2015}
{Sander} A.,  {Shenar} T.,  {Hainich} R.,  {G{\'\i}menez-Garc{\'\i}a} A.,  {Todt} H.,   {Hamann} W.~R.,  2015, \mn@doi [\aap] {10.1051/0004-6361/201425356}, \href {https://ui.adsabs.harvard.edu/abs/2015A&A...577A..13S} {577, A13}

\bibitem[\protect\citeauthoryear{{Scepi}, {Dubus}  \& {Lesur}}{{Scepi} et~al.}{2019}]{scepi2019}
{Scepi} N.,  {Dubus} G.,   {Lesur} G.,  2019, \mn@doi [\aap] {10.1051/0004-6361/201834781}, \href {https://ui.adsabs.harvard.edu/abs/2019A&A...626A.116S} {626, A116}

\bibitem[\protect\citeauthoryear{Shakura \& Sunyaev}{Shakura \& Sunyaev}{1973}]{shakura_black_1973}
Shakura N.~I.,  Sunyaev R.~A.,  1973, \aap, 24, 337

\bibitem[\protect\citeauthoryear{{Shingles} et~al.,}{{Shingles} et~al.}{2020}]{shingles2020}
{Shingles} L.~J.,  et~al., 2020, \mn@doi [\mnras] {10.1093/mnras/stz3412}, \href {https://ui.adsabs.harvard.edu/abs/2020MNRAS.492.2029S} {492, 2029}

\bibitem[\protect\citeauthoryear{{Shlosman} \& {Vitello}}{{Shlosman} \& {Vitello}}{1993}]{shlosman93}
{Shlosman} I.,  {Vitello} P.,  1993, \mn@doi [\apj] {10.1086/172670}, \href {https://ui.adsabs.harvard.edu/abs/1993ApJ...409..372S} {409, 372}

\bibitem[\protect\citeauthoryear{Silk \& Rees}{Silk \& Rees}{1998}]{silk_quasars_1998}
Silk J.,  Rees M.~J.,  1998, \aap, 331, L1

\bibitem[\protect\citeauthoryear{Sim}{Sim}{2005}]{sim_modelling_2005}
Sim S.~A.,  2005, \mn@doi [\mnras] {10.1111/j.1365-2966.2004.08471.x}, 356, 531

\bibitem[\protect\citeauthoryear{{Sim}, {Drew}  \& {Long}}{{Sim} et~al.}{2005}]{sim05}
{Sim} S.~A.,  {Drew} J.~E.,   {Long} K.~S.,  2005, \mn@doi [\mnras] {10.1111/j.1365-2966.2005.09472.x}, \href {https://ui.adsabs.harvard.edu/abs/2005MNRAS.363..615S} {363, 615}

\bibitem[\protect\citeauthoryear{Sim, Long, Miller  \& Turner}{Sim et~al.}{2008}]{sim_multidimensional_2008}
Sim S.~A.,  Long K.~S.,  Miller L.,   Turner T.~J.,  2008, \mn@doi [\mnras , Volume 388, Issue 2, pp. 611-624.] {10.1111/j.1365-2966.2008.13466.x}, 388, 611

\bibitem[\protect\citeauthoryear{Sim, Proga, Miller, Long  \& Turner}{Sim et~al.}{2010}]{sim_multidimensional_2010}
Sim S.~A.,  Proga D.,  Miller L.,  Long K.~S.,   Turner T.~J.,  2010, \mn@doi [\mnras] {10.1111/j.1365-2966.2010.17215.x}, 408, 1396

\bibitem[\protect\citeauthoryear{Sim, Proga, Kurosawa, Long, Miller  \& Turner}{Sim et~al.}{2012}]{sim_synthetic_2012}
Sim S.~A.,  Proga D.,  Kurosawa R.,  Long K.~S.,  Miller L.,   Turner T.~J.,  2012, \mn@doi [\mnras] {10.1111/j.1365-2966.2012.21816.x}, 426, 2859

\bibitem[\protect\citeauthoryear{Sobolev}{Sobolev}{1957}]{sobolev_diffusion_1957}
Sobolev V.~V.,  1957, \sovast, 1, 678

\bibitem[\protect\citeauthoryear{Sobolev}{Sobolev}{1960}]{sobolev_moving_1960}
Sobolev V.~V.,  1960, Moving envelopes of stars

\bibitem[\protect\citeauthoryear{{Stone} \& {Norman}}{{Stone} \& {Norman}}{1992}]{zeus}
{Stone} J.~M.,  {Norman} M.~L.,  1992, \mn@doi [\apjs] {10.1086/191680}, \href {https://ui.adsabs.harvard.edu/abs/1992ApJS...80..753S} {80, 753}

\bibitem[\protect\citeauthoryear{{Strubbe} \& {Quataert}}{{Strubbe} \& {Quataert}}{2009}]{Strubbe2009}
{Strubbe} L.~E.,  {Quataert} E.,  2009, \mn@doi [MNRAS] {10.1111/j.1365-2966.2009.15599.x}, \href {https://ui.adsabs.harvard.edu/abs/2009MNRAS.400.2070S} {400, 2070}

\bibitem[\protect\citeauthoryear{Sutherland}{Sutherland}{1998a}]{sutherland_accurate_1998}
Sutherland R.~S.,  1998a, \mn@doi [\mnras] {10.1046/j.1365-8711.1998.01687.x}, 300, 321

\bibitem[\protect\citeauthoryear{{Sutherland}}{{Sutherland}}{1998b}]{sutherland98}
{Sutherland} R.~S.,  1998b, \mn@doi [\mnras] {10.1046/j.1365-8711.1998.01687.x}, \href {https://ui.adsabs.harvard.edu/abs/1998MNRAS.300..321S} {300, 321}

\bibitem[\protect\citeauthoryear{{Tampo}, {Knigge}, {Long}, {Matthews}  \& {Castro Segura}}{{Tampo} et~al.}{2024}]{tampo2024}
{Tampo} Y.,  {Knigge} C.,  {Long} K.~S.,  {Matthews} J.~H.,   {Castro Segura} N.,  2024, \mn@doi [arXiv e-prints] {10.48550/arXiv.2406.14396}, \href {https://ui.adsabs.harvard.edu/abs/2024arXiv240614396T} {p. arXiv:2406.14396}

\bibitem[\protect\citeauthoryear{{Tanaka} \& {Hotokezaka}}{{Tanaka} \& {Hotokezaka}}{2013}]{tanaka2013}
{Tanaka} M.,  {Hotokezaka} K.,  2013, \mn@doi [\apj] {10.1088/0004-637X/775/2/113}, \href {https://ui.adsabs.harvard.edu/abs/2013ApJ...775..113T} {775, 113}

\bibitem[\protect\citeauthoryear{{Temple} et~al.,}{{Temple} et~al.}{2023}]{temple2023}
{Temple} M.~J.,  et~al., 2023, \mn@doi [\mnras] {10.1093/mnras/stad1448}, \href {https://ui.adsabs.harvard.edu/abs/2023MNRAS.523..646T} {523, 646}

\bibitem[\protect\citeauthoryear{Tomaru, Done, Ohsuga, Nomura  \& Takahashi}{Tomaru et~al.}{2019}]{tomaru_thermal-radiative_2019}
Tomaru R.,  Done C.,  Ohsuga K.,  Nomura M.,   Takahashi T.,  2019, arXiv e-prints, p. arXiv:1905.11763

\bibitem[\protect\citeauthoryear{{Tomaru}, {Done}  \& {Mao}}{{Tomaru} et~al.}{2023}]{tomaru2023}
{Tomaru} R.,  {Done} C.,   {Mao} J.,  2023, \mn@doi [\mnras] {10.1093/mnras/stac3210}, \href {https://ui.adsabs.harvard.edu/abs/2023MNRAS.518.1789T} {518, 1789}

\bibitem[\protect\citeauthoryear{Trump et~al.,}{Trump et~al.}{2006}]{trump_catalog_2006}
Trump J.~R.,  et~al., 2006, \mn@doi [\apjs] {10.1086/503834}, 165, 1

\bibitem[\protect\citeauthoryear{Turing}{Turing}{1948}]{turing}
Turing A.~M.,  1948, \mn@doi [The Quarterly Journal of Mechanics and Applied Mathematics] {10.1093/qjmam/1.1.287}, 1, 287

\bibitem[\protect\citeauthoryear{{Turnshek}}{{Turnshek}}{1984}]{turnshek1984}
{Turnshek} D.~A.,  1984, \mn@doi [\apj] {10.1086/161967}, \href {https://ui.adsabs.harvard.edu/abs/1984ApJ...280...51T} {280, 51}

\bibitem[\protect\citeauthoryear{{Verner} \& {Yakovlev}}{{Verner} \& {Yakovlev}}{1995}]{verner95}
{Verner} D.~A.,  {Yakovlev} D.~G.,  1995, \aaps, \href {https://ui.adsabs.harvard.edu/abs/1995A&AS..109..125V} {109, 125}

\bibitem[\protect\citeauthoryear{{Verner}, {Verner}  \& {Ferland}}{{Verner} et~al.}{1996a}]{verner96_lines}
{Verner} D.~A.,  {Verner} E.~M.,   {Ferland} G.~J.,  1996a, \mn@doi [Atomic Data and Nuclear Data Tables] {10.1006/adnd.1996.0018}, \href {https://ui.adsabs.harvard.edu/abs/1996ADNDT..64....1V} {64, 1}

\bibitem[\protect\citeauthoryear{{Verner}, {Ferland}, {Korista}  \& {Yakovlev}}{{Verner} et~al.}{1996b}]{verner96}
{Verner} D.~A.,  {Ferland} G.~J.,  {Korista} K.~T.,   {Yakovlev} D.~G.,  1996b, \mn@doi [\apj] {10.1086/177435}, \href {https://ui.adsabs.harvard.edu/abs/1996ApJ...465..487V} {465, 487}

\bibitem[\protect\citeauthoryear{{Voit}, {Weymann}  \& {Korista}}{{Voit} et~al.}{1993}]{voit1993}
{Voit} G.~M.,  {Weymann} R.~J.,   {Korista} K.~T.,  1993, \mn@doi [\apj] {10.1086/172980}, \href {https://ui.adsabs.harvard.edu/abs/1993ApJ...413...95V} {413, 95}

\bibitem[\protect\citeauthoryear{Waters \& Proga}{Waters \& Proga}{2019}]{waters_cloud_2019}
Waters T.,  Proga D.,  2019, \mn@doi [\apj] {10.3847/2041-8213/ab12e8}, 876, L3

\bibitem[\protect\citeauthoryear{{Welsh} \& {Horne}}{{Welsh} \& {Horne}}{1991}]{welsh1991}
{Welsh} W.~F.,  {Horne} K.,  1991, \mn@doi [\apj] {10.1086/170530}, \href {https://ui.adsabs.harvard.edu/abs/1991ApJ...379..586W} {379, 586}

\bibitem[\protect\citeauthoryear{Weymann, Morris, Foltz  \& Hewett}{Weymann et~al.}{1991}]{weymann_comparisons_1991}
Weymann R.~J.,  Morris S.~L.,  Foltz C.~B.,   Hewett P.~C.,  1991, \mn@doi [\apj] {10.1086/170020}, 373, 23

\bibitem[\protect\citeauthoryear{Wu, Coughlin  \& Nixon}{Wu et~al.}{2018}]{wu2018}
Wu S.,  Coughlin E.~R.,   Nixon C.,  2018, \mn@doi [\mnras] {10.1093/mnras/sty971}, 478, 3016

\bibitem[\protect\citeauthoryear{{Yusef-Zadeh}, {Morris}  \& {White}}{{Yusef-Zadeh} et~al.}{1984}]{zadeh84}
{Yusef-Zadeh} F.,  {Morris} M.,   {White} R.~L.,  1984, \mn@doi [\apj] {10.1086/161780}, \href {https://ui.adsabs.harvard.edu/abs/1984ApJ...278..186Y} {278, 186}

\bibitem[\protect\citeauthoryear{{Zhang}, {Claus}, {Watson}  \& {Moran}}{{Zhang} et~al.}{2017}]{zhang2017}
{Zhang} Q.,  {Claus} B.,  {Watson} L.,   {Moran} J.,  2017, \mn@doi [\apj] {10.3847/1538-4357/aa5ea9}, \href {https://ui.adsabs.harvard.edu/abs/2017ApJ...837...53Z} {837, 53}

\bibitem[\protect\citeauthoryear{van Velzen et~al.,}{van Velzen et~al.}{2021}]{vanVelzen2020}
van Velzen S.,  et~al., 2021, \mn@doi [ApJ] {10.3847/1538-4357/abc258}, 908, 4

\makeatother
\end{thebibliography}
\end{document}